\documentclass{nature_incl_fig}

\usepackage[utf8]{inputenc}
\usepackage[T1]{fontenc}
\usepackage{sansmath,amssymb, amsmath, graphicx, grffile}

\renewcommand\thefigure{Figure \arabic{figure}}
\renewcommand\thetable{Table \arabic{table}}
\usepackage{nameref} 

\usepackage[dvipsnames]{xcolor}
\usepackage[colorlinks=true,citecolor=blue,linkcolor=blue,urlcolor=blue,breaklinks,hypertexnames=false]{hyperref} 

\vspace{-5ex}\title{Chromatic periodic activity down to 120\,MHz in a Fast \mbox{Radio} Burst}

\author{%
In{\'e}s~Pastor-Marazuela$^{1, 2}$,
Liam~Connor$^{1, 2, 3}$,
Joeri~van~Leeuwen$^{2, 1 *}$,
Yogesh~Maan$^{2}$,
Sander~ter~Veen$^{2}$,
Anna~Bilous$^{2}$,
Leon~Oostrum$^{2, 1}$,
Emily~Petroff$^{1, 4}$,
Samayra~Straal$^{5, 6}$,
Dany~Vohl$^{2}$,
Jisk~Attema$^{7}$,
Oliver~M.~Boersma$^{1, 2}$,
Eric~Kooistra$^{2}$,
Daniel~van~der~Schuur$^{2}$,
Alessio~Sclocco$^{7}$,
Roy~Smits$^{2}$,
Elizabeth~A.~K.~Adams$^{2, 8}$,
Bj{\"o}rn~Adebahr$^{9}$,
W.~J.~G.~de~Blok$^{2, 10, 8}$,
Arthur~H.~W.~M.~Coolen$^{2}$,
Sieds~Damstra$^{2}$,
~Helga~D{\'e}nes~$^{2}$,
~Kelley~M.~Hess$^{2, 8}$,
~Thijs~van~der~Hulst$^{8}$,
~Boudewijn~Hut$^{2}$, 
\mbox{V.~Marianna} Ivashina$^{11}$,
Alexander~Kutkin$^{2, 12}$,
G.~Marcel~Loose$^{2}$,
Danielle~M.~Lucero$^{13}$,
{\'A}gnes~Mika$^{2}$,
Vanessa~A. Moss$^{14, 15, 2}$,
Henk~Mulder$^{2}$,
Menno~J.~Norden$^{2}$,
Tom~Oosterloo$^{2, 8}$,
Emanuela~Orr{\'u}$^{2}$,
Mark~Ruiter$^{2}$
and
Stefan~J.~Wijnholds$^{2}$
}

\newcommand{\FRB}{FRB\,20180916B }
\newcommand{\FRBnospace}{FRB\,20180916B}

\newcommand{\RI}{FRB\,20121102A }
\newcommand{\RInospace}{FRB\,20121102A}
\newcommand{\errors}[2]{$^{+#1}_{-  #2}$}

\newcommand{\pccm}{\,pc\,cm$^{-3}$ }
\newcommand{\pccmnospace}{\,pc\,cm$^{-3}$}
\newcommand{\jyms}{\,Jy\,ms }
\newcommand{\jymsnospace}{\,Jy\,ms}
\newcommand{\mhzms}{\,MHz\,ms$^{-1}$}
\newcommand{\tscat}{$\tau_{\mathrm{sc}}$}
\newcommand{\skyday}{\,sky$^{-1}$\,day$^{-1}$ }

\newcommand{\tdm}{$\tau_{\mathrm{DM}}$ }
\newcommand{\kdm}{$k_{\text{DM}}$}
\newcommand{\ukdm}{\,GHz$^2$cm$^3$pc$^{-1}$ms}

\newcommand{\dmsnrnospace}{$\mathrm{DM}_{\mathrm{S/N}}$}
\newcommand{\dmstruct}{$\mathrm{DM}_{\mathrm{struct}}$ }
\newcommand{\dml}{$\text{DM}_{\text{LOFAR}}$ }

\newcommand{\dma}{$\text{DM}_{\text{Apertif}}$ }
\newcommand{\dmanospace}{$\text{DM}_{\text{Apertif}}$}

\providecommand\JournalTitle[1]{#1} 

\newcommand{\apj}{The Astrophysical Journal}
\newcommand{\apjl}{The Astrophysical Journal Letters}

\iftrue  
  \def\ipm#1{\textcolor{teal}{\bf[#1 -- IPM]}}
  \def\jvl#1{\textcolor{MidnightBlue}{\bf[#1 -- JVL]}}
  \def\lco#1{\textcolor{orange}{\bf[#1 -- LCO]}}
  \def\sms#1{\textcolor{magenta}{\bf[#1 -- SMS]}}
  \def\ym#1{\textcolor{Salmon}{\bf[#1 -- YM]}}
  \def\ab#1{\textcolor{Salmon}{\bf[#1 -- AB]}}
  \def\stv#1{\textcolor{LimeGreen}{\bf[#1 -- StV]}}
  \def\lc#1{\textcolor{LimeGreen}{\bf[#1 -- Liam]}}
  \def\dv#1{\textcolor{blue}{\bf[#1 -- Dany]}}
  \def\ep#1{\textcolor{red}{\bf[#1 -- Emily]}}

\else
  \def\ipm#1{}
  \def\jvl#1{}
  \def\lco#1{}
  \def\sms#1{}
  \def\ym#1{}
  \def\ab#1{}
  \def\stv#1{}
  \def\lc#1{}
  \def\dv#1{}
  \def\ep#1{}
  \renewcommand{\textbf}[1]{{#1}}
\fi

\begin{document}
\maketitle

\begin{affiliations}
\item  Anton Pannekoek Institute, University of Amsterdam, Postbus 94249, 1090 GE Amsterdam, The Netherlands 
\item  ASTRON, the Netherlands Institute for Radio Astronomy, Oude Hoogeveensedijk 4, 7991 PD Dwingeloo, The Netherlands 
\item  Cahill Center for Astronomy, California Institute of Technology, Pasadena, CA, USA 
\item  Veni Fellow 
\item  NYU Abu Dhabi, PO Box 129188, Abu Dhabi, United Arab Emirates 
\item  Center for Astro, Particle, and Planetary Physics (CAP$^3$), NYU Abu Dhabi, PO Box 129188, Abu Dhabi, United Arab Emirates 
\item  Netherlands eScience Center, Science Park 140, 1098 XG, Amsterdam, The Netherlands 
\item  Kapteyn Astronomical Institute, PO Box 800, 9700 AV Groningen, The Netherlands 
\item  Astronomisches Institut der Ruhr-Universit{\"a}t Bochum (AIRUB), Universit{\"a}tsstrasse 150, 44780 Bochum, Germany 
\item  Dept.\ of Astronomy, Univ.\ of Cape Town, Private Bag X3, Rondebosch 7701, South Africa 
\item  Dept.\ of Electrical Engineering, Chalmers University of Technology, Gothenburg, Sweden 
\item  \resizebox{\textwidth}{!}{Astro Space Center of Lebedev Physical Institute, Profsoyuznaya Str. 84/32, 117997 Moscow, Russia}
\item  Department of Physics, Virginia Polytechnic Institute and State University, 50 West Campus Drive, Blacksburg, VA 24061, USA 
\item  CSIRO Astronomy and Space Science, Australia Telescope National Facility, PO Box 76, Epping NSW 1710, Australia 
\item  Sydney Institute for Astronomy, School of Physics, University of Sydney, Sydney, New South Wales 2006, Australia 
\item[*] ~{\normalfont email: leeuwen@astron.nl}
\end{affiliations}

\newpage

\begin{abstract}
Fast radio bursts (FRBs) are extragalactic astrophysical transients\cite{lorimer_bright_2007}
whose brightness requires emitters that are highly energetic,
yet compact enough to produce the short, millisecond-duration bursts.
FRBs have thus far been detected between 300\,MHz\cite{chawla_detection_2020} and 8\,GHz\cite{gajjar_highest-frequency_2018}, 
but lower-frequency emission has remained elusive.
A subset of FRBs is known to repeat, and one of those sources, FRB\,20180916B, does so with a 16.3\,day activity period\cite{the_chimefrb_collaboration_periodic_2020}. Using simultaneous Apertif and LOFAR data, we show that \FRB emits down to 120\,MHz,
and that its activity window is both narrower and earlier at higher frequencies.
Binary wind interaction models predict a narrower periodic activity window 
at lower frequencies, which is the opposite of our observations. 
Our detections establish that low-frequency FRB emission can escape the local medium. For bursts of the same fluence,
\FRB is more active below 200\,MHz than at 1.4\,GHz. Combining our results with previous upper-limits on the all-sky FRB
rate at 150\,MHz,
we find that there are 3--450 FRBs\skyday above 50\,Jy\,ms at 90$\%$ confidence. We are able to
rule out the scenario in which companion winds cause FRB periodicity.  
We also demonstrate that some FRBs live in 
clean environments that do not absorb or scatter 
low-frequency radiation.
\end{abstract}


Among the $\sim$20 currently known repeating FRB sources\cite{spitler_repeating_2016,the_chimefrb_collaboration_second_2019, fonseca_nine_2020},
\FRBnospace\cite{the_chimefrb_collaboration_chimefrb_2019} is one of the most active.
This activity allowed follow-up localisation, and \FRB (also known as FRB\,180916.J0158+65) was found to reside in a spiral galaxy at a luminosity distance of
149\,Mpc\cite{marcote_repeating_2020}. The FRB displays periodicity, with activity cycles of
$\sim$$16.35$\,days\cite{the_chimefrb_collaboration_periodic_2020}.
Based on this discovery,  numerous instruments observed  \FRB
at the predicted peak activity days\cite{pilia_lowest_2020}. The source was
detected in radio from  2\,GHz down to 300\,MHz, but not below\cite{chawla_detection_2020}.

\begin{figure}
\centering
\includegraphics[width=0.7\linewidth]{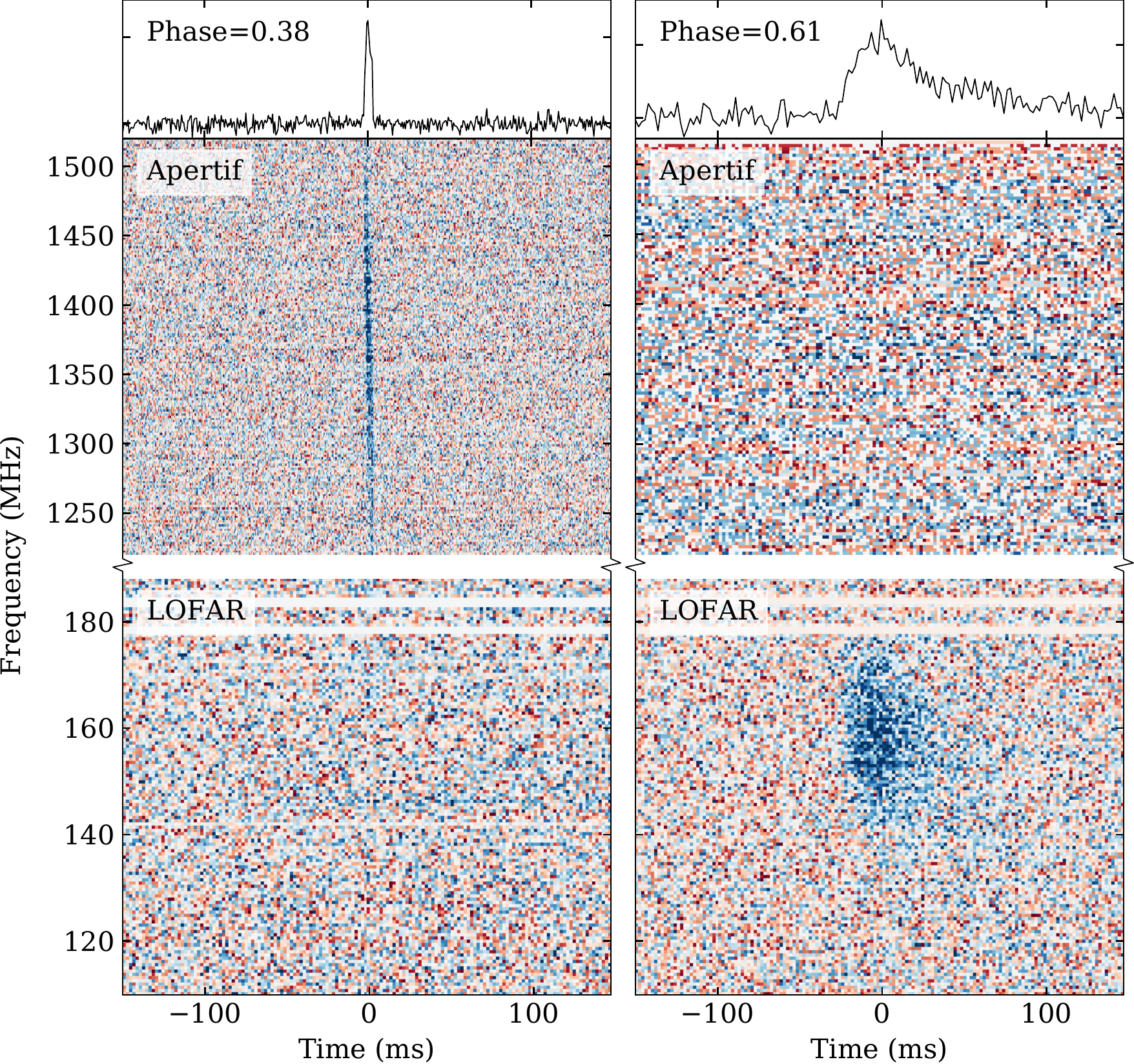}\\
\caption{Composite dynamic spectra of bursts A13 and L06 at Apertif and LOFAR frequency bands. The burst on the left
  (A13) was detected on MJD 58980.52593337 (barycentric), corresponding to a 0.38 activity phase. It is detected only in
  Apertif's frequency band, with no emission below 190\,MHz. The burst on the right (L06) was detected on MJD
  58951.5416274, corresponding to a 0.61 activity phase. It is only detected in the LOFAR frequency band. The top panel of each burst shows their respective pulse profiles. 
\label{fig:arts_lofar_dynspec}
}
\end{figure}

We observed \FRB simultaneously with the Westerbork and LOFAR radio telescopes, and detected multiple bursts with both. 
At the  Westerbork Synthesis Radio Telescope (WSRT),
we used the Apertif Radio Transient System (ARTS\cite{2017arXiv170906104M}) between 1220\,MHz and 1520\,MHz for
388.4\,h. We covered seven activity cycles. 
We recorded 57.6\,h of simultaneous observations with the LOw Frequency ARrray (LOFAR\cite{sha+11}) between 110\,MHz and
190\,MHz during the predicted activity peak days of three activity cycles.
The LOFAR data are public and are also being analyzed independently\cite{pmb+20}.
In the 1.4\,GHz Apertif observations we detected 54 bursts,
whereas the 150\,MHz LOFAR observations led to the detection of nine
bursts. None occurred simultaneously at both frequency bands. 
\ref{fig:arts_lofar_dynspec} shows the composite dynamic spectrum at both frequency bands of an Apertif burst and a LOFAR burst detected during simultaneous observations with the two instruments. The lack of simultaneous emission at both frequencies is visible.

\begin{figure}
\centering
\includegraphics[width=\linewidth]{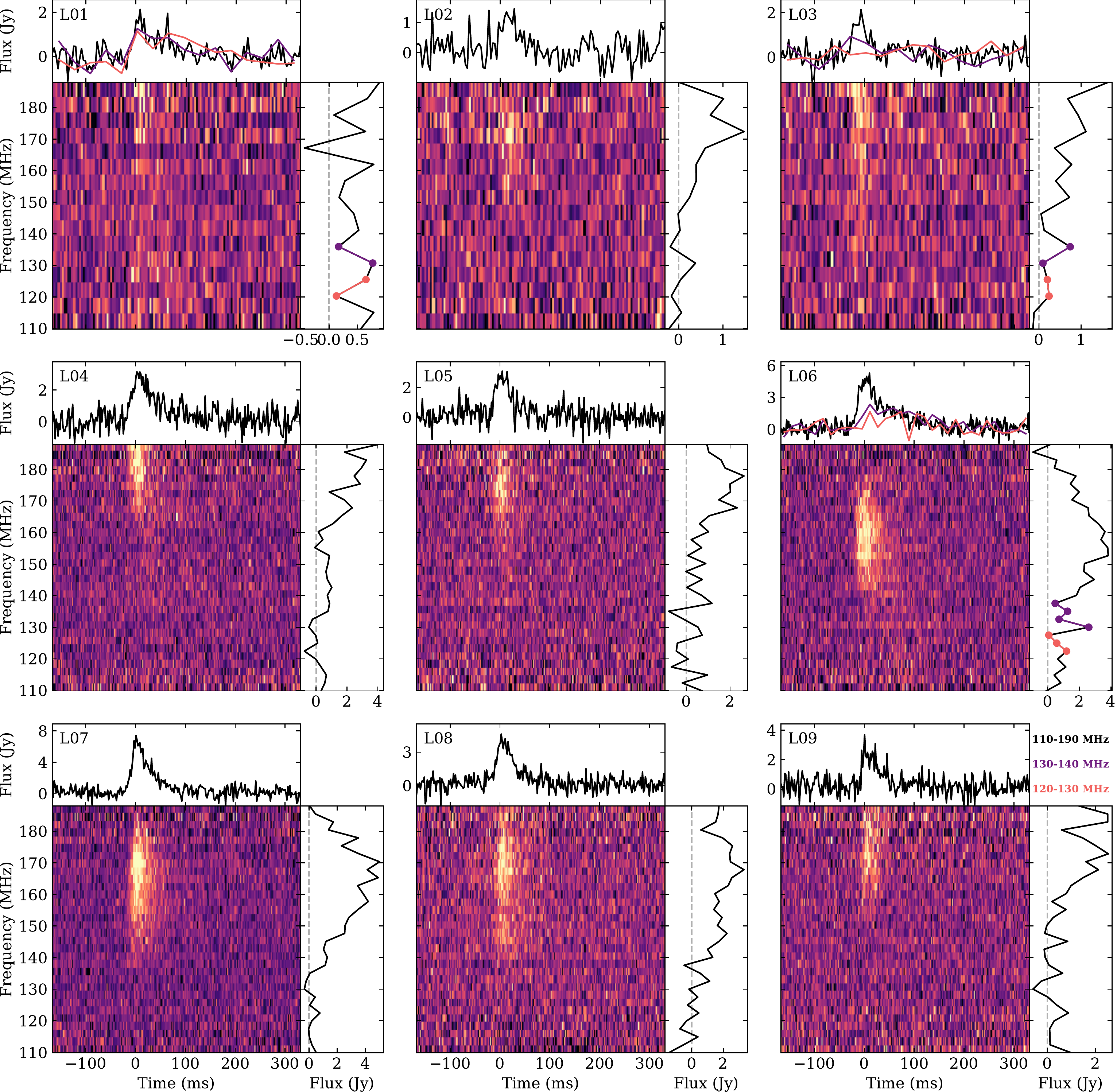}
\caption{Dynamic spectra of the nine bursts from \FRB detected with \mbox{LOFAR}, dedispersed to the S/N maximizing LOFAR DM of 349.00\pccmnospace. For each burst, the top panel shows the calibrated pulse profile, the right panel the spectrum and the bottom left panel the dynamic spectrum. In bursts L01, L03, and L06, where there is evidence of emission below 140\,MHz, the pulse profiles between 130\,MHz and 140\,MHz are plotted in purple, and between 120\,MHz and 130\,MHz are plotted in pink. The dynamic spectra were respectively downsampled by factors 2 and 64 in time and frequency for bursts L01--L03, and by factors 2 and 32 in time and frequency for bursts L04--L09.
  \label{fig:lofar_bursts}
}
\end{figure}

Previous low-frequency searches for FRBs 
have been unsuccessful, whether all-sky\cite{coenen_lofar_2014,karastergiou_limits_2015} 
or  targeted on known repeaters\cite{chawla_detection_2020}.
Such long campaigns (over a thousand hours on sky in total) 
resulted in strict limits on FRB emission below 300\,MHz.
Such upper limits fueled FRB theories in which free-free absorption around the emitter
or strong intervening  scattering was required. 
The nine LOFAR bursts at 120--190\,MHz presented here are the first FRB detections in this frequency range.
The pulse profiles, spectra and dynamic spectra of the nine bursts are presented in \ref{fig:lofar_bursts},
and their properties are summarised in \ref{tab:LOFAR_bursts}.
All had simultaneous Apertif coverage, but no bursts were detected there down to a limit of 0.5\jymsnospace.

\newpage

Remarkably, 
for bursts of the same fluence, \FRB is 
around over an order of magnitude more active at 150\,MHz
than at 1.4\,GHz, as seen in \ref{fig:fluence_cdf}.
Our detections allow for the first bounded constraints on the FRB all-sky rate below 200\,MHz. A lower limit is obtained by assuming \FRB is the only source in the sky emitting at these frequencies. Combining this with previously published upper limits, we find that there are 3--450\skyday above 50\,Jy\,ms at 90$\%$ confidence. Assuming an Euclidean fluence scaling, this is equivalent to 90--14000\skyday above 5\jyms at 150\,MHz, which is 
promising for future 
high-sensitivity low-frequency surveys.

The integrated pulse shapes of the LOFAR bursts in \ref{fig:lofar_bursts} are dominated
by a sharp rise plus a  scattering tail.
We obtain a scattering timescale \tscat=46$\pm$10\,ms at 150\,MHz,
scaling with frequency as \tscat$\propto\nu^{-4.2\pm1.1}$.
This is consistent with the frequency scintillation found for the same source at 1.7\,GHz.
For the  typical $\nu^{-4}$ scaling, the $\sim$60\,kHz
decorrelation bandwidth seen there\cite{marcote_repeating_2020}
translates to a $\sim$45\,ms scattering time 
at 150\,MHz. 
This scatter broadening may explain why none of the millisecond-duration 
frequency-time subcomponents seen at higher frequencies\cite{hessels_frb_2018}
are visible in the dynamic
spectra in \ref{fig:lofar_bursts}. 
The observed scattering time is within a factor of two of the predicted Galactic scattering\cite{cordes_ne2001i_2003}.
Thus, we attribute this 
pulse broadening to scattering in the Milky Way ISM and not plasma in the host galaxy. The fact that 
the ISM scattering is stronger in the Milky Way than in the host galaxy is not surprising, given 
\FRB is at a low galactic latitude, whereas its host is a nearly face-on spiral galaxy \cite{marcote_repeating_2020};
it is, however, notable that the environment local to the source scatters the 
FRB by $\lesssim$\,7\,$\mu$s at 1.4\,GHz. 
The dispersion measure (DM) of the bursts, \dml=349.00$\pm$0.02\pccm, is in excess of previous DM measurements of the same source, which we interpret as an 
additional hint for the presence of unresolved subcomponents and not a frequency-dependent
DM.

The dynamic
spectra in \ref{fig:lofar_bursts}
show emission
from the top of the band at 190\,MHz
down to 120\,MHz for  burst L01 and L07.
As the LOFAR sensitivity decreases towards the bottom of the band, we cannot confidently rule out the presence of emission below 120\,MHz.
Our ability to detect these bursts at frequencies this low
shows that free-free absorption and induced Compton scattering (ICS) do not significantly impact burst propagation for this source.
Below we  discuss the physical constraints for a number of models in greater detail.
Combining these results with the small local RM and DM contribution, as well as the lack of temporal scattering, we do conclude here
that some FRBs reside in clean environments, which is a prerequisite for some FRB applications to cosmology\cite{mcquinn-2014}.\\

In our Apertif campaign, we detected 54 bursts.
Ten of these had LOFAR coverage, but no bursts were detected there down to a fluence of 30\jymsnospace.
Half of the Apertif bursts
have stored polarisation data, giving us access to the Stokes parameters and the polarisation position angle (PA) at multiple cycle phases. The pulse profiles, dynamic spectra of total intensity data and PA, where available, of all 1.4\,GHz bursts are shown in \ref{fig:Apertif_bursts}, and their properties are summarised in \ref{tab:Apertif_bursts}. All bursts are $\sim$100\% linearly polarised, and the PA is constant within a single burst. The PAs are also relatively flat as a function of activity phase, and between cycles. 
This observation can be used to constrain the FRB emission mechanism, as 
well as the origin of periodic activity.
A large fraction of the 1.4\,GHz bursts show multiple subcomponents with a noticeable downward drift in frequency.
We estimated each burst DM by maximising the burst structure \cite{gajjar_highest-frequency_2018, hessels_frb_2018}. This gives similar results as the S/N maximisation technique for single component bursts, but is of particular importance in bursts with complex morphologies. From the brightest bursts (S/N>20), we get the best \dma=348.75$\pm$0.12\pccmnospace. This value is consistent with previous structure maximising DM measurements of the source and limits its DM derivative to less than 0.05\pccmnospace\,yr$^{-1}$.

The downward drift in frequency of the burst subcomponents is a phenomenon that has been previously observed in \FRBnospace, and seems to be common among repeating FRBs\cite{hessels_frb_2018, gajjar_highest-frequency_2018, the_chimefrb_collaboration_second_2019, the_chimefrb_collaboration_chimefrb_2019}.
However, drift rate measurements of \FRB previous to this work had been estimated below 800\,MHz with a value of
$\dot{\nu}=-4.2\pm0.4$\mhzms at 400\,MHz\cite{chawla_detection_2020} and an average of $-21\pm3$\mhzms at 600\,MHz\cite{chamma_shared_2020}.
We obtain an average drift rate at 1370\,MHz of
$-39\pm7$\mhzms. 
This value is nine times larger than the drift rate at 400\,MHz.
The fitted drift rates are consistent with evolving linearly as a function of frequency.
The same linear evolution of the drift rate with frequency has been observed in \RI\cite{josephy_chimefrb_2019}.

\begin{figure}
\centering
\includegraphics[width=\textwidth]{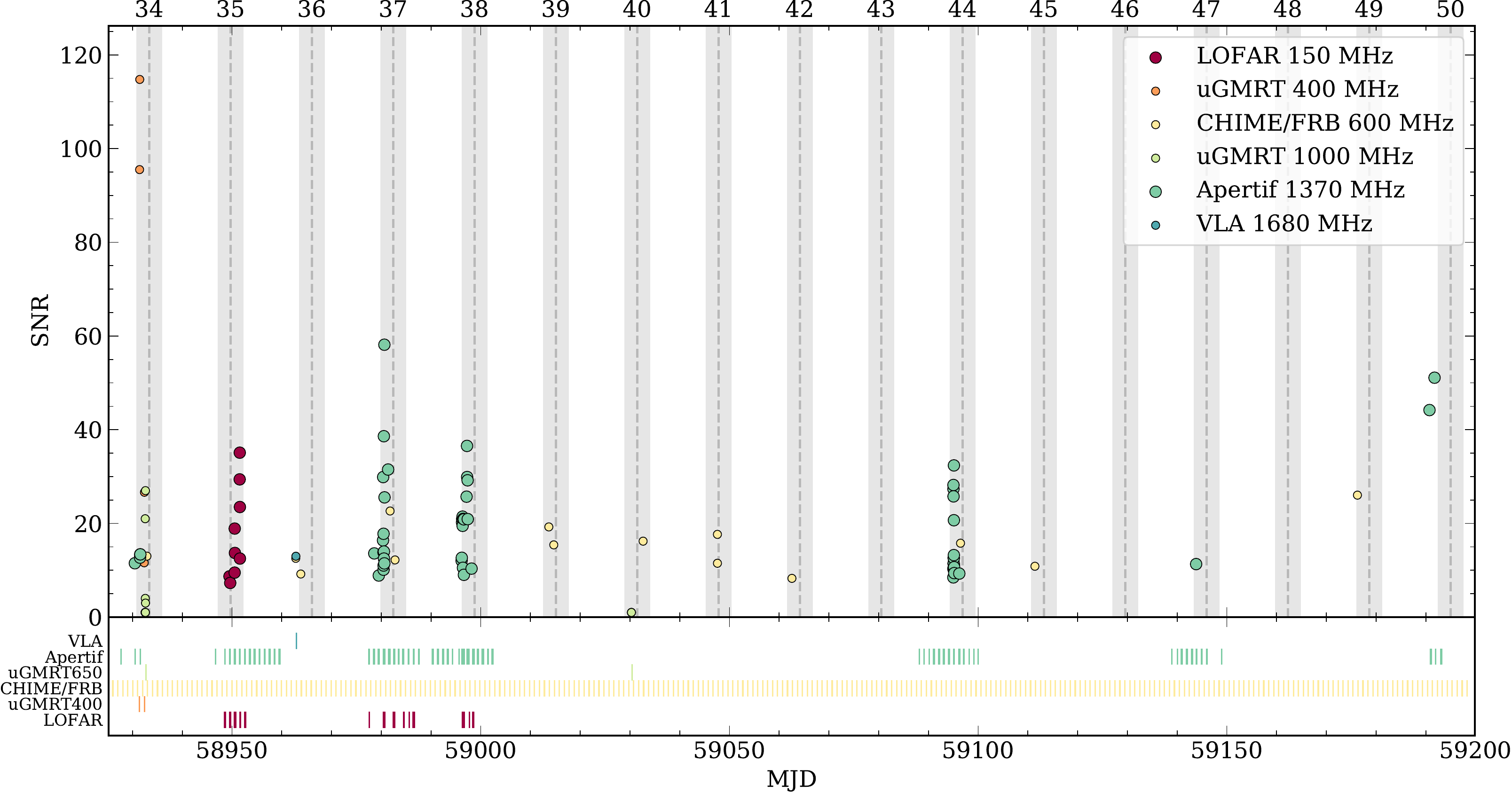}
\caption{Signal-to-noise ratio of the \FRB detections as a function of MJD (top) and duration of the observations of each instrument as a function of MJD (bottom).  Here and in subsequent Figures, Apertif is always shown in green and LOFAR always in dark red.  Detections by other instruments during the covered activity windows are plotted for comparison: uGMRT\cite{sand_low-frequency_2020} at 400 MHz (orange), CHIME/FRB (yellow), uGMRT\cite{marthi_detection_2020} at 650 MHz (lime) and VLA\cite{aggarwal_vlarealfast_2020} (blue). The gray shaded regions correspond to the predicted activity days for a period of 16.35 days, with the predicted peak day as a dashed vertical line. The numbers on top of the plot indicate the cycle number since the first CHIME/FRB detection.
\label{fig:exposure_detections}
}
\end{figure}

Our observations at  1.4\,GHz covered the entire 16.35\,day activity cycle to best investigate the periodicity.
At 150\,MHz we focused on the expected peak active time to maximise the detection probability. 
In \ref{fig:exposure_detections} this coverage is plotted, 
together with the arrival time of the  bursts reported  here and at other
facilities\cite{marthi_detection_2020, sand_low-frequency_2020}.
Our goal with the Apertif observations was to find or rule out any potential aliasing of the period.
That  possibility remained given the short daily source exposures at CHIME/FRB.
Follow-up by other instruments across the predicted activity peak had not been able to rule out this aliasing (See \ref{fig:burst_phase}).
From the arrival MJD of Apertif, CHIME/FRB and all other detections, we built periodograms\cite{aggarwal_vlarealfast_2020} from which we are able to confirm that the best period is 16.29\errors{0.15}{0.17}\,days. This is the only period for which no bursts lie outside of a 6.1\,day activity window including frequencies from 110\,MHz to 1765\,MHz, thus minimising the activity width fraction. 
We searched for short periodicity in both the Apertif and LOFAR observations, but found no significant period between 1\,ms and 80\,s.

\begin{figure}
\centering
\includegraphics[width=0.7\linewidth]{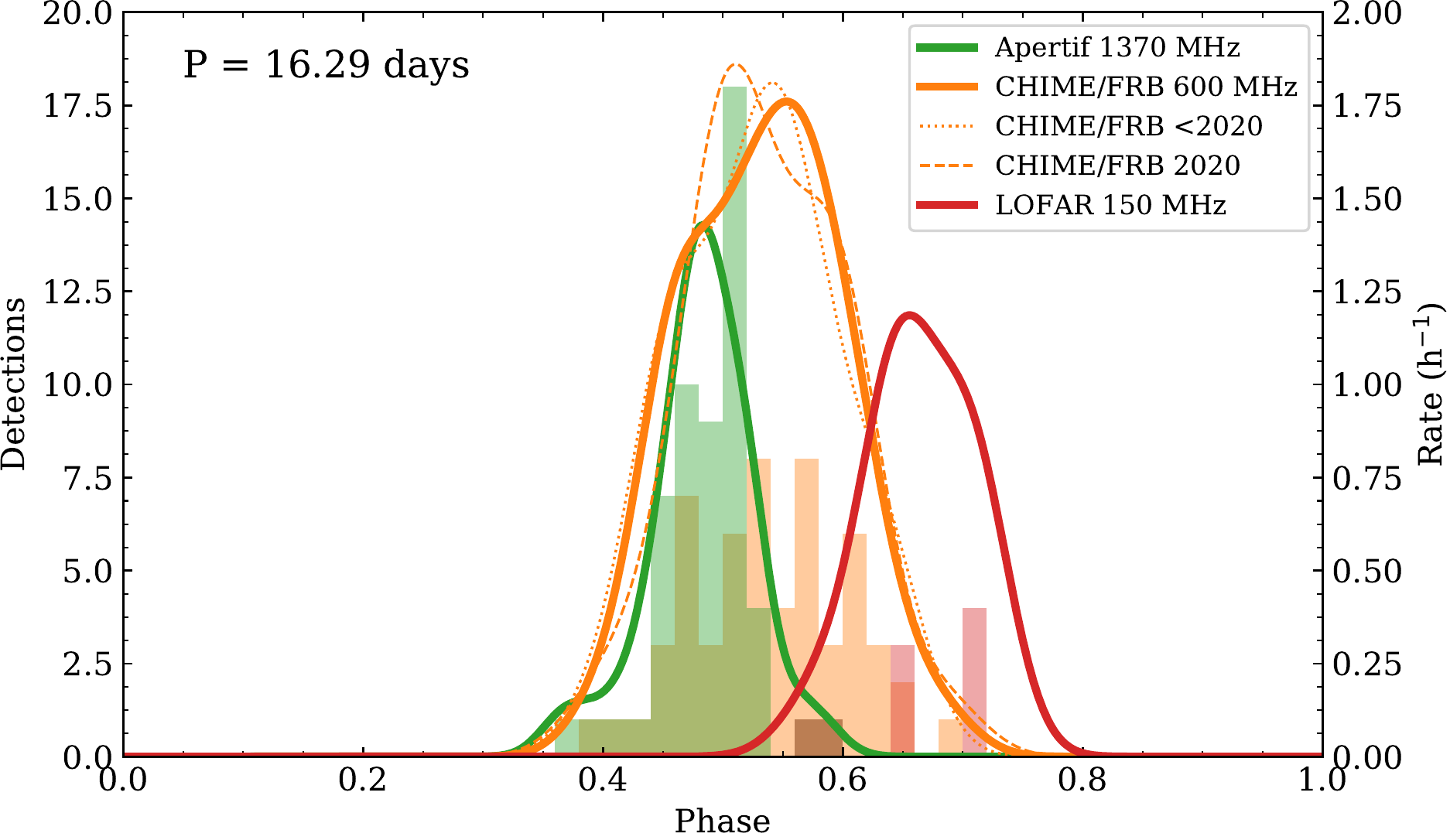}\\
\caption{Activity windows as a function of phase for a period of 16.29 days for Apertif (green), CHIME/FRB (orange) and LOFAR (red). The histograms represent the number of detections and the solid lines the rate obtained with kernel density estimates. The orange dotted line is the KDE for CHIME/FRB bursts before 2020, and the dashed line for CHIME/FRB bursts in 2020, establishing that the wider activity window is not 
due to the longer time baseline for CHIME.
\label{fig:gaussian_kde}
}
\end{figure}

Apertif bursts were found in six out of the seven covered activity cycles, whereas all LOFAR bursts were detected in the activity window with no Apertif detections. However, the Apertif observations during the activity cycle with no Apertif detections started later in phase. 
We observe that most Apertif bursts arrive before CHIME/FRB's activity peak day while LOFAR bursts arrive after. Previous observations of \FRB had hinted at a frequency dependence of the activity window\cite{the_chimefrb_collaboration_periodic_2020}. 
Nevertheless, the scarcity of bursts detected outside of the frequency band covered by CHIME/FRB did not allow for a precise characterisation of the activity window at different frequencies.
Using the Apertif, LOFAR and CHIME/FRB burst samples, we have evaluated the activity windows at 1.4\,GHz, 600\,MHz and 150\,MHz respectively. Using the arrival phase of the bursts with a 16.29\,day period (\ref{fig:gaussian_kde}, see Methods), we calculate the burst rate at each instrument as a function of phase. We find that the activity window is narrower and peaks earlier at 1.4\,GHz than at 600\,MHz. The peak activity at Apertif is $\sim$\,0.7\,days before that of CHIME/FRB and its full-width at half-maximum (FWHM)
is 1.1\,days compared to CHIME/FRB's 2.7\,days. The LOFAR activity cycle appears to peak $\sim$2\,days later than CHIME/FRB's, but the lower number of detections does not allow for a better activity window estimate. It is not yet clear if this effect is discrete, akin to the drifting sub-pulses but on longer timescales, such that for a given frequency range 
the activity window peaks at the same time. Alternatively, it
could be continuous in frequency analogous to dispersion; analyzing the peak frequency of the CHIME/FRB bursts
as a function of activity phase would help answer this question.
We evaluated the likelihood of the bursts being drawn from the same distribution, taking into account the survey strategy. We can discard the Apertif-CHIME/FRB and Apertif-LOFAR burst samples as being drawn from the same distribution with a $>$\,$3\sigma$ confidence, and the CHIME/FRB-LOFAR samples with a $>2\sigma$ confidence (see Methods and \ref{fig:simulated_ks}). 

The initial discovery of periodic activity in \FRB led to many new models to explain this source.
The subsequent detection of a possible 160\,day period in \RI \cite{rajwade_possible_2020}
led to further enthusiasm for periodicity models.
One category of models places the engine of \FRB -- a pulsar or magnetar -- in a 
binary system with a $\sim$16\,day orbital period, where the companion wind obscures the coherent radio emission for most of the orbit via free-free absorption. The
companion can be either a massive star or another neutron star\cite{lyutikov_frb-periodicity_2020, ioka_binary_2020}. 
In these models, a frequency-dependent activity window is predicted but with wider phase ranges at high frequencies because such absorption effects are stronger at longer wavelengths.
Additionally, these models predict a DM evolution due to the dynamic 
absorption column, as well as a low-frequency cutoff. 
Our observations of a smaller phase range at higher frequencies, constant DM, and emission down to 120\,MHz challenge all three predictions of this model.
With the data presented in this work, binary wind models are highly disfavored as an explanation to the periodicity of \FRBnospace. 

Another set of models centers on a precessing magnetar, where the periodic activity of the FRB
follows the precession period. Th precession is  either free if the magnetar is isolated\cite{zanazzi_periodic_2020}
or forced if, for example, it has a fallback disk\cite{tong_periodicity_2020}.
Precessing models predict that a second, shorter periodicity from the neutron
star rotation itself is detectable in the pulse train within an activity window.
We find no such intra-window periodicity.  Mechanisms such as spin noise, pulse profile instability, and dephasing of
the burst beams are to be expected in young magnetars however, and these could conceal the underlying rotation
period. In these models, if the FRB is produced as the neutron star (NS) beam rotates, a PA sweep is
expected\cite{zanazzi_periodic_2020}. We instead observe a flat PA.
Furthermore, free precession models typically require young, hot, 
and highly active magnetars which may still be embedded in their birth environment.
The limits we set on local scattering, absorption, and 
DM variation suggest, however, that \FRB is no longer surrounded by a dense supernova remnant and any remaining magnetar wind is not hampering radio propagation. 

A precessing magnetar could also be responsible for the periodicity of \FRB
if its coherent radio emission is produced farther out.
In synchrotron maser shock models\cite{metzger_fast_2019} 
a flare from the central magnetar causes an
ultra-relativistic shock when colliding with the neighboring medium.
The FRB emission is produced in this magnetized shock. 
This model predicts the flat,
constant intra-burst PAs we observe,
perpendicular to the upstream magnetic field of the surrounding material.
But it is not clear
if such models can power emitters as prolific as \FRB and \RInospace.
The absence of short periodicity and DM variation with phase is consistent 
with the ultra-long period magnetar (ULMP) scenario\cite{beniamini_periodicity_2020}. That model, however, requires expelling enough 
angular momentum to produce a period that is five orders of magnitude larger than any definitively-known neutron star rotation period.

\newpage
\section*{Methods}

A high-level description of the observational and analysis methods is found below.
Further detail following the same order is provided in the \nameref{sec:sm}. 

\section{Observations and burst search}

\subsection{Apertif}

The Westerbork Synthesis Radio Telescope (WSRT) is a radio interferometer located in Drenthe, the Netherlands,
consisting of twelve 25-m dishes in which a new system called Apertif (Aperture Tile in Focus) has recently been 
installed. Single receivers have been replaced by phased array feeds (PAFs), 
increasing its field of view to $\sim$8.7 square degrees\cite{oosterloo_latest_2010, adams_radio_2019}. Apertif can work
in time-domain observing mode to search for new FRBs\cite{connor_bright_2020} and follow-up known
ones\cite{oostrum_repeating_2020} using eight of the WSRT dishes. This capability is provided by a new 
backend, ARTS (the Apertif Radio Transient System\cite{leeu14,2017arXiv170906104M,artsso20}). ARTS covers the full Apertif field-of-view with up to 3000 tied-array beams, each with a typical half-power size of 25' by 25".
In real-time FRB searches, the system records Stokes I data at a central frequency of 1370\,MHz and a 300\,MHz bandwidth with 81.92\,$\mu$s and 195\,kHz time and frequency resolution. The data are then searched in near-real time with our burst search software \textsc{AMBER}\cite{sclocco_real-time_2014, sclocco_amber_2020, sclocco_real-time_2020} and post-processing software \textsc{DARC}\cite{oostrum_darc_2020}. 
Raw FRB candidates are then filtered by a machine learning algorithm that assigns a probability of the candidate being of true astrophysical origin\cite{connor_applying_2018} and later checked by human eyes.
When \textsc{AMBER} identifies an FRB candidate with a duration $<$10\,ms, a signal-to-noise ratio (S/N) >10 and a dispersion measure (DM) 20\% larger than the expected Milky Way contribution to the DM in the pointing direction according to the YMW16 model\cite{yao_new_2017}, the full Stokes IQUV data of the candidate is saved. When following up known sources, the system also stores Stokes IQUV for any candidate with S/N > 10 and a DM within 5\pccm of the source DM.

We carried out observations of \FRB with Apertif, resulting in 388.4\,h on source. The observations
covered seven of the predicted 16.35\,days activity cycles of \FRBnospace, and the exposure times are visualised in the
bottom panel of \ref{fig:exposure_detections}. 
The observations of the three activity cycles after our first detection (numbered 35,37 and 38) ranged over the whole activity phase instead of only at the predicted active days in order to rule out or confirm any potential aliasing of the period\cite{the_chimefrb_collaboration_periodic_2020}. The later observations were scheduled at the confirmed activity peak days.

\subsection{LOFAR}
The LOw Frequency ARray (LOFAR\cite{van_haarlem_lofar_2013, sha+11}) is an interferometric array of
radio telescopes whose core is located in Drenthe, the Netherlands.
LOFAR was used to obtain 57.6\,h of beam-formed data between 110 and 188 MHz,
simultaneous to Apertif observations. LOFAR observations were taken at the predicted active days
to increase the chances of detecting bursts that are broad band from 1.4\,GHz to 150\,MHz, at both telescopes.

The observations were taken during commissioning of the transient buffer boards (TBBs) at LOFAR. In this observing mode, up to five dispersed seconds of raw sub-band data can be saved when a trigger is sent from another instrument. During the simultaneous Apertif-LOFAR observations, if \textsc{AMBER} detected a burst with S/N>10 and a DM within five units of 349.2\pccmnospace, Apertif sent a trigger to LOFAR. The dispersive delay between 1220\,MHz (the bottom of the Apertif band) and 188\,MHz (the top of the HBA band) gives enough time for the pipeline to find the candidate and send the alert, so that LOFAR can freeze the raw data in time. 

All the LOFAR data were also searched offline for FRBs and periodic emission.
After subbanding and RFI cleaning\cite{Maan2020}, data were dedispersed and searched for single pulses. Candidates were clustered in DM and time, visualized, and examined by eye. 

\section{Data analysis}

\subsection{Detected bursts}

During our observing campaign, we detected a total of 63 bursts, 54 with Apertif and 9 with LOFAR. None of these detections took place simultaneously at both instruments. \ref{fig:exposure_detections} shows the S/N of each detection as a function of modified julian day (MJD). It includes the detections by other instruments during the same time span for comparison, and the observation times in the bottom panel. The predicted activity days for a period of 16.35 days are illustrated as shaded regions in order to guide the eye, and the cycle number since the first CHIME/FRB detection are indicated on top.

\subsection{Bursts detected with Apertif}
We detected a total of 54 bursts with an S/N above 8 in 388.4\,h of observations with Apertif. All Apertif bursts are
given an identifier AXX, where XX is the burst number ordered by time of arrival within the Apertif bursts, from A01 to
A54. Twenty-five of those bursts triggered a dump of the full-Stokes data. Eight of the bursts were not detected in real
time, but in the later search of the filterbank observations with \textsc{PRESTO}. The number of IQUV triggers during
cycle 44 is lower due to the incremented RFI environment that triggered IQUV dumps on RFI and avoided saving IQUV data
on later real bursts. \ref{tab:Apertif_bursts} summarises the main properties of the detected bursts.
The burst fluence distribution is further analyzed in the \nameref{sec:sm}.
All detections took place in six out of the seven predicted activity cycles that our observations covered. In spite of observing \FRB during five days centered at the predicted Apertif peak day during cycle 47, only one burst was detected, revealing that the burst rate can fluctuate from cycle to cycle.
\ref{fig:Apertif_bursts} shows the dynamic spectra and pulse profile of all bursts. Additionally, Stokes L and V are plotted for the bursts with full-Stokes data, together with the polarisation position angle (PA).

As shown in \ref{fig:exposure_detections}, all Apertif bursts were detected in a four-day window before the predicted peak day of the corresponding activity cycle, with none of the detections happening after the peak. 
There were no detections outside of a six-day activity window, even though they were largely covered by our observations.
The late start of the observations around MJD 58950 with respect to the beginning of the predicted activity window could explain the non-detections in that cycle. However, the lack of emission at 1.4\,GHz during that cycle cannot be discarded.
After our detections and non-detections during the first four cycles, we refined the expected active window time at 1.4\,GHz and scheduled the observations of the last three cycles accordingly, in five-day windows centered at the predicted Apertif peak day.

The detected bursts present a large variety of properties. Some display a single component, others show rich
time-frequency structure  with up to five components.

\subsection{Bursts detected with LOFAR}
We detected a total of nine LOFAR bursts above a S/N of 7 in $\sim$58\,h of observations
The bursts occurred on 10, 11 and 12 April 2020.
Each burst is given an
identifier LYY, where YY is the burst number ordered by time of arrival within LOFAR bursts, from L01 to
L09. \ref{tab:LOFAR_bursts} summarises the properties of these bursts.
The fluence scale was derived following previously established procedures\cite{kondratiev_lofar_2016,bilous_lofar_2016},
detailed in the \nameref{sec:sm}. As shown in \ref{fig:exposure_detections}, all detections took place on the same predicted activity cycle in which there were no Apertif detections (cycle 35). The observations where the detections took place were performed in coherent Stokes I mode. Excepting the first two, all bursts arrived after the predicted peak day. There were thus no simultaneous bursts at 1.4\,GHz and 150\,MHz in the beamformed data nor the TBBs.
From the dynamic spectra and pulse profiles displayed in \ref{fig:lofar_bursts}, there is no evidence of complex, resolved time-frequency structure. Nevertheless, a scattering tail is manifest in the pulse profiles of the brightest bursts. We will characterise the scattering timescale below. While the tail of burst L06 in \ref{fig:arts_lofar_dynspec} appears to plateau 25\,ms after the main peak, hinting for a second subburst, a fit for multiple scattered bursts did not confidently identify a second component.

Generally, the LOFAR-detected bursts are brightest in the top of the band (see \ref{fig:lofar_bursts}).
Over the almost 2:1 ratio of frequency from that top of the band to the bottom,
most bursts  gradually become less bright.
Although
some previous targeted LOFAR FRB searches
used wide bandwidths\cite{houben_constraints_2019},
most large-area searches were carried out in the lower part of
the band, e.g., 119$-$151\,MHz\cite{coenen_lofar_2014,2019A&A...626A.104S} where LOFAR is more sensitive.
The behavior we see here was likely a factor in the earlier lack of
detections.

The detections reported here already demonstrate there is no low-frequency cutoff above the LOFAR band.
The individual bursts and
the stacked profile (\ref{fig:scattering}) also do not show a clear cutoff within the band. 
Two of the bursts (L01, L06) emit down to at least 120\,MHz  (see \ref{fig:spectra}) and thus cover the entire frequency range.
Furthermore, if we follow  burst L06 from 150 to 120\,MHz in  decreasing frequency,
the emission is ever more delayed with the respect to the onset of the peak
(see \ref{fig:arts_lofar_dynspec} and \ref{fig:lofar_bursts}).
Such behavior suggests unresolved time-frequency downward drift in the tail of the pulse.
From this we conclude the decrease in pulse peak brightness could be intrinsic,
and is not due to a cutoff by intervening
material. Bursts  L04 and L05 show a similar hint of a delayed tail, at slightly higher frequencies.

\subsection{Ruling out aliasing}
To maximise the chance of
detection, \FRB is generally observed predominantly around its predicted activity
peak\cite{chawla_detection_2020, pilia_lowest_2020, aggarwal_vlarealfast_2020, sand_low-frequency_2020}.
The implied lack of coverage outside this purported peak could bias the derived activity cycle.
The best-fit cycle period could be an alias of the true period.
To break this degeneracy, we scheduled observations covering full activity cycles (see \nameref{sec:sm}).
We find there is no aliasing. We determined a new best activity period of 16.29\,days,
where reference MJD 58369.9  centered the peak activity day at phase 0.5.

\subsection{Activity windows}

By using the aforementioned best period and reference MJD to compute the burst arrival phases, we have generated a histogram of detections versus phase on the top panel of \ref{fig:burst_phase}. The cycle coverage by different instruments can be visualised on the bottom panel of the same figure, where it is manifest that CHIME/FRB and Apertif are the only instruments covering the whole activity cycle which have detected bursts. We have used data from all \FRB observations published thus far\cite{the_chimefrb_collaboration_periodic_2020, marcote_repeating_2020, scholz_simultaneous_2020, pilia_lowest_2020, chawla_detection_2020, marthi_detection_2020, aggarwal_vlarealfast_2020, pearlman_multiwavelength_2020}.

Several theoretical models have suggested
the activity window may be frequency dependent\cite{lyutikov_frb-periodicity_2020, ioka_binary_2020, beniamini_periodicity_2020}. In absorptive wind models, for example, one expects a larger duty cycle at high frequencies due to heightened opacity at long wavelengths. There was also an observational hint that higher frequencies may arrive earlier, based on four EVN detections
at 1.7\,GHz\cite{marcote_repeating_2020}. By taking into account the bursts detected by Apertif, CHIME/FRB and LOFAR at 1.4\,GHz, 600\,MHz and 150\,MHz respectively, we can obtain an estimate of the probability of the bursts being drawn from the same distribution at different frequencies.

To do so, we attempted to estimate the detection rate as a function
of activity phase for the three different frequency bands. We estimate
these activity windows by computing the probability density function (PDF)
of detection rate for Apertif, CHIME/FRB and LOFAR using a weighted kernel
density estimator (KDE, see \nameref{sec:sm}).

We applied the KDE separately to the Apertif, CHIME/FRB and LOFAR burst datasets.
Based on the KDE estimation shown in \ref{fig:gaussian_kde}, we find that  higher frequencies appear to arrive earlier in phase, i.e. the activity peaks at a lower phase with larger frequencies.
Additionally, the width of the activity window appears to be larger with CHIME/FRB.
The KDE is useful for estimating probability distributions with a
small number of samples, but it is non-parametric and does not easily allow us to
compare the activity window widths between frequencies. For this we fit a Gaussian
to the detection rate of \FRB for each instrument and find a full-width at half maximum (FWHM) of 1.2$\pm$0.1\,days from the Apertif data at 1370\,MHz and 2.7$\pm$0.2\,days at 400--800\,MHz using the CHIME/FRB bursts. The best-fit peak activity phase for Apertif is 0.494$\pm$0.002 and 0.539$\pm$0.005 for CHIME/FRB. The source activity window is therefore roughly two times wider at CHIME/FRB than
at Apertif and its peak is 0.7 days later at CHIME/FRB. We do not attempt to fit a Gaussian to the LOFAR bursts because of the small number of detections and our limited coverage in phase.
However, we note that four out of the nine detected LOFAR bursts arrive later in phase than every previously detected CHIME/FRB burst.
Therefore, the activity of \FRB at 150\,MHz likely peaks later than at higher frequencies and the activity window may be wider as well. This is in stark contrast with the predictions of simple absorptive wind models where the activity ought to be
wider at higher frequencies. 

By applying a Kolmogorov-Smirnov test to the burst samples of Apertif, CHIME/FRB and LOFAR comparing them two by two and taking into account the different observing strategies, we can discard the Apertif and CHIME/FRB burst samples as being drawn from the same distribution with a three-sigma confidence level, as well as the Apertif and LOFAR burst samples. For the CHIME/FRB and LOFAR bursts, the confidence level of the samples being drawn from different distributions is greater than two sigma. This method is expanded in the  \nameref{sec:sm} section.

Taking all observational biases into account, a dependence of the activity window with frequency must exist in order to get the observed burst distribution, which is narrower and peaks earlier in phase at higher frequencies. This is opposite to the predictions made by binary wind model predictions in which free-free absorption would make the lower frequency emission have a narrower activity width\cite{lyutikov_frb-periodicity_2020, ioka_binary_2020}, and thus disfavors the cause of the periodicity to be free-free absorption in a binary system. 

\subsection{Polarisation}
Monitoring the polarisation position angle (PA) of \FRB over time and across cycles with Apertif is made easier by the
fact that Westerbork is a steerable equatorial mount telescope.
This stability of the system’s response allows us to investigate the polarisation properties of \FRB within each pulse,
within an activity cycle, and even between multiple periods.
After calibration (see \nameref{sec:sm}) 
we find the PA of \FRB to be flat within each burst, with $\Delta$PA<20\,deg, in agreement with other polarised studies of the source\cite{the_chimefrb_collaboration_chimefrb_2019, chawla_detection_2020, nimmo_microsecond_2020}. This is in contrast to most pulsars whose PAs swing across the pulse, in many cases with the S-shaped functional form predicted by the rotating vector model (RVM). In the classic picture, PA varies with the arctangent of pulse longitude and the amount of 
swing is proportional to the emission height but inversely proportional to the star's rotation period\cite{blaskiewicz_relativistic_1991}. However, the flat PAs of \FRB are similar to other FRBs, notably \RI whose intra-burst polarisation exhibits less than 11\,deg of rotation\cite{gajjar_highest-frequency_2018}. They are also similar to radio magnetars. FRB\,181112 was the first source to show significant variation in the polarisation state within a burst and between sub-components of an FRB with temporal structure\cite{cho_spectropolarimetric_2020}. 

While the flat PAs within each \FRB burst are in line with previous measurements, we have found that its PA is also
stable in average within an activity cycle and even between periods, with $\Delta$PA<40\,deg. In models that invoke precession as the origin of 
periodicity and magnetospheric emission as the origin 
of the FRBs, one generically expects a PA change as a function of activity phase. However, the amount of PA swing depends on the geometry of the system and well as the fraction of a precession period that is observable\cite{zanazzi_periodic_2020}, so we cannot rule out precession with our polarisation measurements. In relativistic shock models, the synchrotron 
maser mechanism provides a natural path for flat PAs within a burst, but 
it is not clear how or if the polarisation state could be 
nearly constant within a cycle and over multiple months\cite{metzger_fast_2019, beloborodov_2019}. Given the duty cycle of \FRB appears to be just $\sim$10$\%$ in the Apertif band, it will be useful to 
observe the PAs of \FRB over at lower frequencies with 
a steerable telescope that can cover a full periodic cycle.

\subsection{Dispersion}
The Apertif real-time detection pipeline finds the dispersion measure that maximises the S/N of a burst (\dmsnrnospace).
Any frequency-swept structure intrinsic to the pulse,
as seen in a number of  repeating FRBs, will be absorbed in this value. 
By first fitting to such structure\cite{gajjar_highest-frequency_2018, hessels_frb_2018}
the interstellar dispersion measure can be isolated, and reported as \dmstruct.
We thus determined \dmstruct{ }for all Apertif bursts (see \nameref{sec:sm})
and find it to be  \dma=348.75$\pm$0.12\pccmnospace,
consistent with previous findings.
There is no evidence for a variation of DM with phase (\ref{fig:Apertif_burst_properties}, top panel).

The LOFAR bursts lack detectable time-frequency structure,
but require separating the frequency-dependent scattering tails from the DM fit (see \nameref{sec:sm}).
For the final LOFAR DM, obtained
by averaging over all bursts with S/N>20 (\ref{tab:LOFAR_bursts}),
we find \dml$ = 349.00 \pm 0.02$\pccm .

\subsection{Sub-pulse drift rate} 

Several of our detections at Apertif show downward drifting sub-pulses, enabling us to make the first measurement of the
drift rate  $\dot{\nu}$ of \FRB above 1\,GHz (\ref{fig:drift_rate}).
We obtain an average sub-pulse drift rate of $-39 \pm 7$\mhzms\  at 1370\,MHz.
This is nine times larger than e.g. the previously reported drift rate of $\sim-$4.2\mhzms at
400\,MHz\cite{chawla_detection_2020}. \ref{fig:freq_drift_rate} shows how the downward drift amplifies towards higher
frequencies.
As in \RI\cite{josephy_chimefrb_2019}, the drift rate evolution appears linear.
As these two FRBs reside in significantly different environments, the behavior may be common across FRBs.
The frequency-dependence and consistent sign of the drifting phenomenon will likely offer clues to the FRB emission mechanism\cite{metzger_fast_2019, wang_time-frequency_2019, rajabi_simple_2020}.

\subsection{Scattering}
Most of the LOFAR bursts (\ref{fig:lofar_bursts}) exhibit an exponential tail, indicating the pulse-broadening due to scattering in the intervening medium. To quantify the scatter broadening timescale (\tscat), we divided the dedispersed spectrograms of a few high S/N bursts into 4 or 8 sub-bands 
The burst profiles obtained from the individual sub-bands were modelled as a single Gaussian component convolved with a one-sided exponential function\cite{krishnakumar_multi-frequency_2017, maan_distinct_2019}. The \tscat thus obtained are presented in \ref{tab:LOFAR_bursts}.

In order to obtain a more precise estimate of scatter-broadening timescale, we first divided the bandwidth of the stacked LOFAR bursts dedispersed to their \dml=349.0\pccm 
into eight frequency bands, for which we obtained separate pulse profiles and fitted each to a scattering tail as above. The results are shown in \ref{fig:scattering}.
We obtain the scattering timescale of 45.7$\pm$9.5\,ms at 150\,MHz, which is consistent with the measurements using individual bursts. We also characterize the scatter broadening variation with frequency as \tscat$\propto\nu^{-\alpha}$ and obtain the frequency scaling index $\alpha = -4.2$\errors{1.1}{1.0}. 
This scatter-broadening is consistent with the upper limit of 50\,ms at 150\,MHz that was derived from GBT detections at 350\,MHz\cite{chawla_detection_2020}. By scaling the scatter broadening of LOFAR bursts to Apertif frequencies, we expect \tscat\,$\sim 6.6 \mu$s at 1370\,MHz, which is an order of magnitude smaller than Apertif's temporal resolution.

\subsection{Rates}

Before our LOFAR detections, there existed 
only upper limits on the 
FRB sky rate below 200\,MHz.
Blind searches for fast transients 
at these low frequencies are difficult due to the deleterious 
smearing effects of intra-channel dispersion and scattering, which 
scale as $\nu^{-3}$ and $\nu^{-4}$, respectively. This is 
amplified by the large sky brightness temperatures at long wavelengths, 
due to the red spectrum of Galactic synchrotron emission; pulsars are 
detectable at low frequencies because of their
steeply rising negative spectra, but the spectral 
index of the FRB event rate is not known. We first consider 
the repetition rate of \FRB from our LOFAR and Apertif detections 
to determine its activity as function of frequency. 
We then convert that into a lower-limit 
on the all-sky FRB rate at 150\,MHz and combine it with previous 
upper-limits at those frequencies to derive the first ever 
bounded constraints on FRB rates below 200\,MHz.

We detected nine bursts in 58\,hours of LOFAR observing, 
giving a rate of 0.16\,$\pm0.05$\,h$^{-1}$. Since we only targeted LOFAR during 
simultaneous Apertif observations during the presumed activity window
whose duty cycle is $\sim$0.25, we divide this rate by 4 to get its
repetition rate averaged over time. Assuming a fluence threshold of 50\,Jy\,ms 
and noting that the duration of 
all bursts from this source at 150\,MHz is set by 
scattering and does not vary, we find $R_{150}(\geq\,50\,\rm{\jyms})\approx(3.9\pm1.3)\times10^{-2}$\,h$^{-1}$. At 1370\,MHz, Apertif detected 54 pulses in 388 hours of observing. 
Our coverage of \FRB was deliberately more uniform in activity phase, so only $\sim$149\,h took place during the active days. The phase range in which Apertif detected bursts gives a duty cycle of 0.22.
This results in $R_{1370}(\geq 1\,\rm{\jyms})\approx(8.0\pm1.1)\times10^{-2}$\,h$^{-1}$.
While the absolute detection rates by Apertif and LOFAR are similar, we note
that the fluence threshold was much lower for Apertif than LOFAR. Scaling 
by the known fluence distribution of \FRBnospace, 
$N(\geq\mathcal{F}) \propto \mathcal{F}^{-1.5}$, we find
$R_{1370}(\geq 50\,\rm{\jyms})\approx2.3\times10^{-4}$\,h$^{-1}$. 
We come to the remarkable 
conclusion that the FRB is more active at 150\,MHz than at 1370\,MHz
at the relevant fluences. 

The all-sky FRB event rate is a difficult quantity to determine 
for a myriad of reasons\cite{rane_search_2016}. Beam effects result in a pointing-dependent 
sensitivity threshold, which in turn is affected by the unknown 
source-count slope\cite{lawrence_non-homogeneous_2017,vedantham_fluence_2016}; Each survey 
has back-end dependent incompleteness, including 
in flux density and fluence\cite{keane_fast_2015} as 
well as in pulse duration and DM\cite{connor_interpreting_2019}. Nonetheless, 
meaningful constraints can be made if one is explicit about the 
region of parameter space to which the rate applies. 

As the LOFAR bursts are the sole unambiguous FRB detections below 200 MHz, we 
and other teams\cite{pmb+20} can now provide the first bounded limits on the all-sky event rate at low frequencies. A lower limit on the FRB rate at 150\,MHz can be obtained 
by assuming \FRB is the only source in the sky emitting at these wavelengths.
This lower bound can be combined with previous upper bounds from 
non detections by blind searches at LOFAR and
MWA\cite{karastergiou_limits_2015, coenen_lofar_2014, ter_veen_frats_2019, tingay_search_2015, rowlinson_limits_2016, sokolowski_no_2018}.
The repetition 
rate of \FRB implies that there are at least 
0.6\skyday above 50\,Jy\,ms at 
110--190\,MHz at 95$\%$ confidence. Assuming a Euclidean 
scaling in the brightness distribution that continues 
down to lower fluences, this is equivalent to more than 
90\skyday above 5\jymsnospace. An earlier blind LOFAR
search\cite{karastergiou_limits_2015} placed an upper limit of 29\skyday above 62\,Jy pulses with 5\,ms duration. Combining these two limits, 
we obtain a 90$\%$ confidence region of 3--450\skyday above 
50\,Jy\,ms.

The lower-limit value may be conservative, 
as \FRB is in the Galactic plane at a latitude of just 3.7\,$\deg$, 
which is why its scattering time is 50\,ms at 150\,MHz. 
If the burst width were 5\,ms before entering the Milky Way, then 
a factor of $\sim$3 was lost in S/N due to the low Galactic 
latitude of \FRBnospace. Therefore, a similar FRB at 
a more typical offset from the plane would, in this example, be 
$\sim$3$^\gamma$ times more active, where $\gamma$ is the 
cumulative energy distribution power-law index, because 
the Galactic scattering timescale would only be a few milliseconds.

\section{Data availability}
Raw data were generated by the Apertif system on the Westerbork Synthesis Radio Telescope and by the International LOFAR
Telescope.
The Apertif data that support the findings of this study
are available through the ALERT archive, \url{http://www.alert.eu/FRB20180916B}.
The LOFAR data are available through the LOFAR Long Term Archive, \url{https://lta.lofar.eu/},
by searching for “Observations” at J2000 coordinates RA=01:57:43.2000, DEC=+65:42:01.020.


\newpage
\section*{References}

\begin{thebibliography}{10}
\urlstyle{rm}
\expandafter\ifx\csname url\endcsname\relax
  \def\url#1{\texttt{#1}}\fi
\expandafter\ifx\csname urlprefix\endcsname\relax\def\urlprefix{URL }\fi
\expandafter\ifx\csname doiprefix\endcsname\relax\def\doiprefix{DOI: }\fi
\providecommand{\bibinfo}[2]{#2}
\providecommand{\eprint}[2][]{\href{https://dx.doi.org/#2}{#2}}

\bibitem{lorimer_bright_2007}
\bibinfo{author}{Lorimer, D.~R.}, \bibinfo{author}{Bailes, M.},
  \bibinfo{author}{McLaughlin, M.~A.}, \bibinfo{author}{Narkevic, D.~J.} \&
  \bibinfo{author}{Crawford, F.}
\newblock \bibinfo{journal}{\bibinfo{title}{A bright millisecond radio burst of
  extragalactic origin}}.
\newblock {\emph{\JournalTitle{Science}}} \textbf{\bibinfo{volume}{318}},
  \bibinfo{pages}{777--780}, \doiprefix\href{https://dx.doi.org/10.1126/science.1147532}{10.1126/science.1147532}
  (\bibinfo{year}{2007}).
\newblock \bibinfo{note}{ArXiv: 0709.4301}.

\bibitem{chawla_detection_2020}
\bibinfo{author}{Chawla, P.} \emph{et~al.}
\newblock \bibinfo{journal}{\bibinfo{title}{Detection of {Repeating} {FRB}
  180916.{J0158}+65 {Down} to {Frequencies} of 300 {MHz}}}.
\newblock {\emph{\JournalTitle{arXiv:2004.02862 [astro-ph]}}}
  (\bibinfo{year}{2020}).
\newblock \bibinfo{note}{ArXiv: 2004.02862}.

\bibitem{gajjar_highest-frequency_2018}
\bibinfo{author}{Gajjar, V.} \emph{et~al.}
\newblock \bibinfo{journal}{\bibinfo{title}{Highest-frequency detection of
  {FRB} 121102 at 4-8 {GHz} using the {Breakthrough} {Listen} {Digital}
  {Backend} at the {Green} {Bank} {Telescope}}}.
\newblock {\emph{\JournalTitle{The Astrophysical Journal}}}
  \textbf{\bibinfo{volume}{863}}, \bibinfo{pages}{2},
  \doiprefix\href{https://dx.doi.org/10.3847/1538-4357/aad005}{10.3847/1538-4357/aad005} (\bibinfo{year}{2018}).
\newblock \bibinfo{note}{ArXiv: 1804.04101}.

\bibitem{the_chimefrb_collaboration_periodic_2020}
\bibinfo{author}{{The CHIME/FRB Collaboration}} \emph{et~al.}
\newblock \bibinfo{journal}{\bibinfo{title}{Periodic activity from a fast radio
  burst source}}.
\newblock {\emph{\JournalTitle{Nature}}} \textbf{\bibinfo{volume}{582}},
  \bibinfo{pages}{351--355}, \doiprefix\href{https://dx.doi.org/10.1038/s41586-020-2398-2}{10.1038/s41586-020-2398-2}
  (\bibinfo{year}{2020}).
\newblock \bibinfo{note}{ArXiv: 2001.10275}.

\bibitem{spitler_repeating_2016}
\bibinfo{author}{Spitler, L.~G.} \emph{et~al.}
\newblock \bibinfo{journal}{\bibinfo{title}{A repeating fast radio burst}}.
\newblock {\emph{\JournalTitle{Nature}}} \textbf{\bibinfo{volume}{531}},
  \bibinfo{pages}{202--205}, \doiprefix\href{https://dx.doi.org/10.1038/nature17168}{10.1038/nature17168}
  (\bibinfo{year}{2016}).

\bibitem{the_chimefrb_collaboration_second_2019}
\bibinfo{author}{{The CHIME/FRB Collaboration}}.
\newblock \bibinfo{journal}{\bibinfo{title}{A second source of repeating fast
  radio bursts}}.
\newblock {\emph{\JournalTitle{Nature}}} \textbf{\bibinfo{volume}{566}},
  \bibinfo{pages}{235--238}, \doiprefix\href{https://dx.doi.org/10.1038/s41586-018-0864-x}{10.1038/s41586-018-0864-x}
  (\bibinfo{year}{2019}).

\bibitem{fonseca_nine_2020}
\bibinfo{author}{Fonseca, E.} \emph{et~al.}
\newblock \bibinfo{journal}{\bibinfo{title}{Nine {New} {Repeating} {Fast}
  {Radio} {Burst} {Sources} from {CHIME}/{FRB}}}.
\newblock {\emph{\JournalTitle{The Astrophysical Journal}}}
  \textbf{\bibinfo{volume}{891}}, \bibinfo{pages}{L6},
  \doiprefix\href{https://dx.doi.org/10.3847/2041-8213/ab7208}{10.3847/2041-8213/ab7208} (\bibinfo{year}{2020}).
\newblock \bibinfo{note}{ArXiv: 2001.03595}.

\bibitem{the_chimefrb_collaboration_chimefrb_2019}
\bibinfo{author}{{The CHIME/FRB Collaboration}} \emph{et~al.}
\newblock \bibinfo{journal}{\bibinfo{title}{{CHIME}/{FRB} {Detection} of
  {Eight} {New} {Repeating} {Fast} {Radio} {Burst} {Sources}}}.
\newblock {\emph{\JournalTitle{arXiv:1908.03507 [astro-ph]}}}
  (\bibinfo{year}{2019}).
\newblock \bibinfo{note}{ArXiv: 1908.03507}.

\bibitem{marcote_repeating_2020}
\bibinfo{author}{Marcote, B.} \emph{et~al.}
\newblock \bibinfo{journal}{\bibinfo{title}{A repeating fast radio burst source
  localised to a nearby spiral galaxy}}.
\newblock {\emph{\JournalTitle{Nature}}} \textbf{\bibinfo{volume}{577}},
  \bibinfo{pages}{190--194}, \doiprefix\href{https://dx.doi.org/10.1038/s41586-019-1866-z}{10.1038/s41586-019-1866-z}
  (\bibinfo{year}{2020}).
\newblock \bibinfo{note}{ArXiv: 2001.02222}.

\bibitem{pilia_lowest_2020}
\bibinfo{author}{Pilia, M.} \emph{et~al.}
\newblock \bibinfo{journal}{\bibinfo{title}{The lowest frequency {Fast} {Radio}
  {Bursts}: {Sardinia} {Radio} {Telescope} detection of the periodic {FRB}
  180916 at 328 {MHz}}}.
\newblock {\emph{\JournalTitle{arXiv:2003.12748 [astro-ph]}}}
  (\bibinfo{year}{2020}).
\newblock \bibinfo{note}{ArXiv: 2003.12748}.

\bibitem{2017arXiv170906104M}
\bibinfo{author}{{Maan}, Y.} \& \bibinfo{author}{{van Leeuwen}, J.}
\newblock \bibinfo{journal}{\bibinfo{title}{{Real-time searches for fast
  transients with Apertif and LOFAR}}}.
\newblock {\emph{\JournalTitle{IEEE Proc. URSI GASS}}}
  \doiprefix\href{https://dx.doi.org/10.23919/URSIGASS.2017.8105320}{10.23919/URSIGASS.2017.8105320} (\bibinfo{year}{2017}).
\newblock \eprint{1709.06104}.

\bibitem{sha+11}
\bibinfo{author}{{Stappers}, B.~W.} \emph{et~al.}
\newblock \bibinfo{journal}{\bibinfo{title}{{Observing pulsars and fast
  transients with LOFAR}}}.
\newblock {\emph{\JournalTitle{Astron. \& Astrophys.}}}
  \textbf{\bibinfo{volume}{530}}, \bibinfo{pages}{A80+},
  \doiprefix\href{https://dx.doi.org/10.1051/0004-6361/201116681}{10.1051/0004-6361/201116681} (\bibinfo{year}{2011}).
\newblock \eprint{1104.1577}.

\bibitem{pmb+20}
\bibinfo{author}{{Pleunis}, Z.} \emph{et~al.}
\newblock \bibinfo{journal}{\bibinfo{title}{{LOFAR Detection of 110--188 MHz
  Emission and Frequency-Dependent Activity from FRB~20180916B}}}.
\newblock (\bibinfo{year}{2020}).
\newblock \bibinfo{note}{{\it submitted}}.

\bibitem{coenen_lofar_2014}
\bibinfo{author}{Coenen, T.} \emph{et~al.}
\newblock \bibinfo{journal}{\bibinfo{title}{The {LOFAR} {Pilot} {Surveys} for
  {Pulsars} and {Fast} {Radio} {Transients}}}.
\newblock {\emph{\JournalTitle{Astronomy \& Astrophysics}}}
  \textbf{\bibinfo{volume}{570}}, \bibinfo{pages}{A60},
  \doiprefix\href{https://dx.doi.org/10.1051/0004-6361/201424495}{10.1051/0004-6361/201424495} (\bibinfo{year}{2014}).
\newblock \bibinfo{note}{ArXiv: 1408.0411}.

\bibitem{karastergiou_limits_2015}
\bibinfo{author}{Karastergiou, A.} \emph{et~al.}
\newblock \bibinfo{journal}{\bibinfo{title}{Limits on {Fast} {Radio} {Bursts}
  at 145 {MHz} with {ARTEMIS}, a real-time software backend}}.
\newblock {\emph{\JournalTitle{Monthly Notices of the Royal Astronomical
  Society}}} \textbf{\bibinfo{volume}{452}}, \bibinfo{pages}{1254--1262},
  \doiprefix\href{https://dx.doi.org/10.1093/mnras/stv1306}{10.1093/mnras/stv1306} (\bibinfo{year}{2015}).
\newblock \bibinfo{note}{ArXiv: 1506.03370}.

\bibitem{hessels_frb_2018}
\bibinfo{author}{Hessels, J. W.~T.} \emph{et~al.}
\newblock \bibinfo{journal}{\bibinfo{title}{{FRB} 121102 {Bursts} {Show}
  {Complex} {Time}-{Frequency} {Structure}}}.
\newblock {\emph{\JournalTitle{arXiv:1811.10748 [astro-ph]}}}
  (\bibinfo{year}{2018}).
\newblock \bibinfo{note}{ArXiv: 1811.10748}.

\bibitem{cordes_ne2001i_2003}
\bibinfo{author}{Cordes, J.~M.} \& \bibinfo{author}{Lazio, T. J.~W.}
\newblock \bibinfo{journal}{\bibinfo{title}{{NE2001}.{I}. {A} {New} {Model} for
  the {Galactic} {Distribution} of {Free} {Electrons} and its {Fluctuations}}}.
\newblock {\emph{\JournalTitle{arXiv:astro-ph/0207156}}}
  (\bibinfo{year}{2003}).
\newblock \bibinfo{note}{ArXiv: astro-ph/0207156}.

\bibitem{mcquinn-2014}
\bibinfo{author}{{McQuinn}, M.}
\newblock \bibinfo{journal}{\bibinfo{title}{{Locating the ``Missing'' Baryons
  with Extragalactic Dispersion Measure Estimates}}}.
\newblock {\emph{\JournalTitle{\apjl}}} \textbf{\bibinfo{volume}{780}},
  \bibinfo{pages}{L33}, \doiprefix\href{https://dx.doi.org/10.1088/2041-8205/780/2/L33}{10.1088/2041-8205/780/2/L33}
  (\bibinfo{year}{2014}).
\newblock \eprint{1309.4451}.

\bibitem{chamma_shared_2020}
\bibinfo{author}{Chamma, M.~A.}, \bibinfo{author}{Rajabi, F.},
  \bibinfo{author}{Wyenberg, C.~M.}, \bibinfo{author}{Mathews, A.} \&
  \bibinfo{author}{Houde, M.}
\newblock \bibinfo{journal}{\bibinfo{title}{A shared law between sources of
  repeating fast radio bursts}}.
\newblock {\emph{\JournalTitle{arXiv:2010.14041 [astro-ph]}}}
  (\bibinfo{year}{2020}).
\newblock \bibinfo{note}{ArXiv: 2010.14041}.

\bibitem{josephy_chimefrb_2019}
\bibinfo{author}{Josephy, A.} \emph{et~al.}
\newblock \bibinfo{journal}{\bibinfo{title}{{CHIME}/{FRB} {Detection} of the
  {Original} {Repeating} {Fast} {Radio} {Burst} {Source} {FRB} 121102}}.
\newblock {\emph{\JournalTitle{The Astrophysical Journal}}}
  \textbf{\bibinfo{volume}{882}}, \bibinfo{pages}{L18},
  \doiprefix\href{https://dx.doi.org/10.3847/2041-8213/ab2c00}{10.3847/2041-8213/ab2c00} (\bibinfo{year}{2019}).
\newblock \bibinfo{note}{ArXiv: 1906.11305}.

\bibitem{marthi_detection_2020}
\bibinfo{author}{Marthi, V.~R.} \emph{et~al.}
\newblock \bibinfo{journal}{\bibinfo{title}{Detection of 15 bursts from {FRB}
  180916.{J0158}+65 with the {uGMRT}}}.
\newblock {\emph{\JournalTitle{arXiv:2007.14404 [astro-ph]}}}
  (\bibinfo{year}{2020}).
\newblock \bibinfo{note}{ArXiv: 2007.14404}.

\bibitem{sand_low-frequency_2020}
\bibinfo{author}{Sand, K.~R.} \emph{et~al.}
\newblock \bibinfo{journal}{\bibinfo{title}{Low-frequency detection of
  {FRB180916} with the {uGMRT}}}.
\newblock {\emph{\JournalTitle{ATel}}}  (\bibinfo{year}{2020}).
\newblock \bibinfo{note}{Library Catalog: www.astronomerstelegram.org}.

\bibitem{aggarwal_vlarealfast_2020}
\bibinfo{author}{Aggarwal, K.} \emph{et~al.}
\newblock \bibinfo{journal}{\bibinfo{title}{{VLA}/realfast detection of burst
  from {FRB180916}.{J0158}+65 and {Tests} for {Periodic} {Activity}}}.
\newblock {\emph{\JournalTitle{arXiv:2006.10513 [astro-ph]}}}
  (\bibinfo{year}{2020}).
\newblock \bibinfo{note}{ArXiv: 2006.10513}.

\bibitem{rajwade_possible_2020}
\bibinfo{author}{Rajwade, K.~M.} \emph{et~al.}
\newblock \bibinfo{journal}{\bibinfo{title}{Possible periodic activity in the
  repeating {FRB} 121102}}.
\newblock {\emph{\JournalTitle{Monthly Notices of the Royal Astronomical
  Society}}} \textbf{\bibinfo{volume}{495}}, \bibinfo{pages}{3551--3558},
  \doiprefix\href{https://dx.doi.org/10.1093/mnras/staa1237}{10.1093/mnras/staa1237} (\bibinfo{year}{2020}).
\newblock \bibinfo{note}{ArXiv: 2003.03596}.

\bibitem{lyutikov_frb-periodicity_2020}
\bibinfo{author}{Lyutikov, M.}, \bibinfo{author}{Barkov, M.} \&
  \bibinfo{author}{Giannios, D.}
\newblock \bibinfo{journal}{\bibinfo{title}{{FRB}-periodicity: mild pulsar in
  tight {O}/{B}-star binary}}.
\newblock {\emph{\JournalTitle{arXiv:2002.01920 [astro-ph]}}}
  (\bibinfo{year}{2020}).
\newblock \bibinfo{note}{ArXiv: 2002.01920}.

\bibitem{ioka_binary_2020}
\bibinfo{author}{Ioka, K.} \& \bibinfo{author}{Zhang, B.}
\newblock \bibinfo{journal}{\bibinfo{title}{A {Binary} {Comb} {Model} for
  {Periodic} {Fast} {Radio} {Bursts}}}.
\newblock {\emph{\JournalTitle{The Astrophysical Journal}}}
  \textbf{\bibinfo{volume}{893}}, \bibinfo{pages}{L26},
  \doiprefix\href{https://dx.doi.org/10.3847/2041-8213/ab83fb}{10.3847/2041-8213/ab83fb} (\bibinfo{year}{2020}).
\newblock \bibinfo{note}{ArXiv: 2002.08297}.

\bibitem{zanazzi_periodic_2020}
\bibinfo{author}{Zanazzi, J.~J.} \& \bibinfo{author}{Lai, D.}
\newblock \bibinfo{journal}{\bibinfo{title}{Periodic {Fast} {Radio} {Bursts}
  with {Neutron} {Star} {Free}/{Radiative} {Precession}}}.
\newblock {\emph{\JournalTitle{The Astrophysical Journal}}}
  \textbf{\bibinfo{volume}{892}}, \bibinfo{pages}{L15},
  \doiprefix\href{https://dx.doi.org/10.3847/2041-8213/ab7cdd}{10.3847/2041-8213/ab7cdd} (\bibinfo{year}{2020}).
\newblock \bibinfo{note}{ArXiv: 2002.05752}.

\bibitem{tong_periodicity_2020}
\bibinfo{author}{Tong, H.}, \bibinfo{author}{Wang, W.} \&
  \bibinfo{author}{Wang, H.~G.}
\newblock \bibinfo{journal}{\bibinfo{title}{Periodicity in fast radio bursts
  due to forced precession by a fallback disk}}.
\newblock {\emph{\JournalTitle{arXiv:2002.10265 [astro-ph]}}}
  (\bibinfo{year}{2020}).
\newblock \bibinfo{note}{ArXiv: 2002.10265}.

\bibitem{metzger_fast_2019}
\bibinfo{author}{Metzger, B.~D.}, \bibinfo{author}{Margalit, B.} \&
  \bibinfo{author}{Sironi, L.}
\newblock \bibinfo{journal}{\bibinfo{title}{Fast radio bursts as synchrotron
  maser emission from decelerating relativistic blast waves}}.
\newblock {\emph{\JournalTitle{Monthly Notices of the Royal Astronomical
  Society}}} \textbf{\bibinfo{volume}{485}}, \bibinfo{pages}{4091--4106},
  \doiprefix\href{https://dx.doi.org/10.1093/mnras/stz700}{10.1093/mnras/stz700} (\bibinfo{year}{2019}).
\newblock \bibinfo{note}{ArXiv: 1902.01866}.

\bibitem{beniamini_periodicity_2020}
\bibinfo{author}{Beniamini, P.}, \bibinfo{author}{Wadiasingh, Z.} \&
  \bibinfo{author}{Metzger, B.~D.}
\newblock \bibinfo{journal}{\bibinfo{title}{Periodicity in recurrent fast radio
  bursts and the origin of ultra long period magnetars}}.
\newblock {\emph{\JournalTitle{arXiv:2003.12509 [astro-ph]}}}
  (\bibinfo{year}{2020}).
\newblock \bibinfo{note}{ArXiv: 2003.12509}.

\bibitem{oosterloo_latest_2010}
\bibinfo{author}{Oosterloo, T.}, \bibinfo{author}{Verheijen, M.} \&
  \bibinfo{author}{van Cappellen, W.}
\newblock \bibinfo{journal}{\bibinfo{title}{The latest on {Apertif}}}.
\newblock {\emph{\JournalTitle{Proceedings of Science}}}
  \textbf{\bibinfo{volume}{ISKAF2010 Science Meeting}} (\bibinfo{year}{2010}).

\bibitem{adams_radio_2019}
\bibinfo{author}{Adams, E. A.~K.} \& \bibinfo{author}{van Leeuwen, J.}
\newblock \bibinfo{journal}{\bibinfo{title}{Radio surveys now both deep and
  wide}}.
\newblock {\emph{\JournalTitle{Nature Astronomy}}}
  \textbf{\bibinfo{volume}{3}}, \bibinfo{pages}{188--188},
  \doiprefix\href{https://dx.doi.org/10.1038/s41550-019-0692-4}{10.1038/s41550-019-0692-4} (\bibinfo{year}{2019}).

\bibitem{connor_bright_2020}
\bibinfo{author}{Connor, L.} \emph{et~al.}
\newblock \bibinfo{journal}{\bibinfo{title}{A bright, high rotation-measure
  {FRB} that skewers the {M33} halo}}.
\newblock {\emph{\JournalTitle{arXiv:2002.01399 [astro-ph]}}}
  (\bibinfo{year}{2020}).
\newblock \bibinfo{note}{ArXiv: 2002.01399}.

\bibitem{oostrum_repeating_2020}
\bibinfo{author}{Oostrum, L.~C.} \emph{et~al.}
\newblock \bibinfo{journal}{\bibinfo{title}{Repeating fast radio bursts with
  {WSRT}/{Apertif}}}.
\newblock {\emph{\JournalTitle{Astronomy \& Astrophysics}}}
  \textbf{\bibinfo{volume}{635}}, \bibinfo{pages}{A61},
  \doiprefix\href{https://dx.doi.org/10.1051/0004-6361/201937422}{10.1051/0004-6361/201937422} (\bibinfo{year}{2020}).
\newblock \bibinfo{note}{ArXiv: 1912.12217}.

\bibitem{leeu14}
\bibinfo{author}{{van Leeuwen}, J.}
\newblock \bibinfo{title}{{ARTS -- the Apertif Radio Transient System}}.
\newblock In \bibinfo{editor}{{Wozniak}, P.~R.}, \bibinfo{editor}{{Graham},
  M.~J.}, \bibinfo{editor}{{Mahabal}, A.~A.} \& \bibinfo{editor}{{Seaman}, R.}
  (eds.) \emph{\bibinfo{booktitle}{"The Third Hot-wiring the Transient Universe
  Workshop"}}, \bibinfo{pages}{79} (\bibinfo{year}{2014}).

\bibitem{artsso20}
\bibinfo{author}{{van Leeuwen}, J.} \emph{et~al.}
\newblock \bibinfo{journal}{\bibinfo{title}{{ARTS System Overview}}}.
\newblock {\emph{\JournalTitle{A\&A, in prep}}}  (\bibinfo{year}{2020}).

\bibitem{sclocco_real-time_2014}
\bibinfo{author}{Sclocco, A.}, \bibinfo{author}{Van~Nieuwpoort, R.} \&
  \bibinfo{author}{Bal, H.~E.}
\newblock \bibinfo{journal}{\bibinfo{title}{Real-{Time} {Pulsars} {Pipeline}
  {Using} {Many}-{Cores}}}.
\newblock \bibinfo{pages}{3} (\bibinfo{year}{2014}).
\newblock \bibinfo{note}{Conference Name: Exascale Radio Astronomy}.

\bibitem{sclocco_amber_2020}
\bibinfo{author}{Sclocco, A.}, \bibinfo{author}{Heldens, S.} \&
  \bibinfo{author}{van Werkhoven, B.}
\newblock \bibinfo{journal}{\bibinfo{title}{{AMBER}: {A} real-time pipeline for
  the detection of single pulse astronomical transients}}.
\newblock {\emph{\JournalTitle{SoftwareX}}} \textbf{\bibinfo{volume}{12}},
  \bibinfo{pages}{100549}, \doiprefix\href{https://dx.doi.org/10.1016/j.softx.2020.100549}{10.1016/j.softx.2020.100549}
  (\bibinfo{year}{2020}).

\bibitem{sclocco_real-time_2020}
\bibinfo{author}{{Sclocco}, A.}, \bibinfo{author}{{Vohl}, D.} \&
  \bibinfo{author}{{van Nieuwpoort}, R.~V.}
\newblock \bibinfo{title}{Real-time rfi mitigation for the apertif radio
  transient system}.
\newblock In \emph{\bibinfo{booktitle}{2019 RFI Workshop - Coexisting with
  Radio Frequency Interference (RFI)}}, \bibinfo{pages}{1--8},
  \doiprefix\href{https://dx.doi.org/10.23919/RFI48793.2019.9111826}{10.23919/RFI48793.2019.9111826} (\bibinfo{year}{2019}).
\newblock \bibinfo{note}{ArXiv: 2001.03389}.

\bibitem{oostrum_darc_2020}
\bibinfo{author}{{Oostrum}, L.~C.}
\newblock \bibinfo{title}{Darc: Data analysis of real-time candidates},
  \doiprefix\href{https://dx.doi.org/10.5281/zenodo.3784870}{10.5281/zenodo.3784870} (\bibinfo{year}{2020}).
\newblock \bibinfo{note}{\href{https://dx.doi.org/https://doi.org/10.5281/zenodo.3784870}{https://doi.org/10.5281/zenodo.3784870}}.

\bibitem{connor_applying_2018}
\bibinfo{author}{Connor, L.} \& \bibinfo{author}{van Leeuwen, J.}
\newblock \bibinfo{journal}{\bibinfo{title}{Applying {Deep} {Learning} to
  {Fast} {Radio} {Burst} {Classification}}}.
\newblock {\emph{\JournalTitle{The Astronomical Journal}}}
  \textbf{\bibinfo{volume}{156}}, \bibinfo{pages}{256},
  \doiprefix\href{https://dx.doi.org/10.3847/1538-3881/aae649}{10.3847/1538-3881/aae649} (\bibinfo{year}{2018}).

\bibitem{yao_new_2017}
\bibinfo{author}{Yao, J.~M.}, \bibinfo{author}{Manchester, R.~N.} \&
  \bibinfo{author}{Wang, N.}
\newblock \bibinfo{journal}{\bibinfo{title}{A {New} {Electron} {Density}
  {Model} for {Estimation} of {Pulsar} and {FRB} {Distances}}}.
\newblock {\emph{\JournalTitle{The Astrophysical Journal}}}
  \textbf{\bibinfo{volume}{835}}, \bibinfo{pages}{29},
  \doiprefix\href{https://dx.doi.org/10.3847/1538-4357/835/1/29}{10.3847/1538-4357/835/1/29} (\bibinfo{year}{2017}).
\newblock \bibinfo{note}{ArXiv: 1610.09448}.

\bibitem{van_haarlem_lofar_2013}
\bibinfo{author}{van Haarlem, M.~P.} \emph{et~al.}
\newblock \bibinfo{journal}{\bibinfo{title}{{LOFAR}: {The} {LOw}-{Frequency}
  {ARray}}}.
\newblock {\emph{\JournalTitle{Astronomy \& Astrophysics}}}
  \textbf{\bibinfo{volume}{556}}, \bibinfo{pages}{A2},
  \doiprefix\href{https://dx.doi.org/10.1051/0004-6361/201220873}{10.1051/0004-6361/201220873} (\bibinfo{year}{2013}).

\bibitem{Maan2020}
\bibinfo{author}{{Maan}, Y.}, \bibinfo{author}{{van Leeuwen}, J.} \&
  \bibinfo{author}{{Vohl}, D.}
\newblock \bibinfo{journal}{\bibinfo{title}{Fourier domain excision of periodic
  radio frequency interference}}.
\newblock {\emph{\JournalTitle{in prep.}}}  (\bibinfo{year}{2020}).

\bibitem{kondratiev_lofar_2016}
\bibinfo{author}{Kondratiev, V.~I.} \emph{et~al.}
\newblock \bibinfo{journal}{\bibinfo{title}{A {LOFAR} census of millisecond
  pulsars}}.
\newblock {\emph{\JournalTitle{Astronomy \& Astrophysics}}}
  \textbf{\bibinfo{volume}{585}}, \bibinfo{pages}{A128},
  \doiprefix\href{https://dx.doi.org/10.1051/0004-6361/201527178}{10.1051/0004-6361/201527178} (\bibinfo{year}{2016}).

\bibitem{bilous_lofar_2016}
\bibinfo{author}{Bilous, A.~V.} \emph{et~al.}
\newblock \bibinfo{journal}{\bibinfo{title}{A {LOFAR} census of non-recycled
  pulsars: average profiles, dispersion measures, flux densities, and
  spectra}}.
\newblock {\emph{\JournalTitle{Astronomy \& Astrophysics}}}
  \textbf{\bibinfo{volume}{591}}, \bibinfo{pages}{A134},
  \doiprefix\href{https://dx.doi.org/10.1051/0004-6361/201527702}{10.1051/0004-6361/201527702} (\bibinfo{year}{2016}).

\bibitem{houben_constraints_2019}
\bibinfo{author}{Houben, L. J.~M.} \emph{et~al.}
\newblock \bibinfo{journal}{\bibinfo{title}{Constraints on the low frequency
  spectrum of {FRB} 121102}}.
\newblock {\emph{\JournalTitle{Astronomy \& Astrophysics}}}
  \textbf{\bibinfo{volume}{623}}, \bibinfo{pages}{A42},
  \doiprefix\href{https://dx.doi.org/10.1051/0004-6361/201833875}{10.1051/0004-6361/201833875} (\bibinfo{year}{2019}).

\bibitem{2019A&A...626A.104S}
\bibinfo{author}{{Sanidas}, S.} \emph{et~al.}
\newblock \bibinfo{journal}{\bibinfo{title}{{The LOFAR Tied-Array All-Sky
  Survey (LOTAAS): Survey overview and initial pulsar discoveries}}}.
\newblock {\emph{\JournalTitle{Astronomy \& Astrophysics}}} \textbf{\bibinfo{volume}{626}},
  \bibinfo{pages}{A104}, \doiprefix\href{https://dx.doi.org/10.1051/0004-6361/201935609}{10.1051/0004-6361/201935609}
  (\bibinfo{year}{2019}).
\newblock \eprint{1905.04977}.

\bibitem{scholz_simultaneous_2020}
\bibinfo{author}{Scholz, P.} \emph{et~al.}
\newblock \bibinfo{journal}{\bibinfo{title}{Simultaneous {X}-ray and {Radio}
  {Observations} of the {Repeating} {Fast} {Radio} {Burst} {FRB}
  180916.{J0158}+65}}.
\newblock {\emph{\JournalTitle{arXiv:2004.06082 [astro-ph]}}}
  (\bibinfo{year}{2020}).
\newblock \bibinfo{note}{ArXiv: 2004.06082}.

\bibitem{pearlman_multiwavelength_2020}
\bibinfo{author}{Pearlman, A.~B.} \emph{et~al.}
\newblock \bibinfo{journal}{\bibinfo{title}{Multiwavelength {Radio}
  {Observations} of {Two} {Repeating} {Fast} {Radio} {Burst} {Sources}: {FRB}
  121102 and {FRB} 180916.{J0158}+65}}.
\newblock {\emph{\JournalTitle{arXiv:2009.13559 [astro-ph]}}}
  (\bibinfo{year}{2020}).
\newblock \bibinfo{note}{ArXiv: 2009.13559}.

\bibitem{nimmo_microsecond_2020}
\bibinfo{author}{Nimmo, K.} \emph{et~al.}
\newblock \bibinfo{journal}{\bibinfo{title}{Microsecond polarimetry of the
  repeating {FRB} {20180916B}}}.
\newblock {\emph{\JournalTitle{arXiv:2010.05800 [astro-ph]}}}
  (\bibinfo{year}{2020}).
\newblock \bibinfo{note}{ArXiv: 2010.05800}.

\bibitem{blaskiewicz_relativistic_1991}
\bibinfo{author}{Blaskiewicz, M.}, \bibinfo{author}{Cordes, J.~M.} \&
  \bibinfo{author}{Wasserman, I.}
\newblock \bibinfo{journal}{\bibinfo{title}{A relativistic model of pulsar
  polarization}}.
\newblock {\emph{\JournalTitle{The Astrophysical Journal}}}
  \textbf{\bibinfo{volume}{370}}, \bibinfo{pages}{643},
  \doiprefix\href{https://dx.doi.org/10.1086/169850}{10.1086/169850} (\bibinfo{year}{1991}).

\bibitem{cho_spectropolarimetric_2020}
\bibinfo{author}{Cho, H.} \emph{et~al.}
\newblock \bibinfo{journal}{\bibinfo{title}{Spectropolarimetric analysis of
  {FRB} 181112 at microsecond resolution: {Implications} for {Fast} {Radio}
  {Burst} emission mechanism}}.
\newblock {\emph{\JournalTitle{arXiv:2002.12539 [astro-ph]}}}
  \doiprefix\href{https://dx.doi.org/10.3847/2041-8213/ab7824}{10.3847/2041-8213/ab7824} (\bibinfo{year}{2020}).
\newblock \bibinfo{note}{ArXiv: 2002.12539}.

\bibitem{beloborodov_2019}
\bibinfo{author}{{Beloborodov}, A.~M.}
\newblock \bibinfo{journal}{\bibinfo{title}{{Blast Waves from Magnetar Flares
  and Fast Radio Bursts}}}.
\newblock {\emph{\JournalTitle{\apj}}} \textbf{\bibinfo{volume}{896}},
  \bibinfo{pages}{142}, \doiprefix\href{https://dx.doi.org/10.3847/1538-4357/ab83eb}{10.3847/1538-4357/ab83eb}
  (\bibinfo{year}{2020}).
\newblock \eprint{1908.07743}.

\bibitem{wang_time-frequency_2019}
\bibinfo{author}{Wang, W.}, \bibinfo{author}{Zhang, B.}, \bibinfo{author}{Chen,
  X.} \& \bibinfo{author}{Xu, R.}
\newblock \bibinfo{journal}{\bibinfo{title}{On the {Time}-{Frequency}
  {Downward} {Drifting} of {Repeating} {Fast} {Radio} {Bursts}}}.
\newblock {\emph{\JournalTitle{The Astrophysical Journal}}}
  \textbf{\bibinfo{volume}{876}}, \bibinfo{pages}{L15},
  \doiprefix\href{https://dx.doi.org/10.3847/2041-8213/ab1aab}{10.3847/2041-8213/ab1aab} (\bibinfo{year}{2019}).
\newblock \bibinfo{note}{ArXiv: 1903.03982}.

\bibitem{rajabi_simple_2020}
\bibinfo{author}{Rajabi, F.}, \bibinfo{author}{Chamma, M.~A.},
  \bibinfo{author}{Wyenberg, C.~M.}, \bibinfo{author}{Mathews, A.} \&
  \bibinfo{author}{Houde, M.}
\newblock \bibinfo{journal}{\bibinfo{title}{A simple relationship for the
  spectro-temporal structure of bursts from {FRB} 121102}}.
\newblock {\emph{\JournalTitle{arXiv:2008.02395 [astro-ph]}}}
  (\bibinfo{year}{2020}).
\newblock \bibinfo{note}{ArXiv: 2008.02395}.

\bibitem{krishnakumar_multi-frequency_2017}
\bibinfo{author}{Krishnakumar, M.~A.}, \bibinfo{author}{Joshi, B.~C.} \&
  \bibinfo{author}{Manoharan, P.~K.}
\newblock \bibinfo{journal}{\bibinfo{title}{Multi-frequency {Scatter}
  {Broadening} {Evolution} of {Pulsars}. {I}}}.
\newblock {\emph{\JournalTitle{The Astrophysical Journal}}}
  \textbf{\bibinfo{volume}{846}}, \bibinfo{pages}{104},
  \doiprefix\href{https://dx.doi.org/10.3847/1538-4357/aa7af2}{10.3847/1538-4357/aa7af2} (\bibinfo{year}{2017}).

\bibitem{maan_distinct_2019}
\bibinfo{author}{Maan, Y.}, \bibinfo{author}{Joshi, B.~C.},
  \bibinfo{author}{Surnis, M.~P.}, \bibinfo{author}{Bagchi, M.} \&
  \bibinfo{author}{Manoharan, P.~K.}
\newblock \bibinfo{journal}{\bibinfo{title}{Distinct properties of the radio
  burst emission from the magnetar {XTE} {J1810}-197}}.
\newblock {\emph{\JournalTitle{The Astrophysical Journal}}}
  \textbf{\bibinfo{volume}{882}}, \bibinfo{pages}{L9},
  \doiprefix\href{https://dx.doi.org/10.3847/2041-8213/ab3a47}{10.3847/2041-8213/ab3a47} (\bibinfo{year}{2019}).
\newblock \bibinfo{note}{ArXiv: 1908.04304}.

\bibitem{rane_search_2016}
\bibinfo{author}{Rane, A.} \emph{et~al.}
\newblock \bibinfo{journal}{\bibinfo{title}{A search for rotating radio
  transients and fast radio bursts in the {Parkes} high-latitude pulsar
  survey}}.
\newblock {\emph{\JournalTitle{Monthly Notices of the Royal Astronomical
  Society}}} \textbf{\bibinfo{volume}{455}}, \bibinfo{pages}{2207--2215},
  \doiprefix\href{https://dx.doi.org/10.1093/mnras/stv2404}{10.1093/mnras/stv2404} (\bibinfo{year}{2016}).
\newblock \bibinfo{note}{ArXiv: 1505.00834}.

\bibitem{lawrence_non-homogeneous_2017}
\bibinfo{author}{Lawrence, E.}, \bibinfo{author}{Wiel, S.~V.},
  \bibinfo{author}{Law, C.~J.}, \bibinfo{author}{Spolaor, S.~B.} \&
  \bibinfo{author}{Bower, G.~C.}
\newblock \bibinfo{journal}{\bibinfo{title}{The {Non}-homogeneous {Poisson}
  {Process} for {Fast} {Radio} {Burst} {Rates}}}.
\newblock {\emph{\JournalTitle{The Astronomical Journal}}}
  \textbf{\bibinfo{volume}{154}}, \bibinfo{pages}{117},
  \doiprefix\href{https://dx.doi.org/10.3847/1538-3881/aa844e}{10.3847/1538-3881/aa844e} (\bibinfo{year}{2017}).
\newblock \bibinfo{note}{ArXiv: 1611.00458}.

\bibitem{vedantham_fluence_2016}
\bibinfo{author}{Vedantham, H.~K.}, \bibinfo{author}{Ravi, V.},
  \bibinfo{author}{Hallinan, G.} \& \bibinfo{author}{Shannon, R.}
\newblock \bibinfo{journal}{\bibinfo{title}{The {Fluence} and {Distance}
  {Distributions} of {Fast} {Radio} {Bursts}}}.
\newblock {\emph{\JournalTitle{The Astrophysical Journal}}}
  \textbf{\bibinfo{volume}{830}}, \bibinfo{pages}{75},
  \doiprefix\href{https://dx.doi.org/10.3847/0004-637X/830/2/75}{10.3847/0004-637X/830/2/75} (\bibinfo{year}{2016}).
\newblock \bibinfo{note}{ArXiv: 1606.06795}.

\bibitem{keane_fast_2015}
\bibinfo{author}{Keane, E.~F.} \& \bibinfo{author}{Petroff, E.}
\newblock \bibinfo{journal}{\bibinfo{title}{Fast radio bursts: search
  sensitivities and completeness}}.
\newblock {\emph{\JournalTitle{Monthly Notices of the Royal Astronomical
  Society}}} \textbf{\bibinfo{volume}{447}}, \bibinfo{pages}{2852--2856},
  \doiprefix\href{https://dx.doi.org/10.1093/mnras/stu2650}{10.1093/mnras/stu2650} (\bibinfo{year}{2015}).
\newblock \bibinfo{note}{ArXiv: 1409.6125}.

\bibitem{connor_interpreting_2019}
\bibinfo{author}{Connor, L.}
\newblock \bibinfo{journal}{\bibinfo{title}{Interpreting the distributions of
  {FRB} observables}}.
\newblock {\emph{\JournalTitle{Monthly Notices of the Royal Astronomical
  Society}}} \textbf{\bibinfo{volume}{487}}, \bibinfo{pages}{5753--5763},
  \doiprefix\href{https://dx.doi.org/10.1093/mnras/stz1666}{10.1093/mnras/stz1666} (\bibinfo{year}{2019}).
\newblock \bibinfo{note}{ArXiv: 1905.00755}.

\bibitem{ter_veen_frats_2019}
\bibinfo{author}{ter Veen, S.} \emph{et~al.}
\newblock \bibinfo{journal}{\bibinfo{title}{The {FRATS} project: real-time
  searches for fast radio bursts and other fast transients with {LOFAR} at 135
  {MHz}}}.
\newblock {\emph{\JournalTitle{Astronomy \& Astrophysics}}}
  \textbf{\bibinfo{volume}{621}}, \bibinfo{pages}{A57},
  \doiprefix\href{https://dx.doi.org/10.1051/0004-6361/201732515}{10.1051/0004-6361/201732515} (\bibinfo{year}{2019}).

\bibitem{tingay_search_2015}
\bibinfo{author}{Tingay, S.~J.} \emph{et~al.}
\newblock \bibinfo{journal}{\bibinfo{title}{A search for {Fast} {Radio}
  {Bursts} at low frequencies with {Murchison} {Widefield} {Array} high time
  resolution imaging}}.
\newblock {\emph{\JournalTitle{The Astronomical Journal}}}
  \textbf{\bibinfo{volume}{150}}, \bibinfo{pages}{199},
  \doiprefix\href{https://dx.doi.org/10.1088/0004-6256/150/6/199}{10.1088/0004-6256/150/6/199} (\bibinfo{year}{2015}).
\newblock \bibinfo{note}{ArXiv: 1511.02985}.

\bibitem{rowlinson_limits_2016}
\bibinfo{author}{Rowlinson, A.} \emph{et~al.}
\newblock \bibinfo{journal}{\bibinfo{title}{Limits on {Fast} {Radio} {Bursts}
  and other transient sources at 182 {MHz} using the {Murchison} {Widefield}
  {Array}}}.
\newblock {\emph{\JournalTitle{Monthly Notices of the Royal Astronomical
  Society}}} \textbf{\bibinfo{volume}{458}}, \bibinfo{pages}{3506--3522},
  \doiprefix\href{https://dx.doi.org/10.1093/mnras/stw451}{10.1093/mnras/stw451} (\bibinfo{year}{2016}).

\bibitem{sokolowski_no_2018}
\bibinfo{author}{Sokolowski, M.} \emph{et~al.}
\newblock \bibinfo{journal}{\bibinfo{title}{No low-frequency emission from
  extremely bright {Fast} {Radio} {Bursts}}}.
\newblock {\emph{\JournalTitle{The Astrophysical Journal}}}
  \textbf{\bibinfo{volume}{867}}, \bibinfo{pages}{L12},
  \doiprefix\href{https://dx.doi.org/10.3847/2041-8213/aae58d}{10.3847/2041-8213/aae58d} (\bibinfo{year}{2018}).
\newblock \bibinfo{note}{ArXiv: 1810.04355}.

\bibitem{huijse_robust_2018}
\bibinfo{author}{Huijse, P.} \emph{et~al.}
\newblock \bibinfo{journal}{\bibinfo{title}{Robust {Period} {Estimation}
  {Using} {Mutual} {Information} for {Multiband} {Light} {Curves} in the
  {Synoptic} {Survey} {Era}}}.
\newblock {\emph{\JournalTitle{The Astrophysical Journal Supplement Series}}}
  \textbf{\bibinfo{volume}{236}}, \bibinfo{pages}{12},
  \doiprefix\href{https://dx.doi.org/10.3847/1538-4365/aab77c}{10.3847/1538-4365/aab77c} (\bibinfo{year}{2018}).

\bibitem{ransom_new_2001}
\bibinfo{author}{Ransom, S.~M.}
\newblock \emph{\bibinfo{title}{New {Search} {Techniques} {For} {Binary}
  {Pulsars}}}.
\newblock Ph.D. thesis, \bibinfo{school}{Harvard University}
  (\bibinfo{year}{2001}).

\bibitem{perley_accurate_2017}
\bibinfo{author}{Perley, R.~A.} \& \bibinfo{author}{Butler, B.~J.}
\newblock \bibinfo{journal}{\bibinfo{title}{An {Accurate} {Flux} {Density}
  {Scale} from 50 {MHz} to 50 {GHz}}}.
\newblock {\emph{\JournalTitle{The Astrophysical Journal Supplement Series}}}
  \textbf{\bibinfo{volume}{230}}, \bibinfo{pages}{7},
  \doiprefix\href{https://dx.doi.org/10.3847/1538-4365/aa6df9}{10.3847/1538-4365/aa6df9} (\bibinfo{year}{2017}).

\bibitem{cordes_searches_2003}
\bibinfo{author}{Cordes, J.~M.} \& \bibinfo{author}{McLaughlin, M.~A.}
\newblock \bibinfo{journal}{\bibinfo{title}{Searches for {Fast} {Radio}
  {Transients}}}.
\newblock {\emph{\JournalTitle{The Astrophysical Journal}}}
  \textbf{\bibinfo{volume}{596}}, \bibinfo{pages}{1142--1154},
  \doiprefix\href{https://dx.doi.org/10.1086/378231}{10.1086/378231} (\bibinfo{year}{2003}).

\bibitem{maan_deep_2014}
\bibinfo{author}{Maan, Y.} \& \bibinfo{author}{Aswathappa, H.~A.}
\newblock \bibinfo{journal}{\bibinfo{title}{Deep searches for
  decametre-wavelength pulsed emission from radio-quiet gamma-ray pulsars}}.
\newblock {\emph{\JournalTitle{Monthly Notices of the Royal Astronomical
  Society}}} \textbf{\bibinfo{volume}{445}}, \bibinfo{pages}{3221--3228},
  \doiprefix\href{https://dx.doi.org/10.1093/mnras/stu1902}{10.1093/mnras/stu1902} (\bibinfo{year}{2014}).

\bibitem{scott_multivariate_2015}
\bibinfo{author}{Scott, D.~W.}
\newblock \emph{\bibinfo{title}{Multivariate {Density} {Estimation}: {Theory},
  {Practice}, and {Visualization}}} (\bibinfo{publisher}{John Wiley \& Sons},
  \bibinfo{year}{2015}).
\newblock \bibinfo{note}{Google-Books-ID: pIAZBwAAQBAJ}.

\bibitem{rm-tools}
\bibinfo{author}{{Purcell}, C.~R.}, \bibinfo{author}{{Van Eck}, C.~L.},
  \bibinfo{author}{{West}, J.}, \bibinfo{author}{{Sun}, X.~H.} \&
  \bibinfo{author}{{Gaensler}, B.~M.}
\newblock \bibinfo{title}{{RM-Tools: Rotation measure (RM) synthesis and Stokes
  QU-fitting}} (\bibinfo{year}{2020}).
\newblock \eprint{2005.003}.

\bibitem{ordog-2019}
\bibinfo{author}{{Ordog}, A.}, \bibinfo{author}{{Booth}, R.~A.},
  \bibinfo{author}{{van Eck}, C.~L.}, \bibinfo{author}{{Brown}, J. A.~C.} \&
  \bibinfo{author}{{Landecker}, T.~L.}
\newblock \bibinfo{journal}{\bibinfo{title}{{VizieR Online Data Catalog:
  Faraday rotation of extended emission (Ordog+, 2019)}}}.
\newblock {\emph{\JournalTitle{VizieR Online Data Catalog (other)}}}
  \textbf{\bibinfo{volume}{0640}}, \bibinfo{pages}{J/other/Galax/7}
  (\bibinfo{year}{2019}).

\bibitem{hotan_psrchive_2004}
\bibinfo{author}{Hotan, A.~W.}, \bibinfo{author}{van Straten, W.} \&
  \bibinfo{author}{Manchester, R.~N.}
\newblock \bibinfo{journal}{\bibinfo{title}{Psrchive and {Psrfits} : {An}
  {Open} {Approach} to {Radio} {Pulsar} {Data} {Storage} and {Analysis}}}.
\newblock {\emph{\JournalTitle{Publications of the Astronomical Society of
  Australia}}} \textbf{\bibinfo{volume}{21}}, \bibinfo{pages}{302--309},
  \doiprefix\href{https://dx.doi.org/10.1071/AS04022}{10.1071/AS04022} (\bibinfo{year}{2004}).

\bibitem{kulkarni_dispersion_2020}
\bibinfo{author}{Kulkarni, S.~R.}
\newblock \bibinfo{journal}{\bibinfo{title}{Dispersion measure: {Confusion},
  {Constants} \& {Clarity}}}.
\newblock {\emph{\JournalTitle{arXiv:2007.02886 [astro-ph]}}}
  (\bibinfo{year}{2020}).
\newblock \bibinfo{note}{ArXiv: 2007.02886}.

\bibitem{caleb_simultaneous_2020}
\bibinfo{author}{Caleb, M.} \emph{et~al.}
\newblock \bibinfo{journal}{\bibinfo{title}{Simultaneous multi-telescope
  observations of {FRB} 121102}}.
\newblock {\emph{\JournalTitle{Monthly Notices of the Royal Astronomical
  Society}}} \textbf{\bibinfo{volume}{496}}, \bibinfo{pages}{4565--4573},
  \doiprefix\href{https://dx.doi.org/10.1093/mnras/staa1791}{10.1093/mnras/staa1791} (\bibinfo{year}{2020}).
\newblock \bibinfo{note}{ArXiv: 2006.08662}.

\end{thebibliography}

\newpage
\begin{addendum}
\item[\emph{Acknowledgements}]
This research was supported by 
the European Research Council under the European Union's Seventh Framework Programme
(FP/2007-2013)/ERC Grant Agreement No. 617199 (`ALERT'), 
and by Vici research programme `ARGO' with project number
639.043.815, financed by the Dutch Research Council (NWO). 
Instrumentation development was supported 
by NWO (grant 614.061.613 `ARTS') and the  
Netherlands Research School for Astronomy (`NOVA4-ARTS', `NOVA-NW3', and `NOVA5-NW3-10.3.5.14').
We further acknowledge funding from an NWO Veni Fellowship to EP;
from Netherlands eScience Center (NLeSC) grant ASDI.15.406 to DV and AS;
from National Aeronautics and Space Administration (NASA) grant number NNX17AL74G issued
through the NNH16ZDA001N Astrophysics Data Analysis Program (ADAP) to SMS;
by the WISE research programme, financed by NWO, to EAKA;
from FP/2007-2013 ERC Grant Agreement No. 291531 (`HIStoryNU') to JMvdH;
and from VINNOVA VINNMER grant 2009-01175 to VMI.\\
This work makes use of data from the Apertif system installed at the Westerbork Synthesis Radio Telescope owned by ASTRON. ASTRON, the Netherlands Institute for Radio Astronomy, is an institute of NWO.\\
This paper is based (in part) on data obtained with the International LOFAR Telescope (ILT) under project code
COM\_ALERT.
These data are accessible through the LOFAR Long Term Archive, \url{https://lta.lofar.eu/},
by searching for ``Observations'' with the
J2000 coordinates RA 01:57:43.2000 and DEC +65:42:01.020, or by selecting COM\_ALERT in ``Other projects'' and downloading
data which includes R3 in the ``Observation Description''.
LOFAR (van Haarlem et al. 2013) is the Low Frequency Array designed and constructed by ASTRON. It has observing, data processing, and data storage facilities in several countries, that are owned by various parties (each with their own funding sources), and that are collectively operated by the ILT foundation under a joint scientific policy. The ILT resources have benefitted from the following recent major funding sources: CNRS-INSU, Observatoire de Paris and Université d'Orléans, France; BMBF, MIWF-NRW, MPG, Germany; Science Foundation Ireland (SFI), Department of Business, Enterprise and Innovation (DBEI), Ireland; NWO, The Netherlands; The Science and Technology Facilities Council, UK; Ministry of Science and Higher Education, Poland.\\
We acknowledge use of the CHIME/FRB Public Database, provided at \url{https://www.chime-frb.ca/} by the
CHIME/FRB Collaboration.

\item[\emph{Author contributions}]
IPM, LC, JVL, YM, STV, AB, LO, EP, SS and DV analysed and interpreted the data. 
IPM, LC, JVL, YM and STV contributed to the LOFAR data acquisition, and to the 
conception, design and creation of LOFAR analysis software.
IPM, LC and JVL conceived and drafted the work, and YM, STV, AB, LO, EP, SS and DV contributed significant revisions. 
LO, JA, OB, EK, DVDS, AS, RS, EAKA, BA, WJDB, AHWMC, SD, HD, KMH, TVDH, BH, VMI, AK, GML, DML, AM, VAM, HM, MJN, TO, EO,
MR and
SJW contributed to the conception, design and creation of the Apertif hardware, software, and firmware used in this
work, and to the Apertif data acquisition.
\item[\emph{Competing Interests}] The authors declare that they have no competing financial interests.

\end{addendum}

\newpage
\section*{Extended data}

\renewcommand\thefigure{Extended Figure \arabic{figure}}
\setcounter{figure}{0}
\makeatletter
\renewenvironment{figure}{\let\caption\NAT@extfigcaption}{}
\makeatother

\renewcommand\thetable{Extended Table \arabic{table}}
\setcounter{table}{0}
\renewcommand{\tablename}{}

\begin{figure}
\centering
\includegraphics[width=\linewidth]{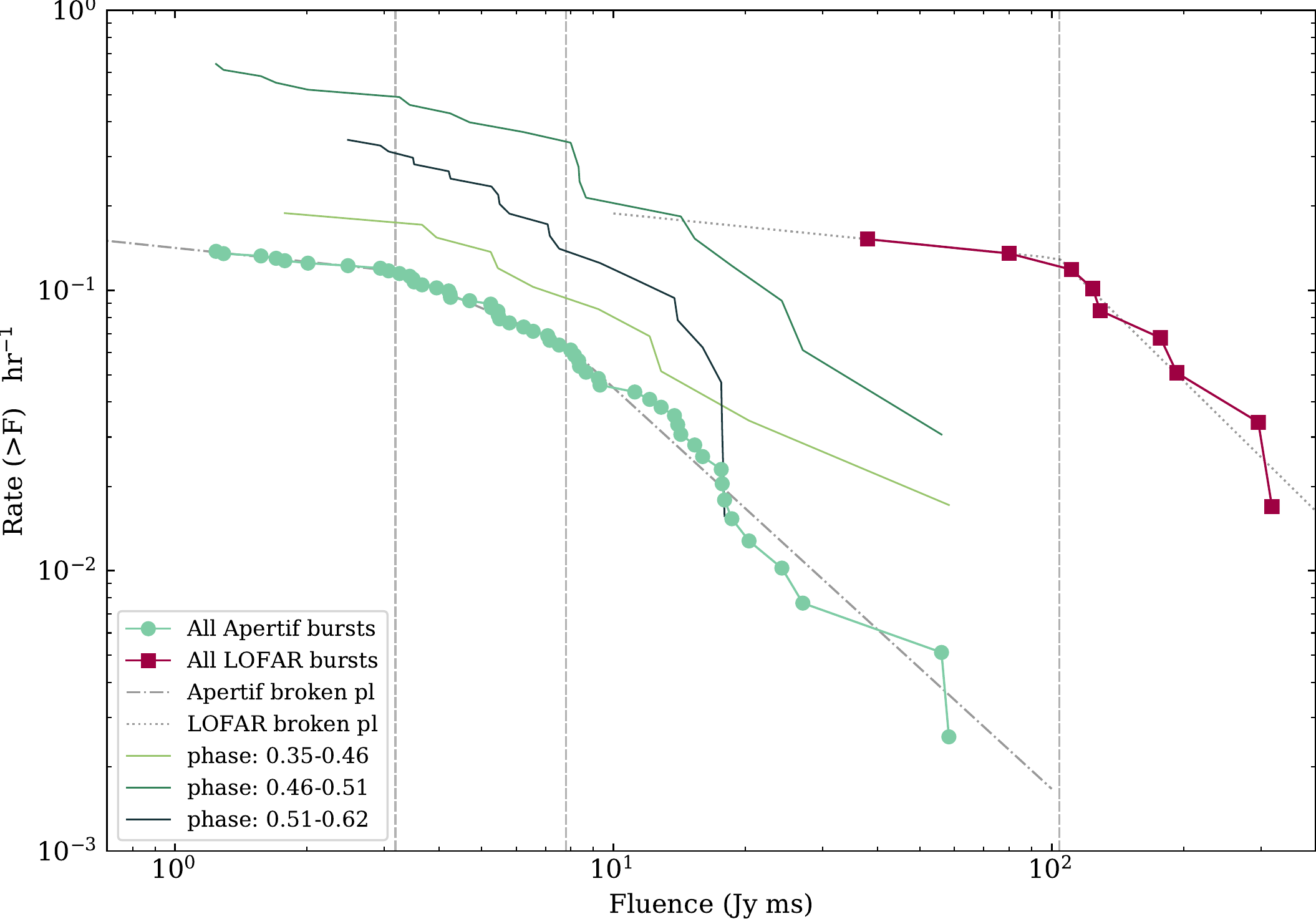}\\
\caption{Cumulative distribution function of burst rate in fluence for both Apertif and LOFAR. The light green markers
  show the CDF of all Apertif bursts, with dash-dotted, dotted and dashed lines giving the power-law fit respectively to
  bursts with fluences lower than 3.2\jymsnospace, between 3.2 and 7.8\jyms and above 7.8\jymsnospace. The coloured
  solid lines correspond to different phase ranges within the active window, with no discernible difference between them
  other than the rate scaling. The LOFAR fluence 
distribution is shown in crimson. The fit to a broken power law with a fluence turnover at 104\jyms is shown as a gray dotted line. For the same fluence, \FRB is more active at 150\,MHz than 1370\,MHz, even at the peak activity phases observed by Apertif.
\label{fig:fluence_cdf}
}
\end{figure}

\begin{figure}
  \centering
\includegraphics[width=0.85\textwidth]{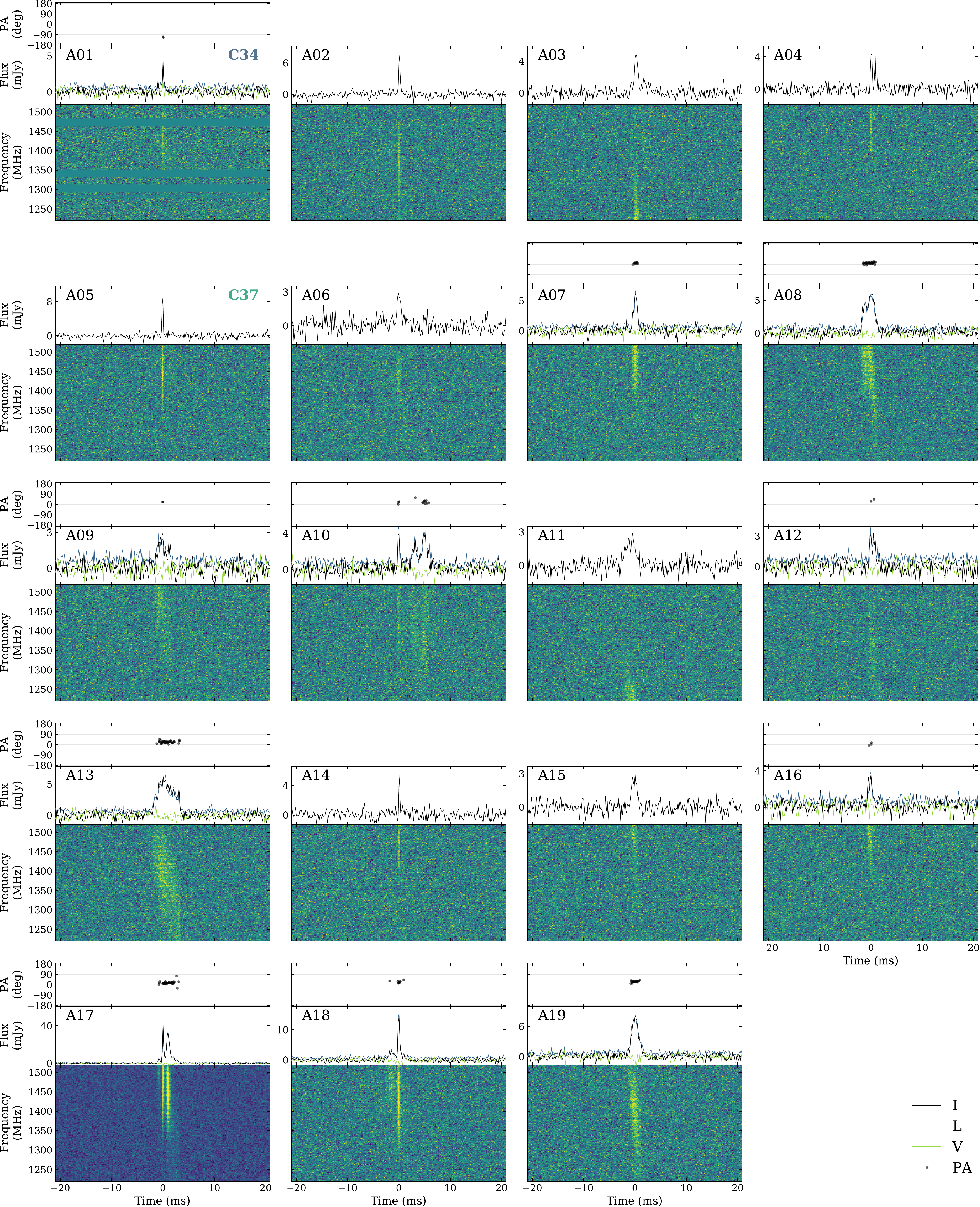}\\
\caption{Dynamic spectra of the bursts (A01-A19) from \FRB detected with Apertif, dedispersed to a DM of
  348.80\pccmnospace. Bursts with full-Stokes data show PA (degrees) in the top panel, ILV in the central panel and dynamic
  spectrum in the bottom one. Bursts with only intensity data show pulse profile in the top panel and dynamic spectrum
  in the bottom one. The burst identifiers are given in the top left corner of the pulse profiles. The dynamic spectra
  have been downsampled by factors 2 and 8 in time and frequency. The activity cycle number is indicated on the top right corner of the first detected burst of each cycle. 
\label{fig:Apertif_bursts}
}
\end{figure}

\clearpage
\addtocounter{figure}{-1}
\begin{figure}
\centering
\includegraphics[width=0.85\textwidth]{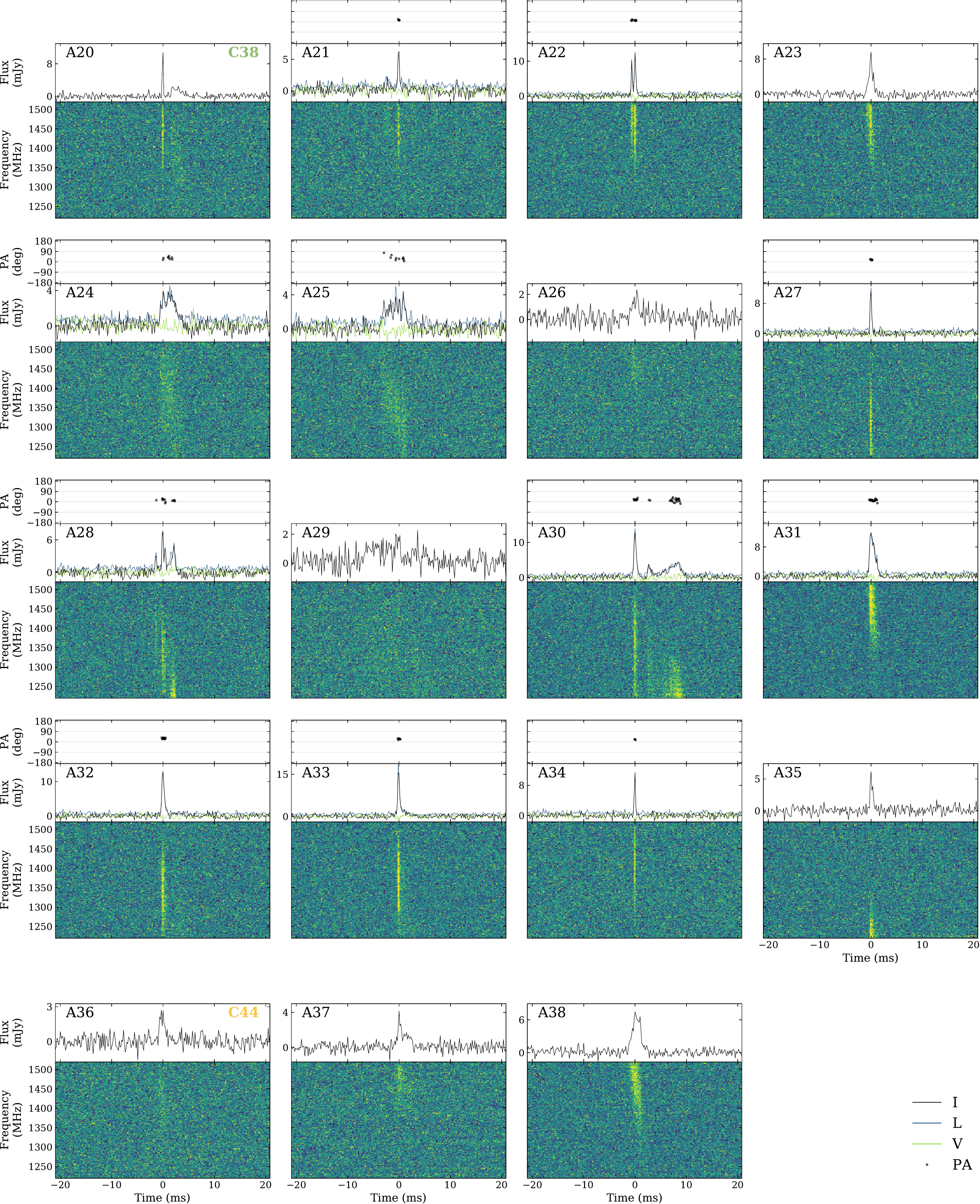}\\
\caption{Continued, bursts (A20-A38).}
\end{figure}

\clearpage
\addtocounter{figure}{-1}
\begin{figure}
\centering
\includegraphics[width=0.85\textwidth]{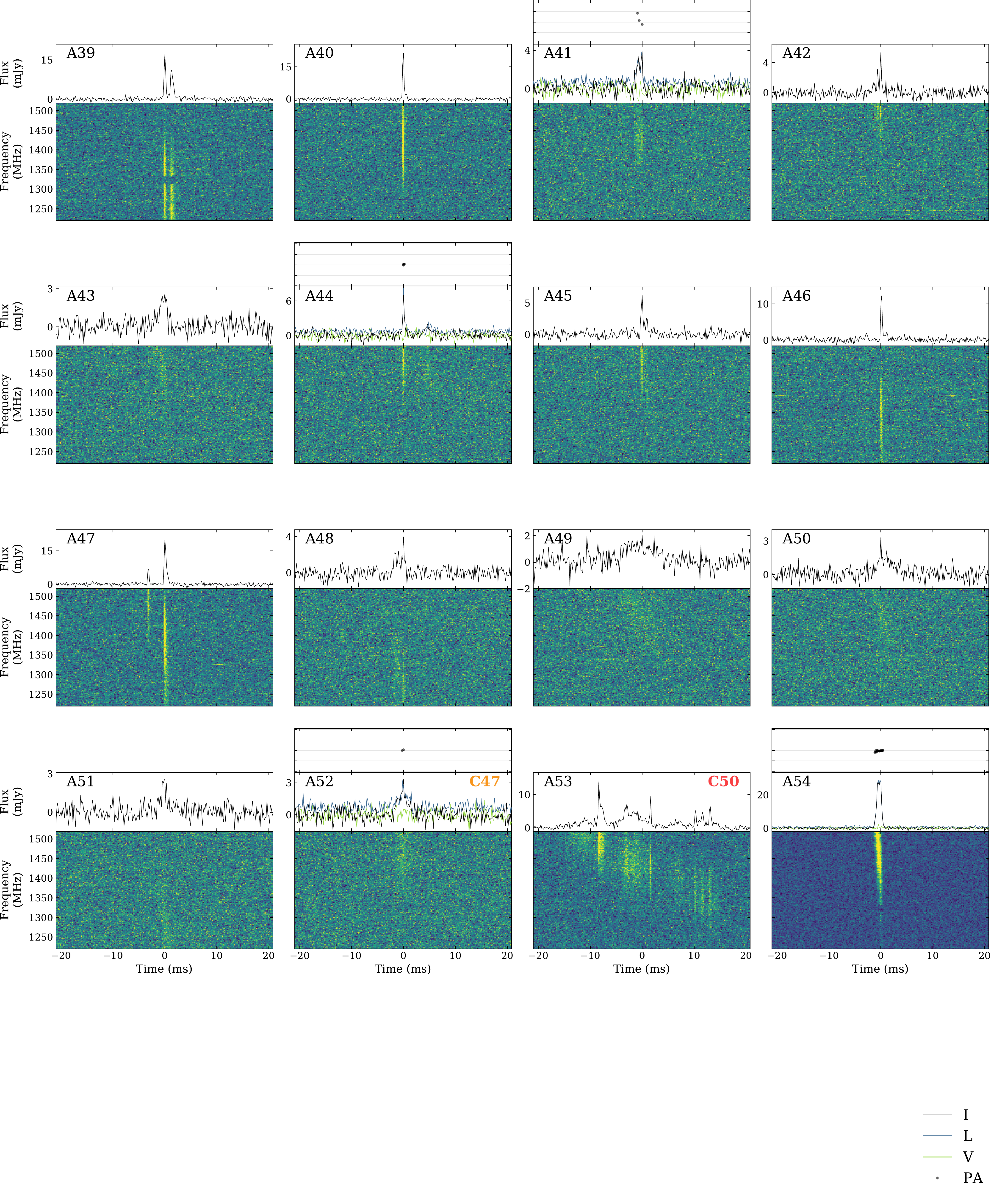}\\
\caption{Continued, bursts (A39-A54).}
\end{figure}

\clearpage
\begin{figure}
\centering
\includegraphics[width=\linewidth]{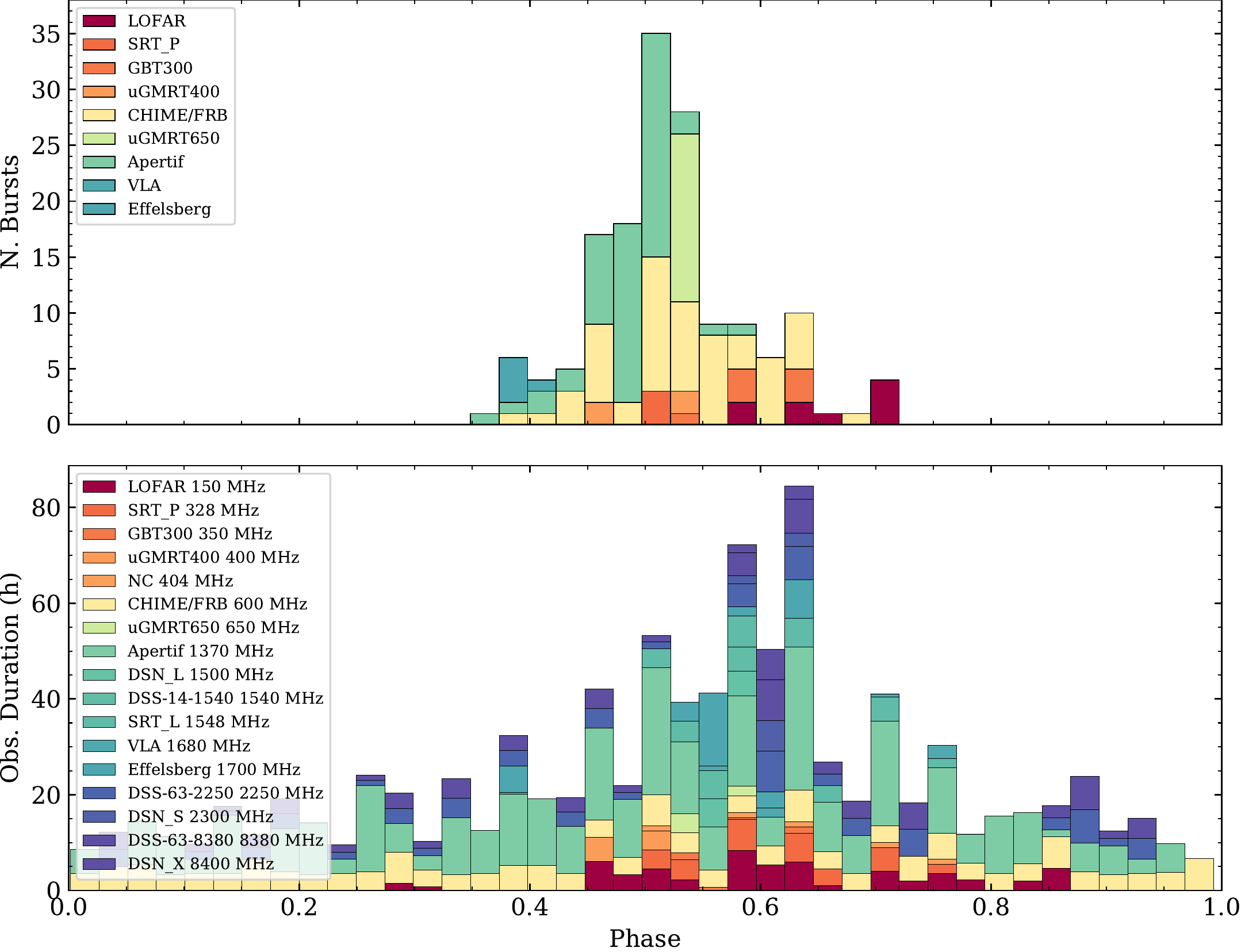}\\
\caption{Histogram of burst detections (top) and observation duration (bottom) as a function of phase for the best
  period fitted to Apertif and CHIME/FRB data (16.29 days). Instruments are color-coded by central frequency, with blue
  for high frequencies and red for low frequencies. Figure was generated using an adaptation of the  \texttt{frbpa}
  package\cite{aggarwal_vlarealfast_2020}.
\label{fig:burst_phase}
}
\end{figure}

\begin{figure}
\centering
\includegraphics[width=0.6\linewidth]{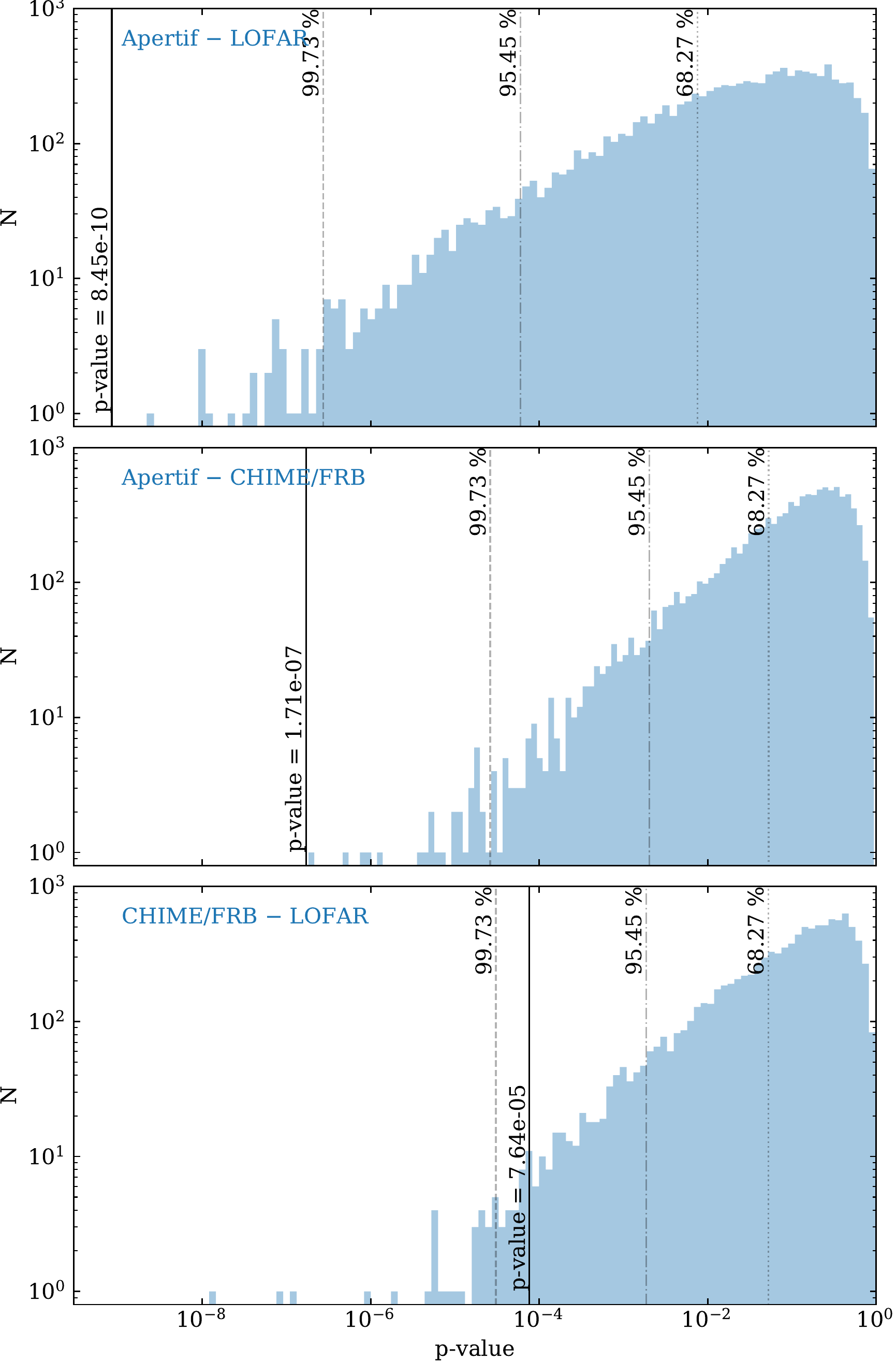}\\
\caption{Comparison of simulated and observed activity window p-values. Each panel compares the p-value obtained through the Kolmogorov-Smirnov statistic on two instrument burst samples. The vertical black lines give the observed p-value, whereas the histograms correspond to 10000 simulations of the p-value that would be obtained if both instrument burst samples were drawn from the same distribution. The top panel compares the burst samples from Apertif and LOFAR, the central panel from Apertif and CHIME/FRB and the bottom panel from CHIME/FRB and LOFAR. The vertical gray dotted, dash-dotted and dashed lines show respectively the p-value where 68.27\% (1$\sigma$), 95.45\% (2$\sigma$) and 99.73\% (3$\sigma$) of the simulations give a larger p-value.
\label{fig:simulated_ks}
}
\end{figure}

\begin{figure}
  \centering
\includegraphics[width=0.8\linewidth]{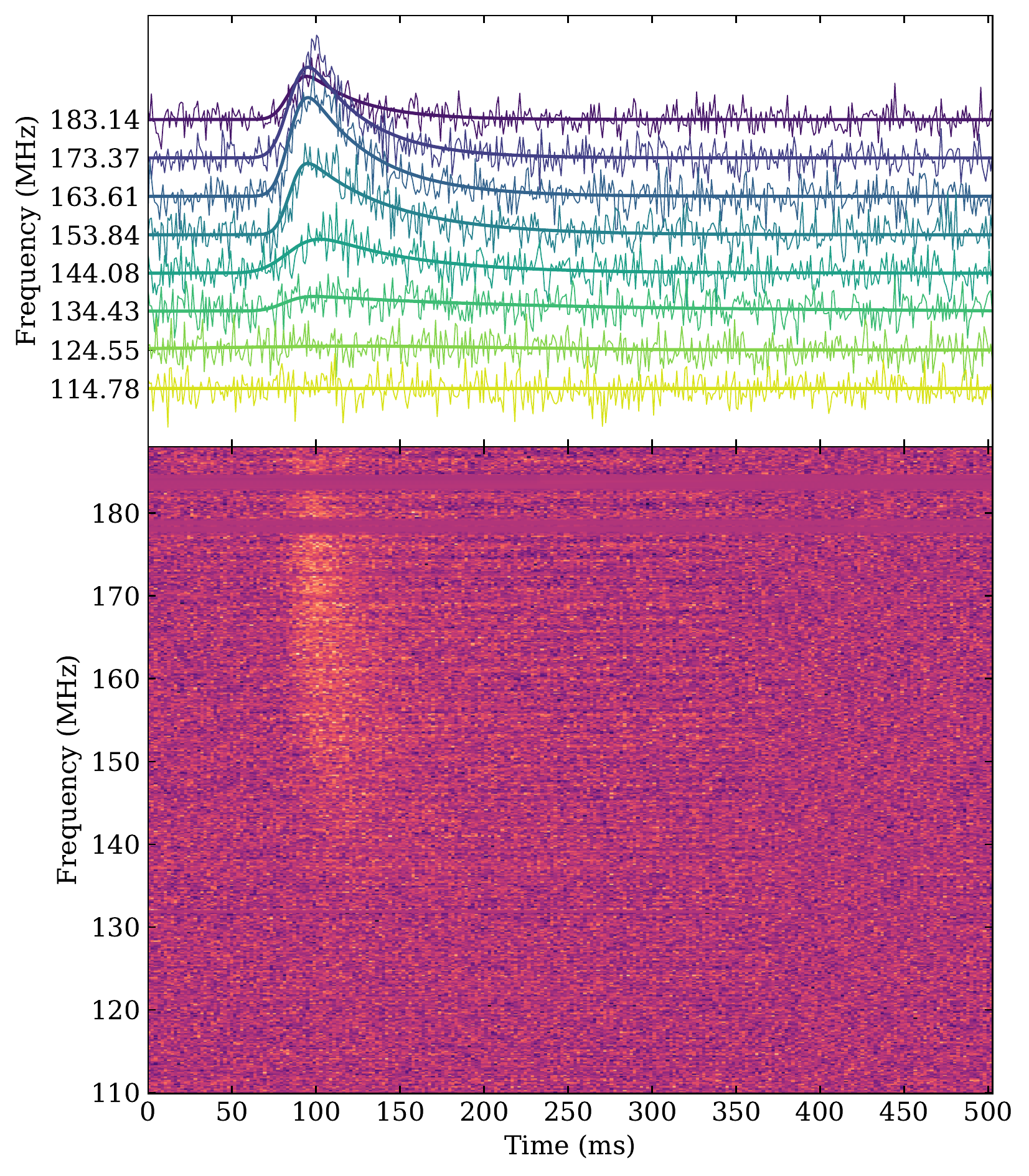}\\
\caption{Stacked LOFAR bursts, dedispersed to the S/N maximising DM of \mbox{349.00\pccmnospace.} The top panel shows
  the pulse profiles in eight different frequency bands, and fits to the scattering tail. The central frequency of the
  band is indicated on the vertical labels. The bottom panel displays the dynamic spectrum of the stacked bursts.
\label{fig:scattering}
}
\end{figure}

\begin{figure}
  \centering
\includegraphics[width=\linewidth]{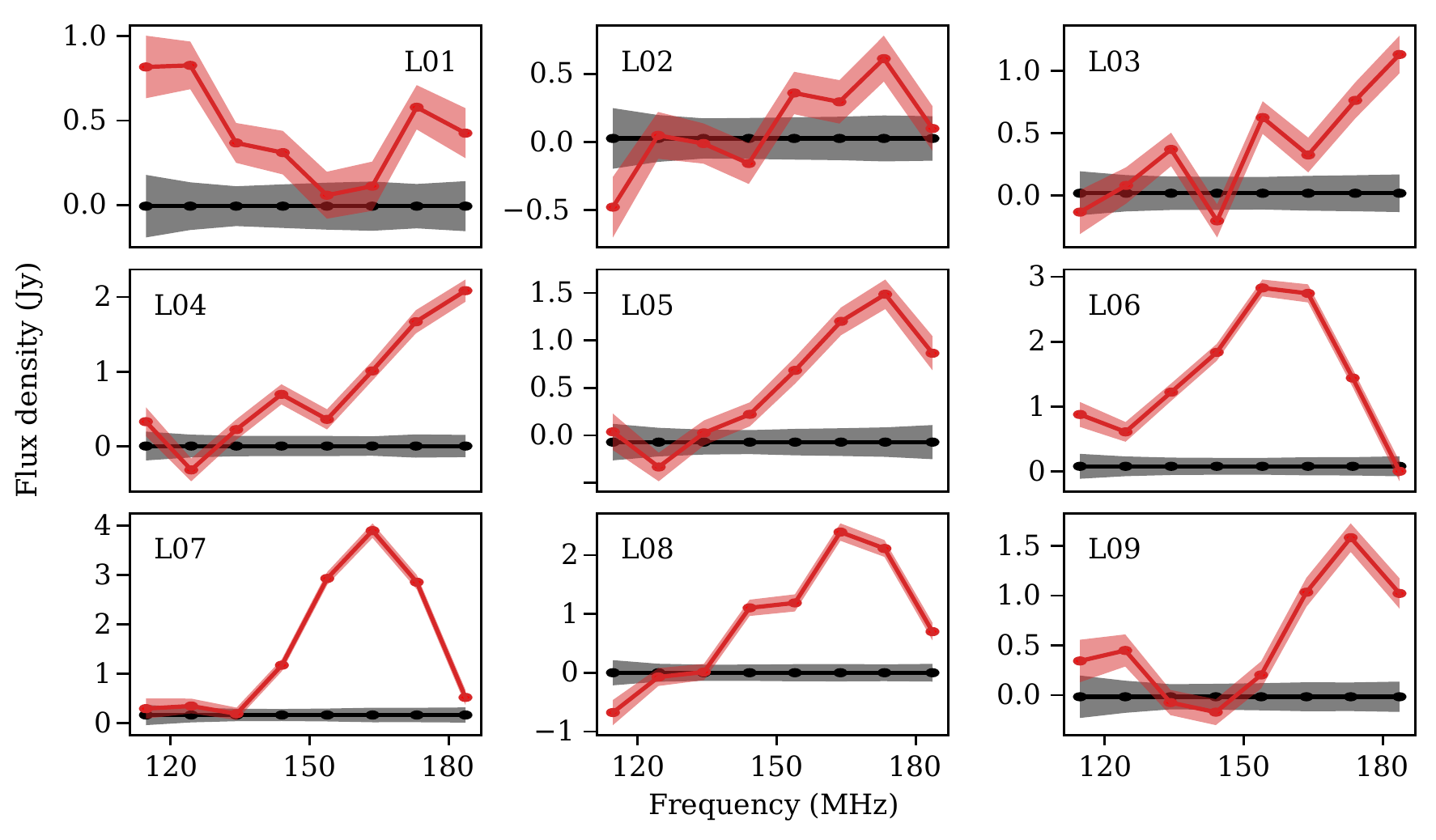}\\
\caption{Comparison of the intrinsic LOFAR burst spectra to the telescope sensitivity limits.
  The red line shows the  flux densities for the bursts,
  averaged over fixed [$-$50, +150]\,ms windows around the burst peak.
  The black line shows the same but for off-burst windows.
  The telescope sensitivity limits ($\pm 1 \sigma$, black and red contours)
  calculated as the standard deviation of 3-s off-burst intervals, scaled to the 200-ms on-burst window. 
  The LOFAR minimum detectable flux varies over the recorded band; it is higher at the band edges.
  The black contours demonstrate the response is relatively flat compared to the burst brightness.
Note that this figure utilizes different frequency/time binning, which explains the  apparent slight differences with \ref{fig:lofar_bursts}.
Seemingly significant negative pulse flux densities at low frequencies in e.g.~L02 and L08 were caused by
slowly-varying, low-level   residual RFI that affected the baseline subtraction.
Nevertheless, bursts L01 and L06 clearly show emission above the noise level, at the lower edge of the LOFAR band.
  \label{fig:spectra}
}
\end{figure}

\begin{figure}
\centering
\includegraphics[width=0.7\linewidth]{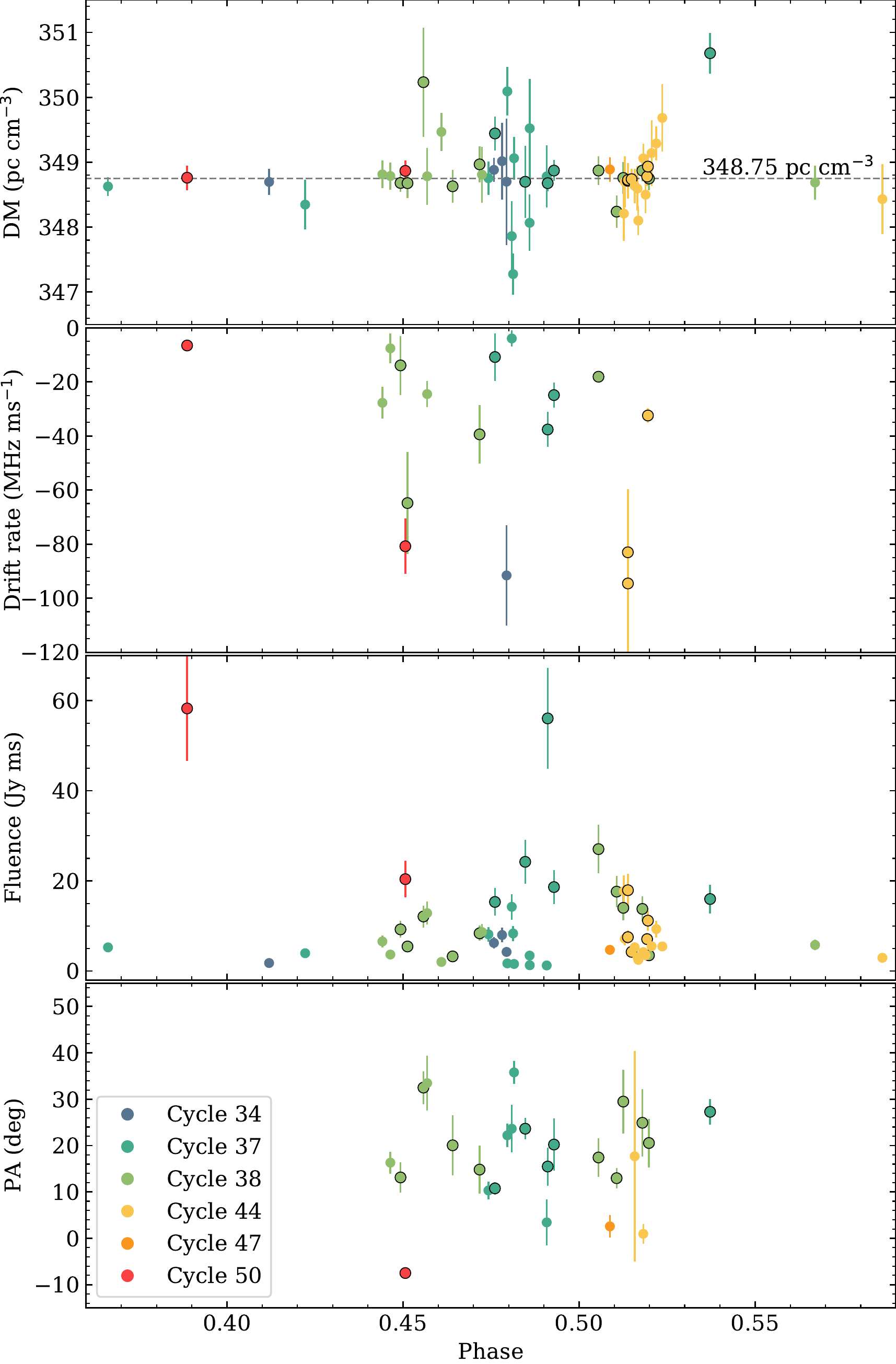}\\
\caption{Apertif burst properties against phase. The top panel shows the structure optimised dispersion measure with the 348.75\pccm average as a reference. The second panel shows the drift rate of bursts with multiple components. The third panel shows the fluence. The bottom panel gives the average polarisation position angle of each burst. Bursts are color-coded by activity cycle. Each color corresponds to a different activity cycle, and the data points with a black edge represent bursts with S/N>20.
\label{fig:Apertif_burst_properties}
}
\end{figure}

\begin{figure}
\centering
\includegraphics[width=0.49\linewidth]{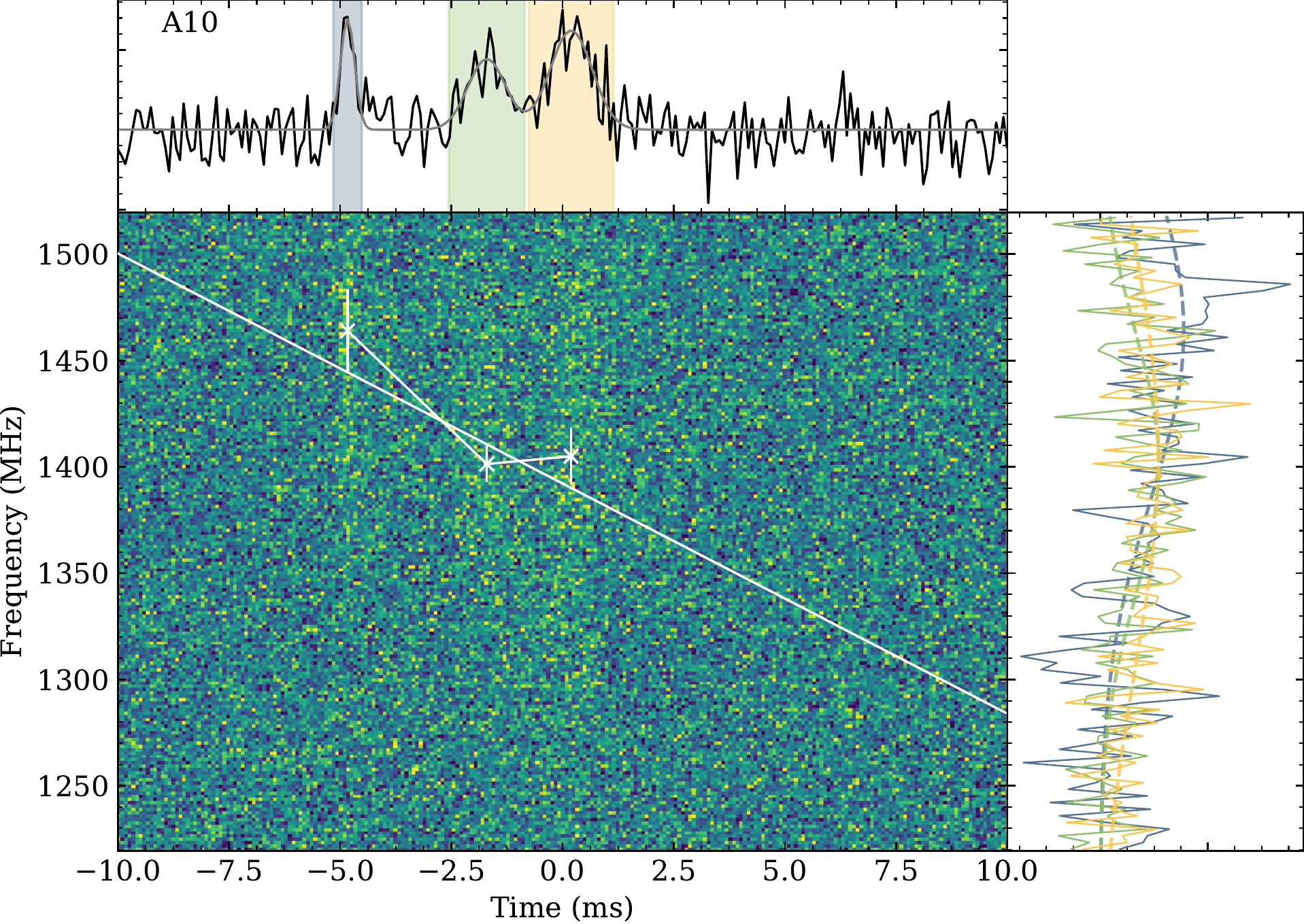}
\includegraphics[width=0.49\linewidth]{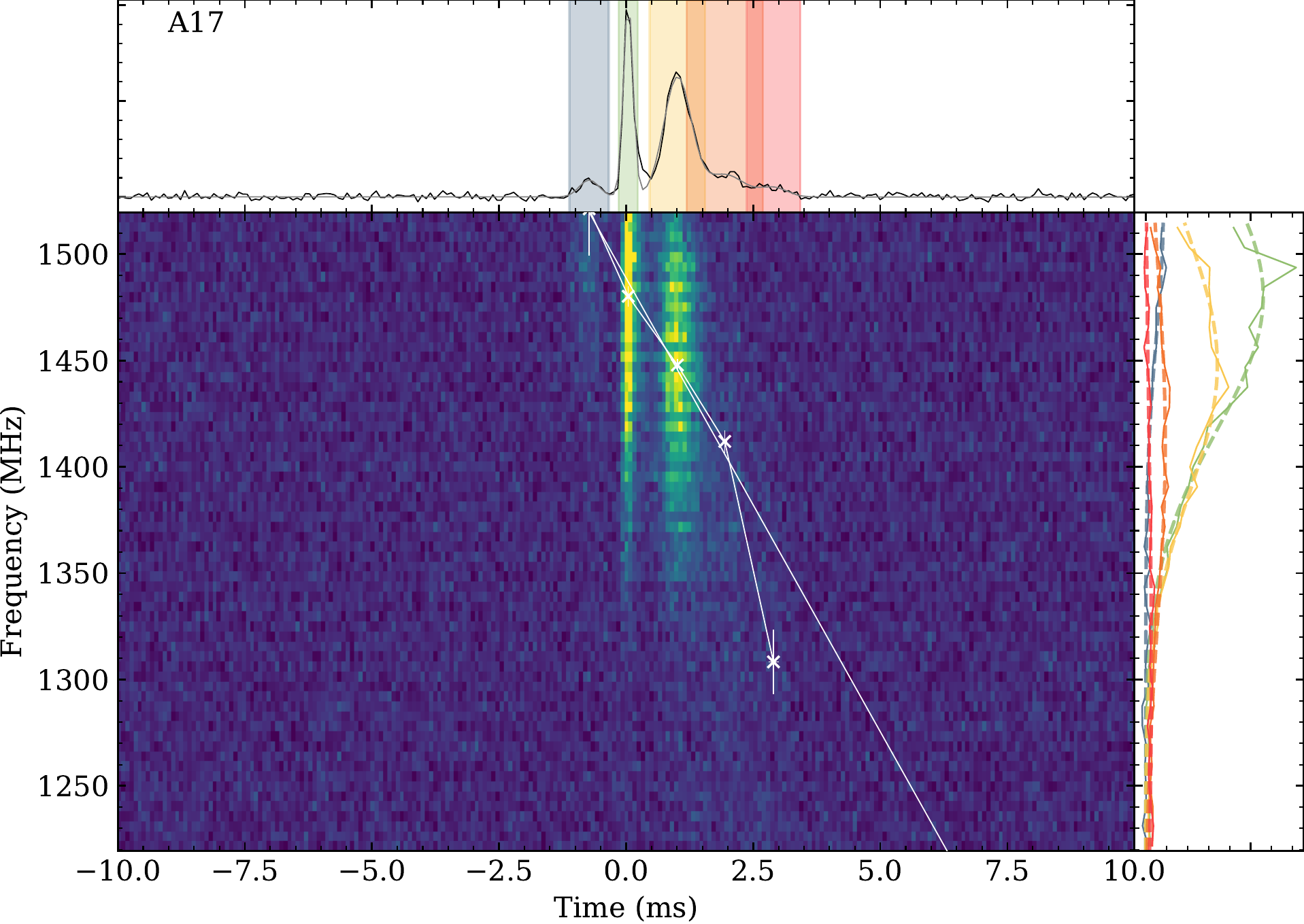}
\includegraphics[width=0.49\linewidth]{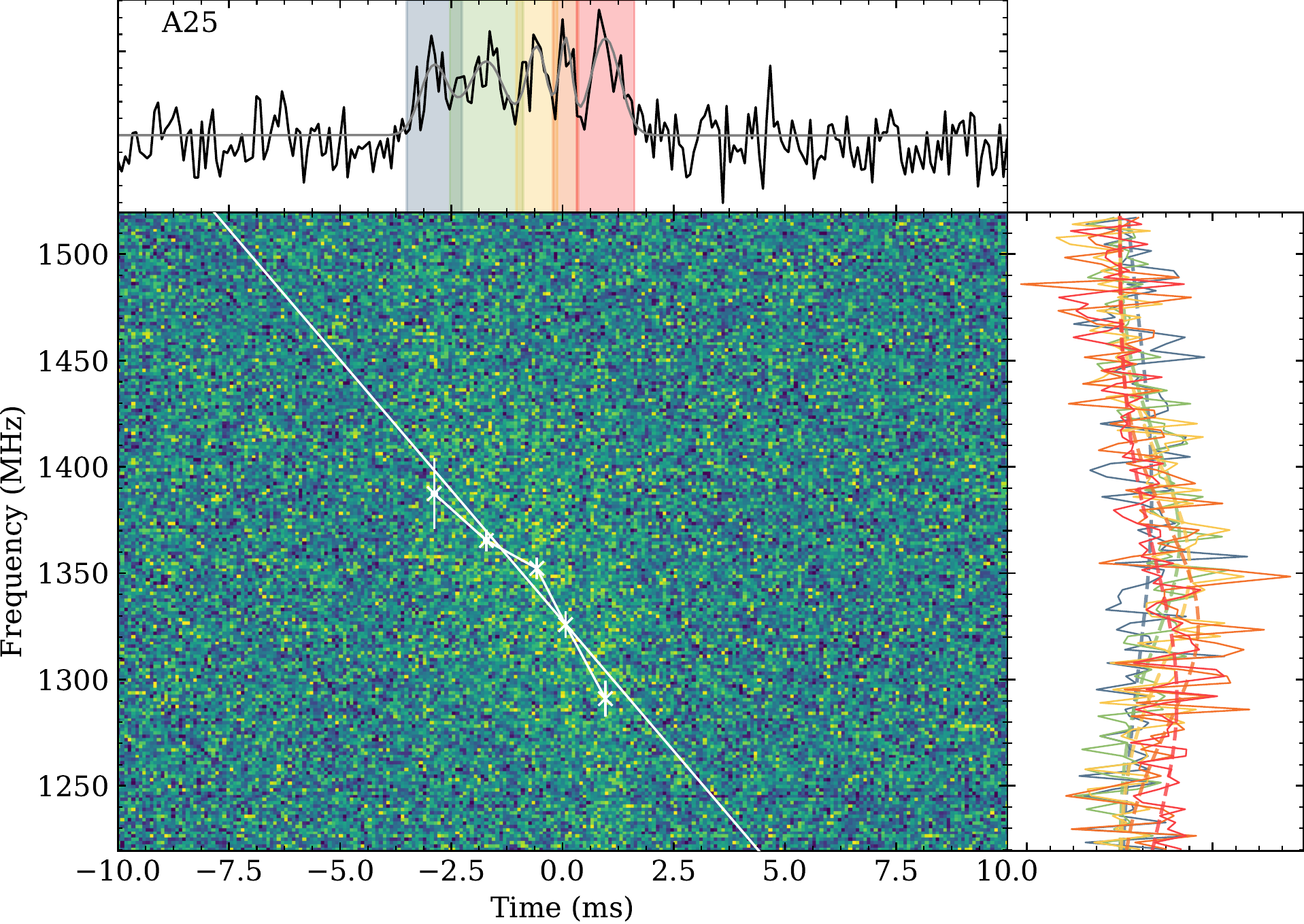}
\includegraphics[width=0.49\linewidth]{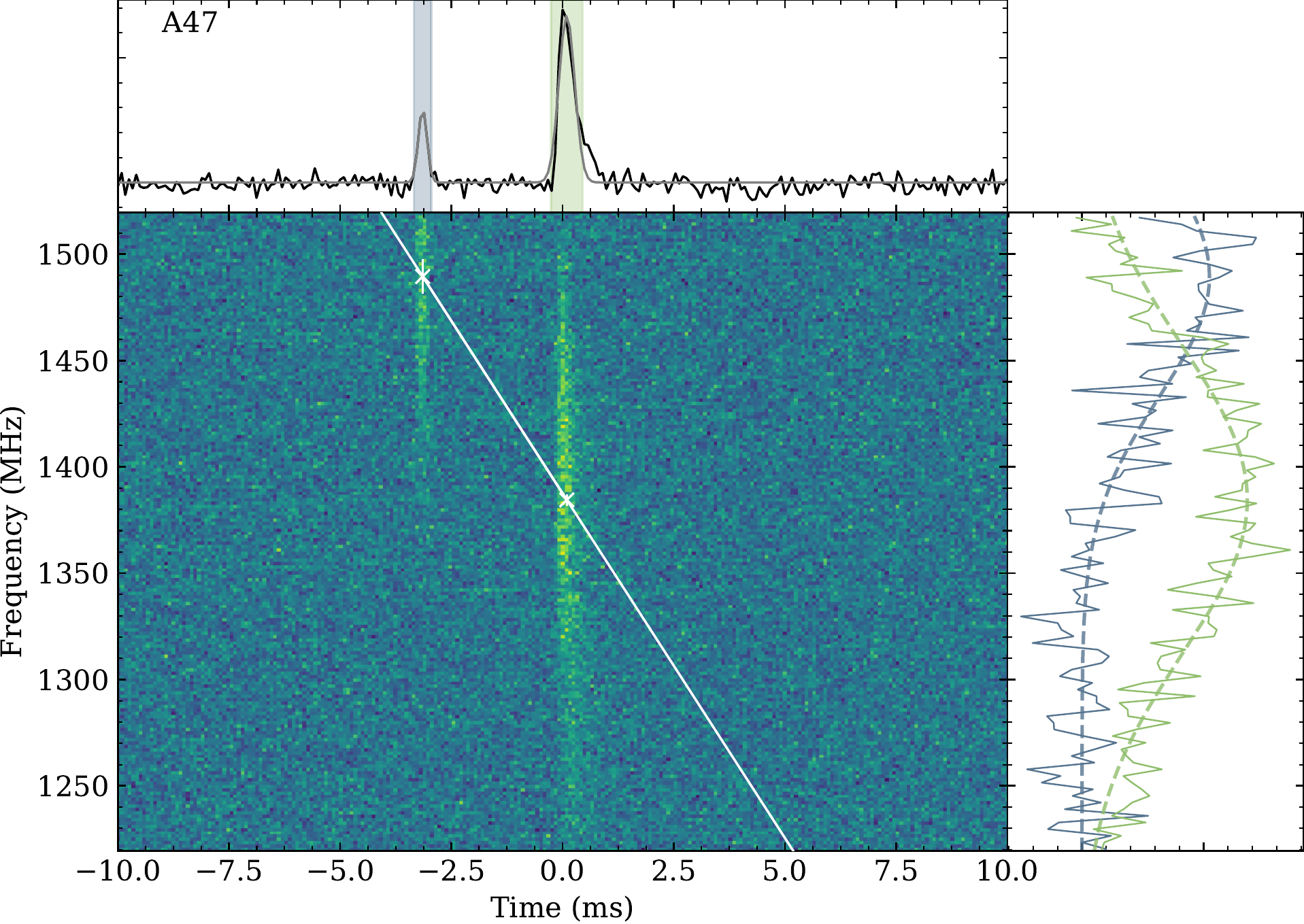}\\
\includegraphics[width=0.98\linewidth]{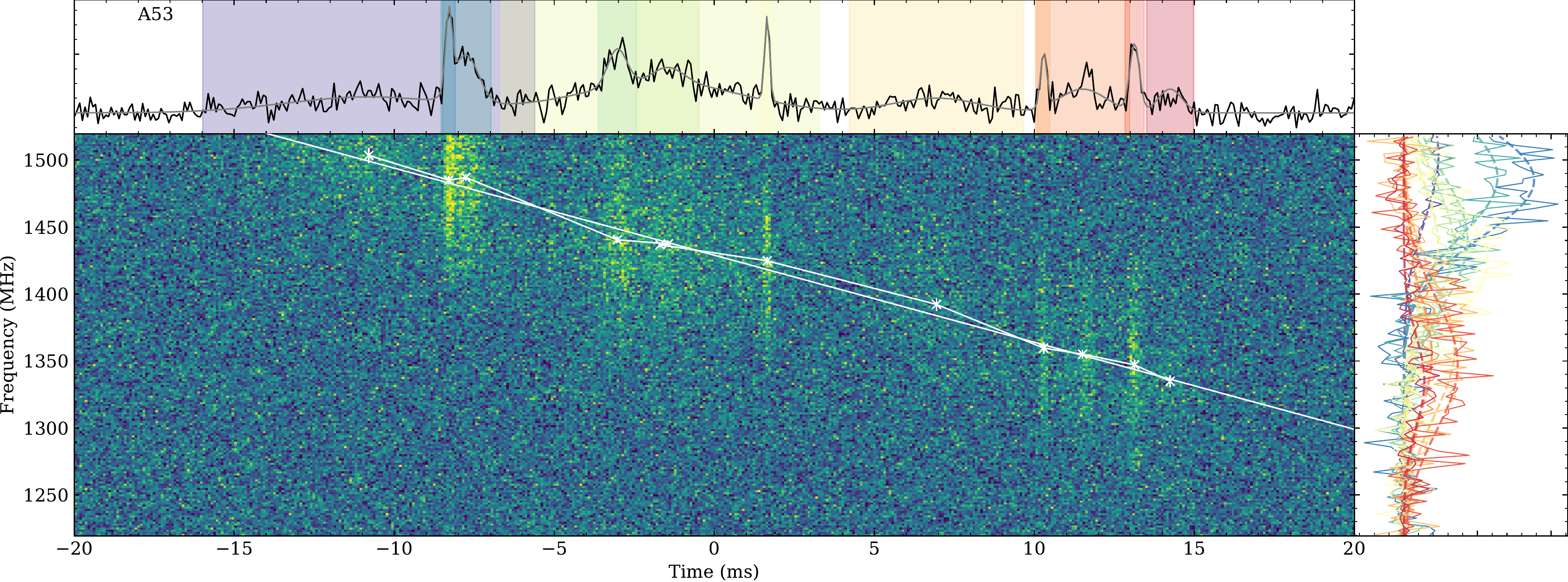}\\
\caption{Five of the bursts with a measurable drift rate. For each burst, the top panel shows the pulse profile as a solid black line and the fitted multi-component Gaussian in gray, with shaded coloured regions indicating the position of the subcomponents. The lower left panels show the dynamic spectra rebinned eight times in frequency, with the centroid of each subcomponent and the fitted drift rate as a white line. The right panels represent the spectra and the fitted Gaussian of each subcomponent, with the same color as the shaded region of the pulse profile.
\label{fig:drift_rate}
}
\end{figure}

\clearpage
\begin{figure}
    \centering
    \includegraphics[width=\textwidth]{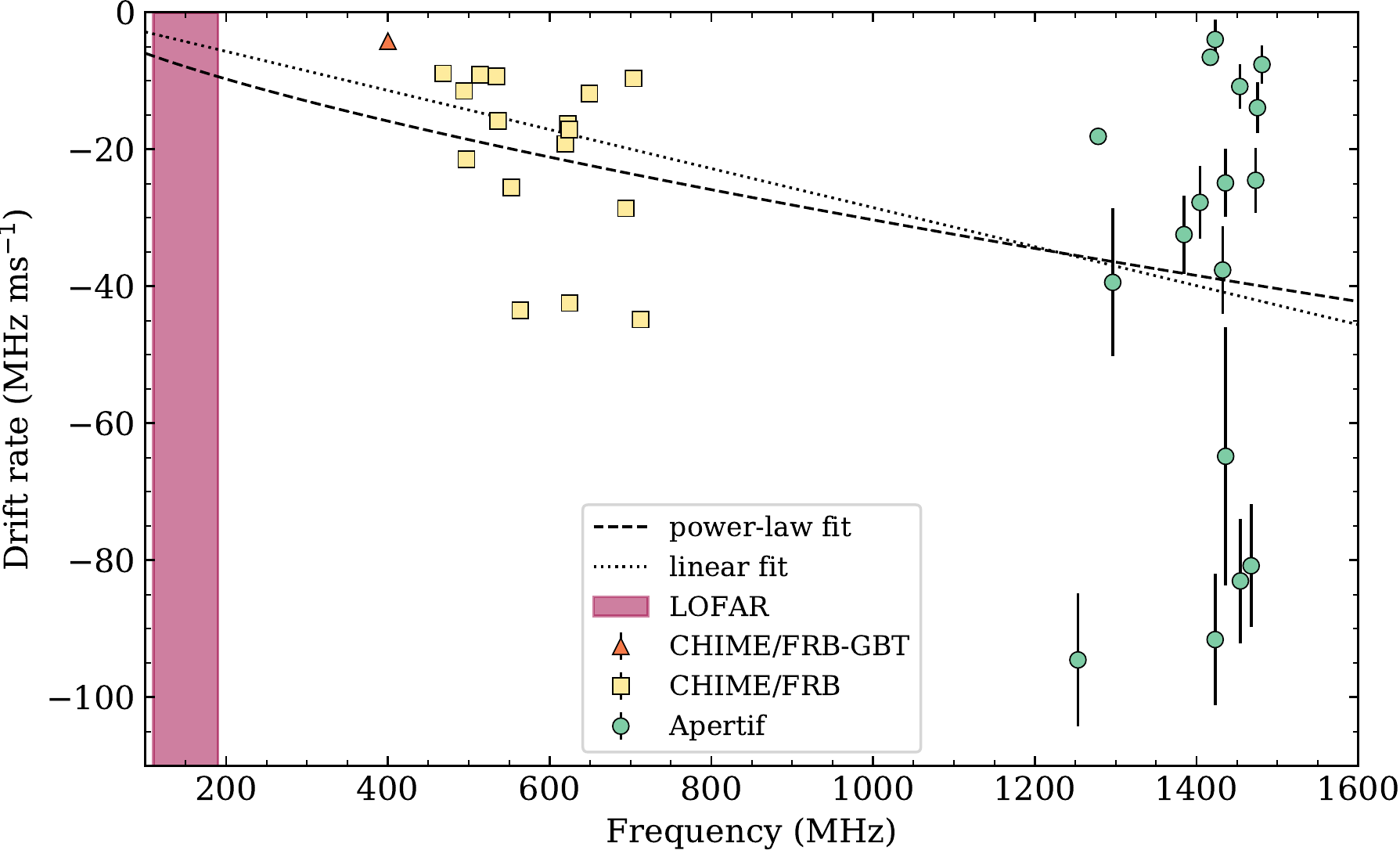}\\
    \caption{Comparison of drift rates at different frequencies. The green circles are the drift rate of bursts presented in this work, detected with Apertif. The yellow squares are drift rates from CHIME/FRB bursts\cite{the_chimefrb_collaboration_periodic_2020,chamma_shared_2020}. The orange triangle is the simultaneous CHIME/FRB-GBT burst where a drift rate was reported\cite{chawla_detection_2020}. The dashed line is a power law fit of the drift rate at different frequencies, $\dot{\nu} = -0.2 \nu^{0.7}$. The dotted line is the linear fit of the drift rate,  $\dot{\nu} = -2.9\times10^{-2} \nu$, and it is almost superposed to the power law fit.
    \label{fig:freq_drift_rate}
    }
\end{figure}

\begin{figure}
\centering
\includegraphics[width=1\linewidth]{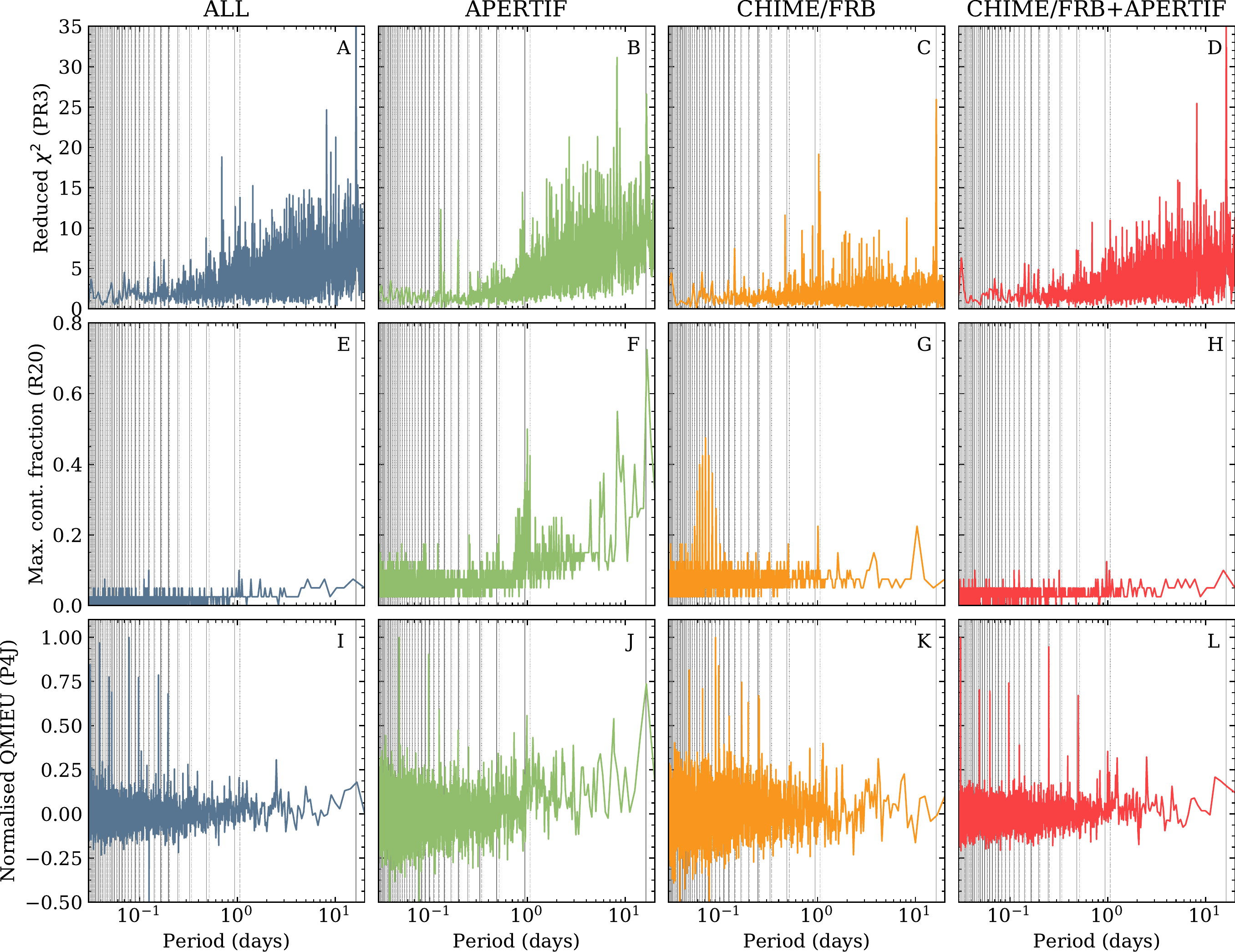}\\
\caption{Periodograms between 0.03\,day and 20\,day periods of four instrument combinations and three different period searching techniques. Each column corresponds, from left to right, to all detections combined (blue), Apertif detections (green), CHIME/FRB detections (yellow) and CHIME/FRB and Apertif detections combined (red). Each column corresponds to a different search technique, with Pearson's $\chi^2$ test at the top, activity width minimisation center, and QMI method at the bottom. The vertical gray lines mark the position of the aliased periods, solid for $f_N = (Nf_{\text{sid}}+f_0)$ and dotted for $f_N = (Nf_{\text{sid}}-f_0)$.
\label{fig:periodograms}
}
\end{figure}

\clearpage
\begin{table}
\centering
\caption{\label{tab:LOFAR_bursts}Summary of LOFAR burst properties.}
{ \scriptsize
\begin{tabular}{cccccccc}
\hline\hline
Burst ID & OBSID & Arrival time & Arrival time & Detection & DM & Fluence & \tscat at 150\,MHz\\
         &       & (MJD) & (UTC) & S/N & (\pccm) & (\jyms) & (ms) \\
\hline
L01 & L775795 & 58949.53491816 & 2020-04-10 12:50:16.929 & 8.7 & 349.03$\pm$0.11 & 111$\pm$55 & ... \\
L02 & L775801 & 58949.63987585 & 2020-04-10 15:21:25.273 & 7.3 & 348.94$\pm$0.08 & 38$\pm$19 & ... \\
L03 & L775977 & 58950.52919335 & 2020-04-11 12:42:02.305 & 9.5 & 349.02$\pm$0.08 & 80$\pm$40 & ... \\
L04 & L775977 & 58950.54130169 & 2020-04-11 12:59:28.466 & 18.9 & 349.41$\pm$0.03 & 177$\pm$88 & ... \\
L05 & L775979 & 58950.58347838 & 2020-04-11 14:00:12.532 & 13.7 & 349.09$\pm$0.04 & 129$\pm$64 & ... \\
L06 & L775953 & 58951.54162736 & 2020-04-12 12:59:56.604 & 29.4 & 349.03$\pm$0.05 & 318$\pm$159 & 48.2$\pm$16.6\\
L07 & L775953 & 58951.55801455 & 2020-04-12 13:23:32.457 & 35.1 & 348.98$\pm$0.02 & 296$\pm$148 & 46.9$\pm$16.0\\
L08 & L775955 & 58951.58470795 & 2020-04-12 14:01:58.767 & 23.5 & 348.99$\pm$0.03 & 193$\pm$96 & 36.2$\pm$19.0\\
L09 & L775955 & 58951.59135120 & 2020-04-12 14:11:32.744 & 12.5 & 348.86$\pm$0.08 & 124$\pm$62 & 42.0$\pm$17.4\\
\hline
\end{tabular}
}
\end{table}

%
\begin{table}
\vspace{-5ex}
\centering
\caption{\label{tab:Apertif_bursts}Summary of Apertif burst properties.}
{\scriptsize
\begin{tabular}{ccccccc}
\hline\hline
Burst ID &  Arrival time & Arrival time & Detection & DM & Fluence & Drift rate\\
         & (MJD) & (UTC) & S/N & (\pccm) & (\jyms) & (\mhzms)\\
\hline
A01 & 58930.47097294 & 2020-03-22T11:18:12.062 & 11.5 & 348.70(20) & 1.8 & ... \\
A02 & 58931.51122577 & 2020-03-23T12:16:09.907 & 12.7 & 348.88(18) & 6.2 & ... \\
A03 & 58931.54877968 & 2020-03-23T13:10:14.564 & 13.4 & 349.02(59) & 8.0 & ... \\
A04 & 58931.56964778 & 2020-03-23T13:40:17.568 & 13.4 & 348.70(97) & 4.2 & ... \\
\hline
A05 & 58978.59561357 & 2020-05-09T14:17:41.012 & 13.6 & 348.63(14) & 5.2 & ... \\
A06 & 58979.50785914 & 2020-05-10T12:11:19.030 & 8.9 & 348.35(38) & 4.0 & ... \\
A07 & 58980.35572077 & 2020-05-11T08:32:14.275 & 16.4 & 348.75(26) & 8.2 & ... \\
A08 & 58980.38590828 & 2020-05-11T09:15:42.475 & 29.9 & 349.44(26) & 15.3 & -9.16 \\
A09 & 58980.44318898 & 2020-05-11T10:38:11.528 & 13.9 & 350.09(37) & 1.7 & ... \\
A10 & 58980.46375074 & 2020-05-11T11:07:48.064 & 17.8 & 347.86(54) & 14.2 & -3.95 \\
A11 & 58980.46995949 & 2020-05-11T11:16:44.500 & 10.1 & 347.28(32) & 8.3 & ... \\
A12 & 58980.47426015 & 2020-05-11T11:22:56.077 & 11.0 & 349.06(33) & 1.6 & ... \\
A13 & 58980.52593337 & 2020-05-11T12:37:20.643 & 38.6 & 348.70(56) & 24.2 & ... \\
A14 & 58980.54629988 & 2020-05-11T13:06:40.310 & 14.0 & 348.07(44) & 3.4 & ... \\
A15 & 58980.54684270 & 2020-05-11T13:07:27.209 & 12.5 & 349.52(76) & 1.3 & ... \\
A16 & 58980.62542392 & 2020-05-11T15:00:36.627 & 11.5 & 348.78(48) & 1.2 & ... \\
A17 & 58980.62998094 & 2020-05-11T15:07:10.353 & 58.1 & 348.68(13) & 56.0 & -26.36 \\
A18 & 58980.65889322 & 2020-05-11T15:48:48.374 & 25.6 & 348.87(16) & 18.6 & -18.12 \\
A19 & 58981.38138907 & 2020-05-12T09:09:12.016 & 31.5 & 350.68(32) & 16.0 & ... \\
\hline
A20 & 58996.15501128 & 2020-05-27T03:43:12.975 & 12.1 & 348.81(21) & 6.6 & -10.01 \\
A21 & 58996.19203445 & 2020-05-27T04:36:31.776 & 12.7 & 348.79(21) & 3.7 & -3.56 \\
A22 & 58996.23898191 & 2020-05-27T05:44:08.037 & 20.2 & 348.68(14) & 9.2 & -8.83 \\
A23 & 58996.27129126 & 2020-05-27T06:30:39.565 & 20.9 & 348.68(23) & 5.5 & -42.74 \\
A24 & 58996.34499129 & 2020-05-27T08:16:47.247 & 21.4 & 350.23(84) & 12.1 & ... \\
A25 & 58996.36224320 & 2020-05-27T08:41:37.812 & 19.5 & 348.78(44) & 12.8 & -25.20 \\
A26 & 58996.42810299 & 2020-05-27T10:16:28.098 & 10.5 & 349.47(29) & 2.0 & ... \\
A27 & 58996.48015176 & 2020-05-27T11:31:25.112 & 21.0 & 348.63(25) & 3.2 & ... \\
A28 & 58996.60480633 & 2020-05-27T14:30:55.267 & 20.9 & 348.97(28) & 8.4 & -52.65 \\
A29 & 58996.61583838 & 2020-05-27T14:46:48.436 & 9.0 & 348.81(43) & 8.7 & ... \\
A30 & 58997.15492630 & 2020-05-28T03:43:05.632 & 25.7 & 348.87(22) & 27.1 & ... \\
A31 & 58997.23883623 & 2020-05-28T05:43:55.450 & 36.5 & 348.24(25) & 17.6 & -25.47 \\
A32 & 58997.26968437 & 2020-05-28T06:28:20.730 & 29.9 & 348.76(25) & 14.0 & ... \\
A33 & 58997.35800780 & 2020-05-28T08:35:31.874 & 29.2 & 348.87(17) & 13.8 & ... \\
A34 & 58997.38837259 & 2020-05-28T09:19:15.392 & 20.9 & 348.75(18) & 3.5 & ... \\
A35 & 58998.15708057 & 2020-05-29T03:46:11.761 & 10.4 & 348.69(26) & 5.8 & ... \\
\hline
A36 & 59095.01258701 & 2020-09-03T00:18:07.518 & 8.5 & 348.21(42) & 17.7 & ... \\
A37 & 59095.01647024 & 2020-09-03T00:23:43.029 & 10.3 & 348.70(40) & 7.2 & ... \\
A38 & 59095.03083630 & 2020-09-03T00:44:24.256 & 27.4 & 348.71(27) & 7.5 & -37.45 \\
A39 & 59095.03119917 & 2020-09-03T00:44:55.608 & 28.2 & 348.73(11) & 17.9 & -29.27 \\
A40 & 59095.04813576 & 2020-09-03T01:09:18.930 & 25.8 & 348.74(16) & 4.2 & ... \\
A41 & 59095.06242878 & 2020-09-03T01:29:53.847 & 11.4 & 348.64(26) & 5.3 & ... \\
A42 & 59095.07525644 & 2020-09-03T01:48:22.156 & 12.7 & 348.59(34) & 3.1 & ... \\
A43 & 59095.07913932 & 2020-09-03T01:53:57.637 & 11.5 & 348.10(23) & 2.5 & ... \\
A44 & 59095.10211216 & 2020-09-03T02:27:02.491 & 12.6 & 349.06(23) & 4.2 & ... \\
A45 & 59095.11289895 & 2020-09-03T02:42:34.469 & 13.2 & 348.50(28) & 3.5 & ... \\
A46 & 59095.11989684 & 2020-09-03T02:52:39.087 & 20.7 & 348.78(14) & 7.1 & ... \\
A47 & 59095.12368446 & 2020-09-03T02:58:06.337 & 32.4 & 348.93(23) & 11.2 & -19.01 \\
A48 & 59095.14075045 & 2020-09-03T03:22:40.839 & 10.0 & 349.14(51) & 5.5 & ... \\
A49 & 59095.16236684 & 2020-09-03T03:53:48.495 & 10.7 & 349.29(26) & 9.3 & ... \\
A50 & 59095.19030365 & 2020-09-03T04:34:02.235 & 9.4 & 349.68(52) & 5.5 & ... \\
A51 & 59096.20840871 & 2020-09-04T05:00:06.513 & 9.3 & 348.43(54) & 2.9 & ... \\
\hline
A52 & 59143.81778929 & 2020-10-21T19:37:36.995 & 11.3 & 348.89(19) & 4.7 & ... \\
\hline
A53 & 59190.73164688 & 2020-12-07T17:33:34.290 & 44.2 & 348.76(19) & 58.3 & -6.55 \\
A54 & 59191.74125466 & 2020-12-08T17:47:24.403 & 51.1 & 348.87(16) & 20.4 & -80.80 \\
\hline
\end{tabular}
}
\end{table}

\begin{table}
    \centering
    \begin{tabular}{lrccc}
        \hline\hline
        Instrument & N.bursts & PR3 & R20 & PJ4 \\
        \hline
        ALL & 154 & 16.34\errors{0.11}{0.15} & 16.29\errors{0.16}{0.18} & 16.30\errors{0.20}{0.24} \\
        Apertif & 54 & 16.38\errors{3.56}{3.69} & 16.41\errors{2.18}{1.70} & 16.35\errors{0.47}{0.31} \\
        CHIME/FRB & 57 & 16.36\errors{0.07}{0.16} & 16.31\errors{0.16}{0.18} & 16.35\errors{0.10}{0.14} \\
        CHIME/FRB+Apertif & 111 & 16.28\errors{0.17}{0.11} & 16.29\errors{0.15}{0.17} & 16.30\errors{0.13}{0.16} \\
        \hline
    \end{tabular}
    \caption{Best periods obtained with different burst combinations using three different techniques; Pearson's $\chi^2$ test (PR3)\cite{the_chimefrb_collaboration_chimefrb_2019}, activity width minimisation (R20)\cite{rajwade_possible_2020}, and quadratic mutual information periodicity search (PJ4)\cite{huijse_robust_2018}.}
    \label{tab:periods}
\end{table}

\clearpage
\section*{Supplementary Methods}
\label{sec:sm}

\setcounter{section}{0}
\section{Observations and burst search}

\subsection{Apertif}

For 165 out of 388 observing hours, the high-resolution data were kept for a deeper offline search with \textsc{PRESTO}\cite{ransom_new_2001}. After masking channels known to be affected by RFI, the data were dedispersed to DMs between 310\pccm and 397\pccm in steps of 0.3\pccmnospace. Each time series was then searched for single pulses with S/N $>$ 8 and width $<$ 100\,ms. After clustering the candidates in DM and time, the candidate with the highest S/N in each cluster was visualised and inspected by eye.
A small fraction of the data were strongly affected by RFI, mainly in cycle 44 (as numbered in \ref{fig:exposure_detections}) during September 3rd and 4th. These data were cleaned with \textsc{rfiClean}\footnote{\url{https://github.com/ymaan4/rfiClean}} and RFI was further masked with \textsc{PRESTO}'s \textsc{rfifind}.
A large fraction of channels was masked completely. Hence we cannot exclude the presence of faint or narrowband bursts that would have been above our sensitivity threshold without RFI.

In addition to the single pulse search, we searched the data for periodic signals with periods between 0.1\,ms and 1\,s. To account for any drift in the pulse frequency due to acceleration of the source in a putative orbit, an acceleration search was performed with a maximum Fourier-drift parameter of $z=200$, corresponding to a maximum line-of-sight acceleration of 0.5$\mathrm{\,m\,s^{-2}}$ for a periodicity of 1\,ms and the typical observation duration of 3 hrs. The implicit assumption of constant acceleration holds as long as the orbit is longer than $\sim$30 hrs. All candidates were inspected visually.

\subsection{LOFAR}
Most LOFAR stations are located across the Netherlands, and 14 are distributed in neighboring countries in order to
increase its spatial resolution. The observations presented here used between 18 and 23 core (Dutch) stations, and used
coherent stokes mode at a time resolution of 983.04\,$\mu$s and a frequency resolution of 3.052 kHz. Most data, including all detections, were recorded in intensity only (Stokes I). Data from May 27/28/29 was recorded in full polarisation (Stokes IQUV).

The offline search for FRBs and periodic emission used \textsc{PRESTO}. The data were first sub-banded using \textsc{Sigproc}. In this process, every 25 consecutive channels were dedispersed using a DM of 349.5\pccm and averaged together, resulting in 1024 sub-bands across the full 78.1\,MHz bandwidth. Strong periodic and other RFI were mitigated using \textsc{rfiClean}\cite{Maan2020}, and any remaining RFI were subsequently masked using  \textsc{PRESTO}'s \texttt{rfifind}. The data were then dedispersed to DMs between 342 and 358 \pccm in steps of 0.03\,\pccmnospace. Each dedispersed time series was searched for single pulses with S/N\,$>$\,7 and pulse-width\,$<$\,250\,ms. Similar to the offline search of the Apertif data, the candidates were clustered in DM and time, and the candidate with the highest S/N in each cluster was visualized and examined by eye. Each of the dedispersed time series was also subjected to a periodicity search using \textsc{PRESTO}'s \texttt{accelsearch}, with a maximum Fourier-drift parameter of $z=128$. This value implies that, for an observing duration of 1\,hour, we have searched for average accelerations of about 2.96 and 296$\mathrm{\,m\,s^{-2}}$ of 1000 and 10\,Hz signals, respectively. We note that our periodicity search is not sensitive to periods shorter than a few tens of milliseconds due to significant scatter-broadening at the LOFAR frequencies. For each observation, all the candidates with periods up to 80\,s were folded and the corresponding diagnostic plots were examined by eye.

\section{Data analysis}

\subsection{Bursts detected with Apertif: fluence distribution}
To estimate the fluence of all Apertif bursts, we obtained the mean pulse profiles using 21\,ms time windows centered at each pulse's peak. This window duration is larger than the widest burst, except for A53 where a 42\,ms window was needed to cover the whole burst duration. We normalised each pulse profile by the standard deviation of an off-burst region in order to convert the time series into SNR units. We determined the system-equivalent flux density (SEFD) by performing drift scans of the calibrator sources 3C147 and 3C286 whose flux densities are known\cite{perley_accurate_2017}. Next we applied the radiometer equation\cite{cordes_searches_2003, maan_deep_2014} to convert the pulse profile into flux units (Jy) using the SEFD, and integrated over the 21\,ms or 42\,ms time windows to obtain the fluence of each burst (Jy\,ms). We applied this technique in order to account for the burst structure. We assume 20\% errors on the fluence based on the instability of the system over several days of observations.

The cumulative distribution function (CDF) of Apertif bursts fluences, presented in \ref{fig:fluence_cdf}, can be fitted to a broken power-law with two turnovers. By applying a least squares minimisation technique and assuming Poissonian errors on the rate, we find the break fluences to be located at 3.2$\pm$0.2\jyms and 7.8$\pm$0.4\jymsnospace. 
For bursts with S/N>10 displaying the typical burst width of 2\,ms, our fluence completeness threshold is $\sim$1.7\jyms . 
The full range of widths for pulses near our S/N detection limit (\ref{fig:Apertif_bursts}) is between 1$-$5\,ms, which leads to a fluence range of 1$-$3\jyms (\ref{tab:Apertif_bursts}).
The  lower-fluence turnover falls right above this range and we will thus assume that it is due to the Apertif sensitivity.  
The 7.8\jyms turnover is however above our completeness threshold and cannot be due to instrumental effects. CHIME/FRB bursts have been observed to show a turnover at 5.3\jyms that was associated to the sensitivity of the instrument\cite{the_chimefrb_collaboration_periodic_2020}. The potential presence of a turnover at 7.8\jyms intrinsic to the fluence distribution of \FRB could have been concealed by the sensitivity turnover.
Each segment of the broken power law of the CDF follows $R(>F)\propto F^{\Gamma}$, where $R$ is the rate (h$^{-1}$), $F$ the fluence (Jy\,ms) and $\Gamma$ the power law index. For $F$>7.8\jymsnospace, we get $\Gamma=-1.4\pm0.1$. This index is consistent with the CDF of CHIME/FRB bursts, where they get $\alpha=\Gamma-1\sim-2.3$. For bursts with $3.2$\jyms$<F<7.8$\jyms we get $\Gamma=-0.7\pm0.1$, and for $F<3.2$\jyms we get $\Gamma=-0.2\pm0.1$. All errors give the standard deviation of the fitted parameters.

\subsection{Bursts detected with LOFAR: fluence distribution}
The flux density scale for LOFAR observations was derived from
 the radiometer equation, using information about frequency-dependent antenna and sky temperatures, models of telescope gain (frequency- and direction-dependent), number of performing stations/tiles, RFI environment, as well as observing bandwidth, integration time, and number of polarization summed. The uncertainty of the flux density measurements was estimated as
 50\% systematic uncertainty on the band integrated flux caused by an imperfect knowledge of the system parameters. 
For the details of the calibration procedure and flux uncertainty estimates we refer the reader to LOFAR censuses of millisecond\cite{kondratiev_lofar_2016} and normal\cite{bilous_lofar_2016} pulsars. 

\ref{fig:fluence_cdf} shows the CDF of LOFAR bursts. It can be fitted to a broken power law with the break fluence located at $104\pm12$\jyms. This fluence falls well within our LOFAR sensitivity limits, and we thus attribute the break to our completeness level. The power law index of bursts with $F>104$\jyms is $\Gamma=-1.5\pm0.2$, consistent with the Apertif and the CHIME/FRB power law indices. However, the burst rate at the same fluence is two orders of magnitude larger for LOFAR bursts than for Apertif bursts. The power law index for bursts with $F<104$\jyms is $\Gamma=-0.2\pm0.2$.

\subsection{Activity windows}
The CHIME/FRB detections span multiple years, while the Apertif and LOFAR detections are all in 2020.
Since an error of 0.05\,days in a period $\sim$16\,days could lead to a phase delay of $\sim$0.15 after two years and thus to a broadening of the resulting activity window, we compared the PDF including all CHIME/FRB bursts with what would be obtained only with the bursts detected before 2020 and during 2020. We observe that the three CHIME/FRB distributions are consistent with each other, and all are both wider and later in arrival phase than the Apertif profile.

\subsection{Activity windows: Kernel density estimation}
\label{sec:sm:kde}

The KDE is a non-parametric smoothing technique in which a kernel is built at each data point from a sample and their contributions are summed in order to estimate an unknown probability density function. With $\{X_i : i = 1, 2, ..., n\}$ the observed data, a sample of $n$ observations drawn from a distribution $f(x)$ with an unknown density, we define its weighted KDE in the general case as
\begin{equation}
   \hat{f}(x) = \frac{1}{h}\sum^{n}_{i=1} p_i K \left(\frac{X_i-x}{h}\right),
\end{equation}
with $K$ the kernel function, $h>0$ the bandwidth, and $\{p_i : i= 1, ..., n\}$ the probability weights of each data sample.
In this case, the input data $\vec{X}$ are the activity phases of each of the $n$ detections and the weights $\vec{p}$ are the inverse of the reciprocal observing time at that phase, and hence $\hat{f}(x)$ is the equivalent of a detection rate.
We used a Gaussian function as kernel $K$ and applied Scott's rule for bandwidth selection\cite{scott_multivariate_2015}, thus having
\begin{equation}
  h = n_{\text{eff}}^{-1/(d+4)} = n_{\text{eff}}^{-1/5},
\end{equation}
with $d=1$ the number of dimensions and $n_{\text{eff}}$ the effective number of datapoints, that differs from $n$ when applying a weighted KDE,
\begin{equation}
  n_{\text{eff}} = \frac{(\sum^{n}_{i=1} p_i)^2}{\sum^{n}_{i=1}p_i^2}.
\end{equation}

When applying the Gaussian KDE to Apertif, CHIME/FRB and LOFAR burst activity phases, we obtain what is shown on \ref{fig:gaussian_kde}.

\subsection{Activity windows: simulations}\label{sec:activity_window_simulations}

Although the KDE of Apertif, CHIME/FRB and LOFAR look different on \ref{fig:gaussian_kde}, we have tested whether the burst samples of the three instruments could be drawn from the same distribution by applying a Kolmogorov-Smirnov (KS) test, comparing them two by two. The p-value obtained by comparing the Apertif and CHIME/FRB samples is $1.71\times10^{-7}$, for Apertif and LOFAR samples $8.45\times10^{-10}$, and for CHIME/FRB and LOFAR samples $7.64\times10^{-5}$. In general, if p-value\,<\,0.01, which is the case here, we can reject the null hypothesis that the samples are drawn from the same distribution.
Though CHIME/FRB and Apertif have similarly uniform coverage in activity phase, the per-cycle sampling function is different between the two surveys. The Westerbork dishes are steerable
and can observe \FRB for approximately half the day while CHIME/FRB is a transit instrument that can only observe a given source for $\sim$20 minutes per day. We therefore wanted to make sure that there were no selection effects involved in the inferred activity window. The different observing strategies used with each instrument could have led to a bias in the observed PDFs in other ways as well, for example jitter in the activity period.
We explore this possibility by simulating the observed population if the samples were drawn from the same intrinsic distribution, modelled as a Gaussian with the same phase centre, width, and average activity rate.

In our simulations, we first generate the number of bursts per cycle $N$. This number will be drawn from a normal distribution centered at CHIME/FRB's average rate, $R\sim 0.32$\,h$^{-1}\sim 125$\,cycle$^{-1}$ for a period $P=16.29$\,days, and with a standard deviation of $R/5\sim 25$\,cycle$^{-1}$. Secondly we generate the $N$ burst arrival phases for the given cycle. The arrival phases will be drawn from a normal distribution centered at CHIME/FRB's phase centre, 0.52, and standard deviation 2.73\,days or $\sim$0.17 in phase. Next we count the number of bursts that Apertif, CHIME/FRB, and LOFAR would have detected with their observing times in the given cycle. This is applied to all the cycles that each instrument covered in their observations. We build the simulated periodograms and apply a KS test to compare them two by two. We perform this simulation 10000 times.

The results of the simulations are shown in \ref{fig:simulated_ks}.
We confirm that the p-value obtained for the Apertif-LOFAR and Apertif-CHIME/FRB sample combinations are well below 99.73\% ($3\sigma$) of the simulated KS p-value obtained for 10000 simulations, indicating that the activity windows are indeed different and are not due to an observational bias. Meanwhile, the CHIME/FRB-LOFAR p-value is below 95.45\% (2$\sigma$) of the simulated p-values, but does not reach the 3$\sigma$ threshold. This could be due either to the lower number of LOFAR detections or to a highest similarity between the activity windows at CHIME/FRB and LOFAR frequency ranges, which are closer that Apertif's. However, we note that four of the LOFAR detections in a single activity cycle arrive later in phase than any previously detected CHIME/FRB burst.

\subsection{Activity windows: ruling out aliasing}

Many instruments followed up \FRB during the predicted activity days in order to increase the chance of detection\cite{chawla_detection_2020, pilia_lowest_2020, aggarwal_vlarealfast_2020, sand_low-frequency_2020} since the discovery of a periodicity in its activity. However, this could lead to a bias in the derived activity cycle due to the lack of coverage out of the predicted activity days.
Although the detection of \FRB with other instruments and different observing strategies put strong constraints on the
allowed $N$ values, the aliasing had not been robustly ruled out until now. Discarding (or confirming) any potential
aliasing was one of our original motivations.

As noted by the authors of the periodicity discovery in the activity cycles of \FRBnospace\cite{the_chimefrb_collaboration_periodic_2020}, the short daily exposure that CHIME/FRB has on source and a regular sampling time of a sidereal day $P_{\text{sid}}=0.99727$\,days could lead to a degeneracy between the reported frequency $f_0 = (16.35$\,days$)^{-1}$ and an aliasing of this frequency at $f_N = (Nf_{\text{sid}}\pm f_0)$, with $N$ a positive integer and $f_{\text{sid}}=P_{\text{sid}}^{-1}$ the inverse of a sidereal day.
The possibility of $N$ larger than 0 prompted some of the proposed periodicity models, mainly the ultra-long period magnetars\cite{beniamini_periodicity_2020}, which would be more comfortably explained by shorter periods. 

In order to confirm the value of the period,
we scheduled our observations with three to nine hours of daily exposures covering \text{five} 16.35\,day activity cycles during the first four covered activity cycles.
We next generated periodograms of the detected bursts using different instrument combinations and the different period search techniques available in the \texttt{frbpa} package\cite{aggarwal_vlarealfast_2020}. These search techniques are a Pearson's $\chi^2$ test (PR3)\cite{the_chimefrb_collaboration_periodic_2020}, an activity width minimisation algorithm (R20)\cite{rajwade_possible_2020}, and a quadratic-mutual-information-based (QMI) periodicity search technique (P4J)\cite{huijse_robust_2018}. We built periodograms for Apertif-only bursts, CHIME/FRB-only bursts, CHIME/FRB+Apertif bursts and all detected bursts from all instruments combined. 
We study the CHIME/FRB+Apertif burst combination because these are the only two instruments that have detections and a coverage of the whole activity phase instead of only at the predicted peak days.
The periodograms that are presented in \ref{fig:periodograms} were computed by searching periods between 0.01\,days and 20\,days to show all the aliased $P_N$ periods for $N$ between 0 and 37.

The periodograms using only CHIME/FRB data (panels C, G, K of \ref{fig:periodograms}) show numerous peaks below eight days that align with the predicted aliasing values (gray vertical lines), as expected for a transit instrument that observes a source with a sampling time of a sidereal day. On the other hand, the periodograms of Apertif bursts (panels B, F, J) show no prominent periods below eight days except for a broad peak with a $\sim$1\,day period in the R20 periodogram, explained by the daily frequency of the observations. By combining CHIME/FRB and Apertif bursts, the effects of the different observing strategies on the periodogram are diminished, and the significance of most aliased peaks is reduced (panels D, H, L). We particularly focus on the activity width minimising plot (R20, panel H). The low values of the maximum continuous fraction indicate that bursts are detected across the whole activity phase for all periods below eight days, allowing us to rule out any potential aliased period. This is further confirmed when adding the bursts from \FRB that were detected by other instruments in panel E.

Additionally, we generated periodograms for periods between 1.57 and 60 days in order to update the value of the period estimate. The results in \ref{tab:periods} give the value of the periodogram peak located at $\sim$16\,days. The error bars are given by the FWHM of the periodogram peaks.
The fewer cycles covered by Apertif observations translate as an uncertainty larger than one day in the period estimation with the PR3 and R20 techniques, in contrast with the period estimation using CHIME/FRB bursts only.
The combination of CHIME/FRB and Apertif bursts gives a period of 16.29\,days with the PR3 test.
The activity width minimisation technique (R20) gives the most consistent period estimates when applied to different instrument combinations. Thus we will hereafter consider the best period to be 16.29\,days computed with respect to a reference MJD 58369.9 to center the peak activity day at a 0.5 phase.

\subsection{Polarisation}
As the Westerbork dishes are 
equatorial mount telescopes, the source is always in the same central beam and the on-sky orientation of the Apertif
feeds do not change with parallactic angle.
This eases the  study of the intrinsic  polarisation position angle (PA). For
surveys at lower frequencies or for sources with higher RMs, such study is made difficult by the covariance between RM and PA, where

\begin{align}
    \Delta\mathrm{PA} &= 2\Delta\mathrm{RM}\lambda^2 \\
    &\approx 5.5^\circ \left ( \frac{\Delta \mathrm{RM}}{1\,\rm{rad}\,\rm{m}^{-2}} \right) \left ( \frac{\nu}{1370\,\rm{MHz}} \right )^{-2}.
\end{align}

We calibrate the polarisation response of Apertif by observing the sources 3C286 and 3C147. The former is roughly 12$\%$ linearly polarised with a stable PA; the latter is known to be unpolarised, which allows us to solve for leakage from I into Q, U, and V. For the analysis, we dedisperse the bursts to 348.75\pccm and we use $\mathrm{RM}=-115$\,rad\,m$^{-2}$ known from previous RM measurements of the source\cite{chawla_detection_2020}. We have done 
Q/U fitting to the Apertif data using \textsc{RM-Tools}\cite{rm-tools} as 
well as our own code, and found RM values consistent 
with this value. Our limited range in $\Delta\lambda^2$ 
compared with CHIME limits our ability to look for RM variation 
in activity phase and across cycles. For the purpose of 
monitoring PA over time, we feel confident using a previously-determined 
RM value given the Galactic Faraday foreground appears to 
be $-115\pm12$\,rad\,m$^{-2}$ and likely dominates the total 
RM of \FRBnospace\cite{ordog-2019, the_chimefrb_collaboration_second_2019}.

\subsection{Dispersion}
\label{sec:sm:dm}
\textsc{AMBER} reports the dispersion measure that maximised  the burst S/N (\dmsnrnospace). This procedure is based on the
assumption that the signal perfectly follows a power law \tdm$\propto \nu^{-2}$, where \tdm is the time delay in the
burst arrival time at frequency $\nu$. However, FRBs often display a range of complex features which can be either
intrinsic to the source or introduced by propagation effects.
The discovery of multiple subcomponents showing a downward drift in frequency in bursts from the first repeating FRB (\RI) -- and later from other repeaters -- motivated the development of methods that maximise DM upon structure (\dmstruct) rather than S/N\cite{gajjar_highest-frequency_2018, hessels_frb_2018}. As $\mathrm{DM}_{\mathrm{S/N}}$ assumes that the signal can be completely described by a $\nu^{-2}$ power law, \dmstruct is more likely to represent the actual dispersive effect\cite{oostrum_repeating_2020}.
If a burst shows a single component, the computed \dmstruct is equivalent to \dmsnrnospace.
Hence, we report \dmstruct for all Apertif bursts, which was determined using a modified version of {\texttt {DM\_PHASE}}\footnote{\url{https://www.github.com/DanieleMichilli/DM_phase}}. 
We define the best Apertif DM by computing the median \dmstruct of bursts that were detected with S/N>20.
We obtain  \dmanospace=348.75$\pm$0.12\pccm, where the errors represent the median absolute deviation. \dma is consistent with values previously reported in the literature, and we use this value to create the dynamic spectra of the bursts shown in \ref{fig:Apertif_bursts}. As shown on the top panel of \ref{fig:Apertif_burst_properties}, the DM appears constant with phase, and the 1$\sigma$ errors of most bursts are consistent with 348.75\pccmnospace, with the exception of bursts A08, A19, A24 and A31.
The difference in DM could be explained by the presence of subbursts that are not resolved in time, as reveals a visual inspection of the pulse profiles dedispersed to 348.75\pccmnospace.

We first computed the best DM for the LOFAR bursts with the \textsc{Psrchive} command \texttt{pdmp}\cite{hotan_psrchive_2004}. This searches the DM that maximises S/N. We can apply this to LOFAR bursts since the bursts at those frequencies do not show any apparent complex time-frequency structure. With this technique we found initial DM values to which we applied later corrections.
By dividing each LOFAR pulse profile into multiple subbands and fitting them to a scattered Gaussian, we recovered the Gaussian centers at each frequency and applied an additional \tdm$\propto\nu^{-2}$ correction to align them in time. The \texttt{pdmp} DM is overestimated with respect to the revised value.
Another aspect we need to take into consideration is the constant that is used to compute a DM from the frequency-dependent burst time delay, \kdm~in \ukdm. In \textsc{Psrchive}, \kdm=4.15, whereas \textsc{PRESTO} and the structure maximising DM algorithm use \kdm=1/0.241. Although both these values are an approximation of the actual \kdm\cite{kulkarni_dispersion_2020}, we corrected the revised LOFAR DM values to have the same \kdm=1/0.241 as the Apertif bursts to be consistent within our reported values.
The final DMs are shown in the last column of \ref{tab:LOFAR_bursts}.
The best LOFAR DM, \dml, is defined as the average DM of all bursts with S/N>20. We find \dml=349.00$\pm$0.02\pccm, with the error reporting the standard deviation. We used \dml to dedisperse all LOFAR bursts and stack them in order to increase the S/N and later compute the average scattering timescale.

\subsection{Sub-pulse drift rate} 
\label{sec:sm:drift}

Several FRBs are now known to exhibit downward drifting sub-pulses in which earlier sub-bursts arrive at higher frequencies. Thus far, the sign of this phenomenon is always the same and there is currently no example of upward drift within a burst.
The rate of the drift, $\dot{\nu}$,
is a function of frequency for the first known repeater, \RI\cite{hessels_frb_2018, josephy_chimefrb_2019, caleb_simultaneous_2020}.

In order to measure the sub-burst drift rate, we selected the Apertif bursts with multiple discernible components. We assumed that each sub-burst profile was well described by a Gaussian due to the lack of observed scattering tails above 350\,MHz\cite{the_chimefrb_collaboration_chimefrb_2019, marcote_repeating_2020, chawla_detection_2020}. After dedispersing each burst to its \dmstruct, we applied a least-squares fitting routine to each sub-burst pulse profile using a multi-component Gaussian with two, three, four, five, or twelve components. We identified the sub-components of each burst by eye and used them as initial parameters in the fit.
Next, we defined each sub-pulse time component by taking $2\sigma$ around the peak of the fitted Gaussian.
We fitted the frequency structure of each component to a Gaussian.
From the Gaussians fitted in time and frequency, we obtained the sub-pulse centroids of each burst, which we next used to compute their drift rate by fitting a linear function. The results of applying this method to bursts A10, A17, A25, A47 and A53 are shown in \ref{fig:drift_rate} to illustrate different burst morphologies and number of components.

With this method, we obtain an average sub-pulse drift rate of $-39 \pm 7$ \mhzms\  at 1370\,MHz,
where we quote the standard  error  on the mean.
The standard deviation of the sample is 31\,\mhzms.
In \ref{fig:freq_drift_rate} we compare the value of the reported \FRB drift rates at 400\,MHz\cite{chawla_detection_2020} and 600\,MHz\cite{chamma_shared_2020} to the central frequency of the burst envelope. We observe that the average drift rate amplitude increases towards lower frequencies.

We quantify the drift rate evolution by fitting the reported values to a power-law $\dot{\nu} = k_p\nu^\gamma $, with $k_p$ a constant and $\gamma$ the power-law index, and to a linear function $\dot{\nu} = k_l\nu$ with $k_l$ a constant through a $\chi^2$ minimisation. The frequency $\nu$ is in MHz and its derivative $\dot{\nu}$ is in \mhzms. Both models are defined so that there is no turnover of the drift rate from negative to positive at a frequency $\nu>0$, since we do not expect this to be physically possible.
A least squares minimisation fit to the power-law gives $\gamma = 0.7\pm0.4$ and $k_p = -0.2\pm0.6$, and a fit to the
linear function gives $k_l=-(2.9\pm0.4)\times10^{-2}$. By scaling the fitted functions to the frequency of the LOFAR HBA, we would expect the drift rate to be around $\sim -6$\mhzms at 150\,MHz, although the apparent lack of multiple components in the LOFAR bursts do not allow us to confirm this.

\end{document}